%% file: main.tex
\documentclass[aps,prl,superscriptaddress,reprint]{revtex4-2}
\usepackage{amsmath}
\usepackage{amsfonts}
\usepackage{amssymb}
\usepackage{physics}
\usepackage{arydshln}
\usepackage{array}
\usepackage{hyperref}
\usepackage{graphicx}
\usepackage{adjustbox}
\usepackage{caption}
\usepackage{subcaption}
\usepackage{tabularx}
\usepackage{booktabs}
\usepackage{color}
\usepackage[T1]{fontenc}
\usepackage[capitalize]{cleveref}
\usepackage{multirow}
\graphicspath{{./figures/}}

\begin{document}
\title{Uncovering Exotic Paired States in the 2D Spin-Imbalanced Fermi Gas with Neural Wave Functions}
\author{Wan Tong Lou}
\author{Gino Cassella}
\author{Andres Perez Fadon}
\author{Halvard Sutterud}
\affiliation{
  Department of Physics,
  Imperial College London,
  South Kensington Campus,
  London SW7 2AZ,
  United Kingdom}
\author{David Pfau}
\affiliation{
  DeepMind,
  6 Pancras Square,
  London N1C 4AG,
  United Kingdom}
\affiliation{
  Department of Physics,
  Imperial College London,
  South Kensington Campus,
  London SW7 2AZ,
  United Kingdom}
\author{James S.\ Spencer}
\affiliation{
  DeepMind,
  6 Pancras Square,
  London N1C 4AG,
  United Kingdom}
\author{Johannes Knolle}
\affiliation{
  Department of Physics,
  Imperial College London,
  South Kensington Campus,
  London SW7 2AZ,
  United Kingdom}
\affiliation{
  Department of Physics TQM,
  Technische Universit\"{a}t M\"{u}nchen,
  James-Franck-Stra{\ss}e 1,
  D-85748 Garching,
  Germany}
\affiliation{
  Munich Center for Quantum Science and Technology (MCQST),
  80799 Munich,
  Germany}
\author{W.M.C.~Foulkes}
\affiliation{
  Department of Physics,
  Imperial College London,
  South Kensington Campus,
  London SW7 2AZ,
  United Kingdom}

\date{\today}

\begin{abstract}
We study the zero-temperature phase diagram of the 2D spin-imbalanced Fermi gas with short-ranged attractive interactions using the recently developed neural network variational Monte Carlo method with the AGPs FermiNet Ansatz~\cite{pfauAbInitioSolutionManyElectron2020,louNeuralWaveFunctions2024}. The Fulde-Ferrell-Larkin-Ovchinnikov phase is observed in the weakly interacting BCS limit and a polarised superfluid is seen in the strongly interacting BEC limit.
When the interactions are strong, the minority-spin momentum density is reduced almost to zero in the momentum-space region occupied by the unpaired majority-spin electrons.
When the interactions are very strong, phase separation occurs, with regions containing bosonic pairs and unpaired regions occupied by the remaining majority-spin particles.
In addition, we observe translational symmetry breaking at intermediate interaction strengths, where the system forms an exotic crystal of Cooper pairs in a Fermi fluid of unpaired majority-spin particles.
We provide a possible explanation for the formation of the crystalline phase, explain the origins of the k-space momentum-density hole when the pairs are tightly bound, and discuss how our approach opens new directions for future work.
\end{abstract}

\maketitle

\paragraph{Introduction---}\label{section:introduction}

\input{sections/introduction}

\paragraph{Neural Network Variational Monte Carlo---}\label{section:nnvmc}
\input{sections/nnvmc}

\paragraph{Results---}\label{section:results}
\input{sections/results}

\paragraph{Discussion---}\label{section:discussion}
\input{sections/discussion}

\paragraph*{Acknowledgments---}
\input{sections/acknowledgments}

\bibliographystyle{apsrev4-2}
\bibliography{main}

\clearpage
\newpage
\onecolumngrid{}
\begin{center}
  \section{End Matter}
\end{center}
\twocolumngrid{}
\label{section:end_matter}
\input{sections/end_matter}

\clearpage
\newpage
\appendix
\onecolumngrid{}
\section{Supplemental Materials}
\input{sections/appendix}

\end{document}

%% file: sections/introduction.tex
Dilute ultracold Fermi gases provide an ideal platform to explore the physics of spin-imbalanced superfluidity.
With their high-precision tunability of both the interaction strength (through Feshbach resonances~\cite{chin2010}) and the population imbalance, they serve as a clean and controllable model system~\cite{chenBCSBECCrossover2005,randeriaBCSBECCrossover2012}.
In particular, the two dimensional spin-imbalanced Fermi gas (2D SIFG), in which fermions of opposite spin feel short-ranged attractive interactions, is a subject of intense theoretical~\cite{fooDiffusionMonteCarlo2019,feiguinPairCorrelationsSpinImbalanced2009,moreoTwodimensionalNegativeUHubbard1991,chenExploringExoticSuperfluidity2007,hePhaseDiagramCold2008,koponenFFLOState12008,wolakPairingTwodimensionalFermi2012} and experimental~\cite{ongSpinImbalancedQuasiTwoDimensionalFermi2015,mitraPhaseSeparationPair2016} interest.
The enhanced quantum fluctuations in two dimensions are expected to stabilize exotic phases such as the Fulde–Ferrell–Larkin–Ovchinnikov (FFLO) state~\cite{fuldeSuperconductivityStrongSpinExchange1964,Larkin:1964wok}, where Cooper pairs carry a finite center-of-mass momentum, and the Sarma phase~\cite{sarmaInfluenceUniformExchange1963}, where gapless fermionic excitations coexist with the superfluid.


The quantitative theoretical description of the 2D SIFG at $T$$=$$0$ presents a formidable challenge of quantum many-body physics, especially in the crossover regime where the Bardeen-Cooper-Schrieffer (BCS) superfluid becomes a Bose-Einstein condensate (BEC).
Mean-field approaches, which are relatively reliable in 3D, are known to be unable to capture the complex correlations in strongly-interacting 2D systems, even in the unpolarized case~\cite{randeriaBoundStatesCooper1989,randeriaSuperconductivityTwodimensionalFermi1990,heQuantumFluctuationsBCSBEC2015,parishQuasiOneDimensionalPolarizedFermi2007,hePhaseDiagramCold2008}.
Several beyond-mean-field techniques have been used to study the 2D SIFG~\cite{fooDiffusionMonteCarlo2019,feiguinPairCorrelationsSpinImbalanced2009,wolakPairingTwodimensionalFermi2012}, but the 1D and 3D cases have been better explored and are better understood~\cite{batrouniExactNumericalStudy2008,taylorSpinpolarizedFermiSuperfluids2007}.
Numerical approaches often have limitations.
For example, density matrix renormalization group (DMRG) calculations are restricted to quasi-1D geometries~\cite{potapovaDimensionalCrossoverMultileg2023,feiguinPairCorrelationsSpinImbalanced2009}, and conventional quantum Monte Carlo (QMC) methods are limited by the expressivity of their Ans\"atze and may be hampered by the fermionic sign problem~\cite{lohSignProblemNumerical1990,troyerComputationalComplexityFundamental2005,yiTwodimensionalPolarizedSuperfluids2024}.
In this context, the recently developed neural network variational Monte Carlo (NNVMC) method provides a promising alternative~\cite{carleoSolvingQuantumManybody2017}.

In this Letter, we extend the NNVMC methodology to allow quantitative studies of spin-imbalanced paired systems and investigate the ground-state properties of the 2D SIFG across the whole BCS-BEC crossover regime.
In particular, we utilize the AGPs FermiNet Ansatz~\cite{louNeuralWaveFunctions2024}, derived from the FermiNet architecture~\cite{pfauAbInitioSolutionManyElectron2020,spencerBetterFasterFermionic2020}, to map out different $T$$=$$0$ phases of the 2D SIFG at fixed polarization by computing its ground-state properties.
Phases are characterized by their particle density, superfluid order parameter, and momentum distribution.
At fixed polarization, we observe a variety of different phenomena as the strength of the interaction is varied.

To help understand the rich variety of results, we divide the systems studied into three regimes, (I), (II), and (III), depending on interaction strength.
In the weakly interacting BCS regime, labeled (I), the FFLO phase is observed for a wide range of interaction strengths, consistent with previous work on lattice models~\cite{vitaliExoticSuperfluidPhases2022,yiTwodimensionalPolarizedSuperfluids2024} and diffusion Monte Carlo results~\cite{fooDiffusionMonteCarlo2019}.
In the strongly interacting BEC regime, labeled (III), we observe a phase-separated Bose-Fermi mixture~\cite{taylorSpinpolarizedFermiSuperfluids2007,pieriTrappedFermionsDensity2006,shinRealizationStronglyInteracting2008}.
The majority-spin momentum density consists of an almost fully occupied Fermi circle, the area of which matches the density of unpaired majority-spin fermions, surrounded by a diffuse halo arising from the tightly bound minority-majority pairs.
The minority-spin momentum density is almost zero within the Fermi circle of the unpaired majority-spin fermions, with the same diffuse halo outside.
We account for these results within a mean-field BCS calculation.
Lastly, in the intermediate crossover regime, labeled (II), the real-space particle density reveals a previously unknown translational-symmetry-broken phase in which the paired fermions form a crystal lattice, surrounded by a sea of excess unpaired fermions.
Our results provide new insight into the intricate phase diagram of this canonical model and demonstrate the potential of neural-network-based methods to unravel challenging open problems in the field of strongly correlated quantum matter.

%% file: sections/nnvmc.tex
The NNVMC method extends the variational Monte Carlo (VMC) method~\cite{foulkesQuantumMonteCarlo2001} by utilizing artificial neural networks as variational trial wave functions.
Such Ans\"atze are often known as neural quantum states~\cite{carleoSolvingQuantumManybody2017} or neural wave functions (NWF)~\cite{pfauAbInitioSolutionManyElectron2020,gao2023}.
The network parameters are optimized using the variational principle in combination with established machine learning algorithms~\cite{carleoSolvingQuantumManybody2017,pfauAbInitioSolutionManyElectron2020,chooFermionicNeuralnetworkStates2020,hermannDeepneuralnetworkSolutionElectronic2020,vonglehnSelfAttentionAnsatzAbinitio2023}.
External data are not required.
By leveraging the power of deep learning architectures, NWF offer a systematically improvable and highly expressive framework capable of capturing complex quantum correlations far beyond those accessible using traditional variational approaches~\cite{louNeuralWaveFunctions2024,cassellaDiscoveringQuantumPhase2023,cassellaNeuralNetworkVariational2024,pesciaNeuralnetworkQuantumStates2022,liFermionicNeuralNetwork2022b,luoBackflowTransformationsNeural2019}.

The optimal network parameters are found by minimizing the expectation value of the Hamiltonian.
This work uses the AGPs FermiNet network architecture~\cite{louNeuralWaveFunctions2024}, which extends the original FermiNet~\cite{pfauAbInitioSolutionManyElectron2020,spencerBetterFasterFermionic2020} and improves its ability to represent paired wave functions~\cite{bajdichPfaffianPairingWave2006,casulaCorrelatedGeminalWave2004,casulaGeminalWavefunctionsJastrow2003,bouchaudPairWaveFunctions1988}.
Detailed discussions of the original (Slater) FermiNet architecture can be found in Ref.~\cite{pfauAbInitioSolutionManyElectron2020} for normal systems and Ref.~\cite{louNeuralWaveFunctions2024} for superfluids.
For \emph{spin-imbalanced} systems with $N^\uparrow > N^\downarrow$, the AGPs FermiNet wave function we use takes the form~\cite{casulaGeminalWavefunctionsJastrow2003,bouchaudPairWaveFunctions1988,louNeuralWaveFunctions2024}:
\begin{widetext}
\begin{equation}
  \Psi^D_\text{AGPs FermiNet}\left(\mathbf{r}^\uparrow_1, \dots, \mathbf{r}^\uparrow_{p+u}, \mathbf{r}^\downarrow_{1}, \dots, \mathbf{r}^\downarrow_{p}\right)
  = \sum^D_k \det \left(
    \begin{array}{cccccc}
      \varphi^k(\mathbf{r}^\uparrow_1, \mathbf{r}^\downarrow_1)
      & \cdots
      & \varphi^k(\mathbf{r}^\uparrow_1, \mathbf{r}^\downarrow_{p})
      & \phi^{k\uparrow}_1(\mathbf{r}^\uparrow_1)
      & \cdots
      & \phi^{k\uparrow}_u(\mathbf{r}^\uparrow_1)
      \\
      \varphi^k(\mathbf{r}^\uparrow_2, \mathbf{r}^\downarrow_1)
      & \cdots
      & \varphi^k(\mathbf{r}^\uparrow_2, \mathbf{r}^\downarrow_{p})
      & \phi^{k\uparrow}_1(\mathbf{r}^\uparrow_2)
      & \cdots
      & \phi^{k\uparrow}_u(\mathbf{r}^\uparrow_2)
      \\
      \vdots
      & \ddots
      & \vdots
      & \vdots
      & \ddots
      & \vdots
      \\
      \varphi^k(\mathbf{r}^\uparrow_{p+u}, \mathbf{r}^\downarrow_1)
      & \cdots
      & \varphi^k(\mathbf{r}^\uparrow_{p+u}, \mathbf{r}^\downarrow_{p})
      & \phi^{k\uparrow}_1(\mathbf{r}^\uparrow_{p+u})
      & \cdots
      & \phi^{k\uparrow}_u(\mathbf{r}^\uparrow_{p+u})
    \end{array}
  \right) ,
  \label{eq:agpsferminet_unpaired}
\end{equation}
\end{widetext}
where $D$ is the number of determinants and we have defined $(N^\uparrow, N^\downarrow)\equiv (p+u, p)$, with $p=N^\downarrow$ being the number of spin-down electrons (and thus the maximum number of pairs) and $u = N^\uparrow - N^\downarrow$ the number of unpaired spin-up particles.
A more detailed discussion of the construction of FermiNet geminals, the architecture of the neural network, and how the AGPs FermiNet differs from the original (Slater) FermiNet can be found in Ref.~\cite{louNeuralWaveFunctions2024}.
The hyperparameters of the neural network and optimization algorithm can be found in the Supplemental Materials.

%% file: sections/results.tex
The Hamiltonian of a 2D SIFG is
\begin{equation}
  \hat{H} = -\frac{1}{2} \sum^{N^\uparrow}_i \nabla^2_{\mathbf{r}^\uparrow_i} - \frac{1}{2} \sum^{N^\downarrow}_i \nabla^2_{\mathbf{r}^{\downarrow}_i} - \sum^{N^\uparrow}_i \sum^{N^\downarrow}_j \frac{2v_0 \mu^2}{\cosh^2\left(\mu \abs{\mathbf{r}^\uparrow_i - \mathbf{r}^{\downarrow}_j}\right)},
  \label{eq:sifg_hamiltonian}
\end{equation}
where we use the modified P\"oschl-Teller function to model the short-range attactive interaction.
The parameters $v_0$ and $\mu$ are chosen to achieve the desired strength and range.
The strength of the interaction at low energies is best characterised by the $s$-wave scattering length, $a_s$, which is always positive because the two-body scattering problem in 2D has a bound state for any attractive interaction, no matter how weak.
More precisely, we characterize the interaction strength by the parameter $\eta = \ln(k_F a_s)$, where $k_F= \sqrt{2\pi n}$ is the Fermi wavevector of a \emph{spin-balanced} 2D Fermi gas with the same total particle number density as the spin-imbalanced system considered.
The weakly interacting (BCS) limit is attained as $a_s\rightarrow +\infty$ and $\eta\rightarrow +\infty$, while the strongly interacting (BEC) limit is attained as $a_s\rightarrow 0^+$ and $\eta\rightarrow -\infty$.
Although analytical expressions for $a_s$ and $\eta$ in terms of $\mu$ and $v_0$ exist in 3D~\cite{fluggePracticalQuantumMechanics1971}, no such expressions are available in 2D.
We therefore solve the 2D scattering problem numerically to find $a_s$ and $\eta$ for each choice of $\mu$ and $v_0$.

We study the SIFG in a 2D periodic square box of side $L$ and a constant density, setting $k_F=1$.
This corresponds to $r_s = \sqrt{2}$, where $r_s$, the radius of a circle that contains one particle on average, provides a convenient measure of the inter-particle distance.
To ensure that the range of the interaction is small compared to the inter-particle separation, we set $\mu = 24 k_F$, whilst varying $v_0$ to drive the system to various scattering lengths.
The chosen values of $v_0$ and the corresponding values of $\eta$ can be found in Table~(\ref{tab:v0_as}) of Appendix I in the End Matter.

\begin{figure}[ht!]
  \centering
  \begin{subfigure}[b]{0.49\columnwidth}
    \centering
    \includegraphics[width=\columnwidth]{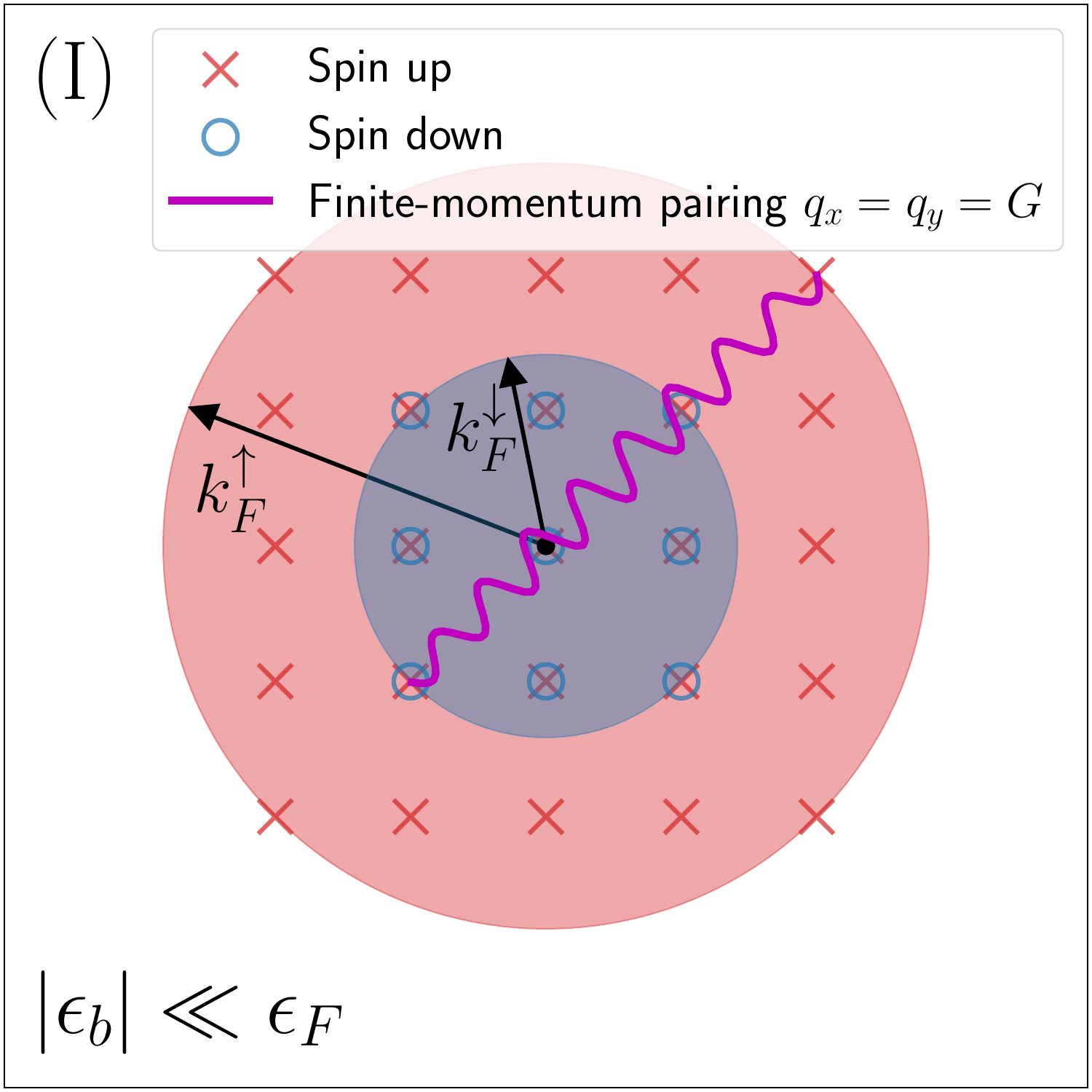}
  \end{subfigure}
  \hfill
  \begin{subfigure}[b]{0.49\columnwidth}
    \centering
    \includegraphics[width=\columnwidth]{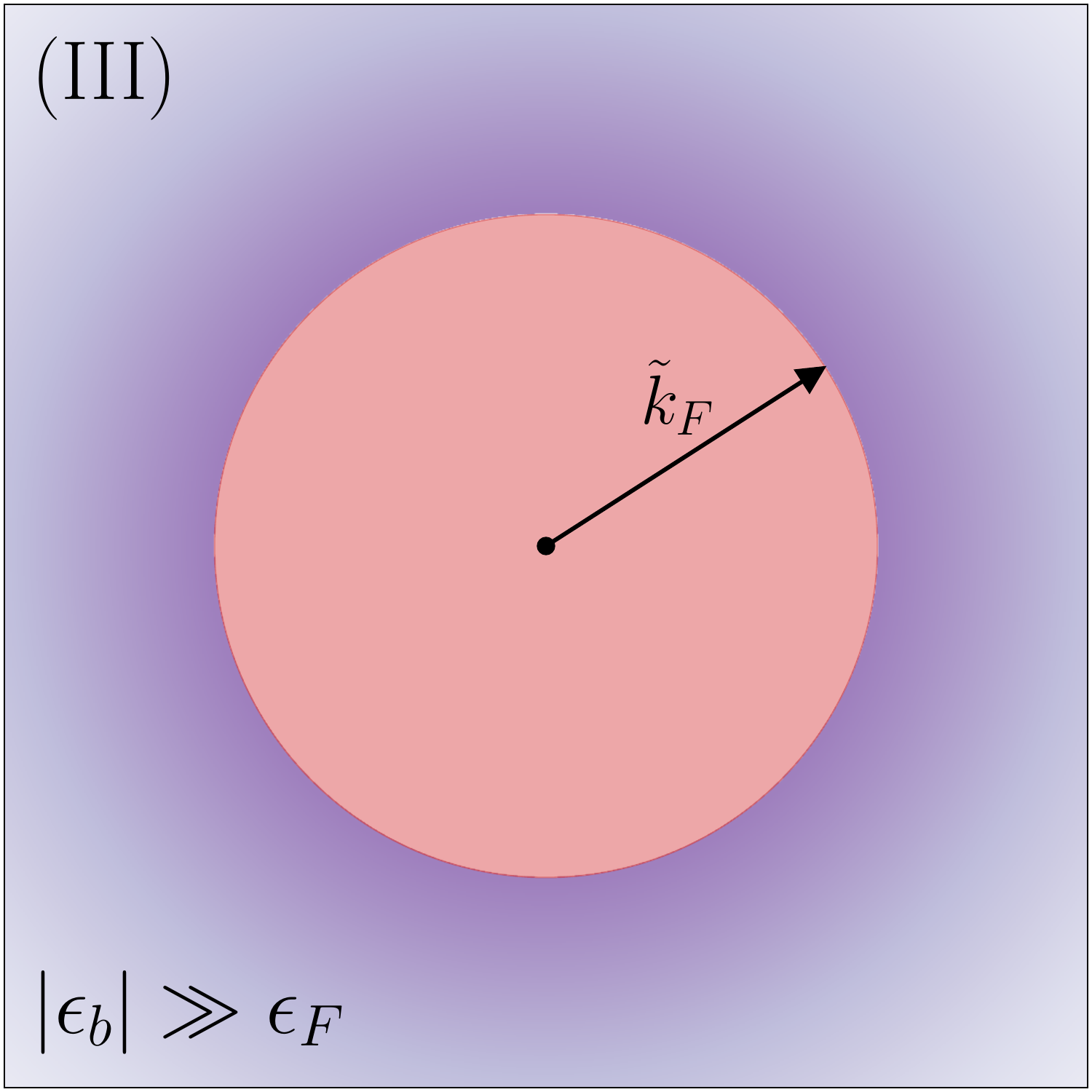}
  \end{subfigure}
  \caption{
    Mean-field (BCS) momentum density of the SIFG.
    \emph{Left}:
      regime (I), weakly interacting SIFG.
      The crosses and circles indicate the occupied spin-up and spin-down momentum states of the $(N^\uparrow, N^\downarrow) = (25, 9)$ SIFG.
      The wavy purple line shows the optimal pairing between opposite spins in the weakly interacting limit. The spin-up and spin-down Fermi circles of the infinite system are colored pink and blue-gray, respectively.
    \emph{Right}: regime (III), strongly interacting SIFG.
      The mauve region indicates the diffuse momentum density arising from the tightly-bound real-space pairs.
      The area of the Fermi circle with radius $\tilde{k}_F$ matches the density of unpaired majority-spin fermions.
  }
  \label{fig:bcs}
\end{figure}

The momentum density in a finite simulation cell with volume $\Omega$ is defined by~\cite{needsVariationalDiffusionQuantum2020}:
\begin{equation}
  n_{\mathbf{q}\alpha}
  = \expval{\hat{c}^\dagger_{\mathbf{q}\alpha} \hat{c}^{\phantom{\dagger}}_{\mathbf{q}\alpha}}
  = \frac{1}{\Omega} \int d\mathbf{r}\ e^{i\mathbf{q \cdot r}} \bar{\rho}_{\alpha}^{(1)}(\mathbf{r}) ,
  \label{eq:momentum_density}
\end{equation}
where $\alpha\in\{\uparrow,\downarrow\}$ is the spin species and $\bar{\rho}_{\alpha}^{(1)}(\mathbf{r})$ is the translationally averaged one-body density matrix (OBDM).
This may be accumulated during the Monte Carlo simulation~\cite{needsVariationalDiffusionQuantum2020}.

The red crosses and blue circles in the left-hand panel of Fig.~\ref{fig:bcs} indicated the occupied spin-up and spin-down momentum states of the periodic system of $(N^{\uparrow},N^{\downarrow}) = (25, 9)$ fermions simulated in this work. In the non-interacting limit, the ground state is a single Slater determinant of the corresponding plane waves. The wavy purple line shows the optimal pairing between opposite spins in the weakly interacting limit. The pink and blue-gray regions are the spin-up and spin-down Fermi circles of the corresponding infinite system.

\begin{figure*}[ht!]
  \centering
  \includegraphics[width=\textwidth]{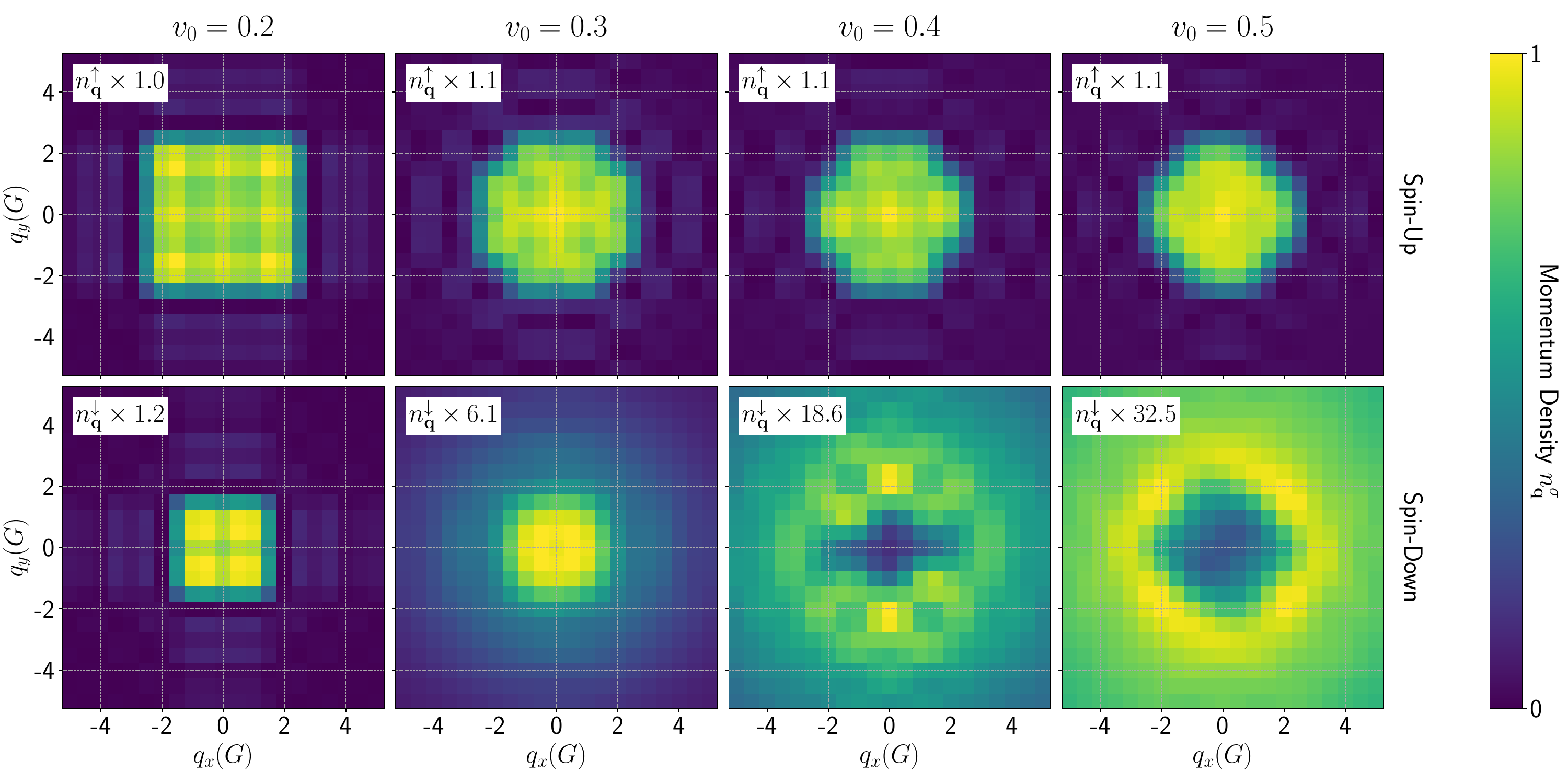}
  \caption{
    Momentum density of the $(N^\uparrow, N^\downarrow) = (25, 9)$ SIFG for a range of different interaction strengths across the BCS-BEC crossover.
    The top and bottom rows show the momentum densities of spin-up and spin-down particles, respectively.
    The strength of the interaction, $v_0$, increases from left to right.
    To aid comparison, all momentum densities are normalized such that their maximum value is equal to one.
    This is accomplished by multiplying the momentum density by the number in the top-left corner in each subfigure.
  }
  \label{fig:25_9-md}
\end{figure*}

Figure~\ref{fig:25_9-md} shows the momentum density obtained from our simulations at interaction strengths ranging from $v_0=0.2$ to $v_0=0.5$.
The minority-spin momentum densities are significantly smaller than the majority-spin momentum densities when the interaction is strong, so the results shown in different panels have been rescaled as shown in the legends to aid visual comparison.
The original numerical values and the full sets of results for this and a smaller system can be found in the Supplemental Material.

For $v_0 \le 0.2$ (region I), we observe slightly smeared out versions of the non-interacting momentum-space occupations from the left panel in Fig.~(\ref{fig:bcs}), consistent with the BCS picture of superfluidity.
At $v_0=0.3$ (region II), the four spin-up particles with the highest energy, originally located at $\mathbf{k}=(\pm 2G, \pm 2G)$, where $G = 2\pi/L$, move out to higher $|\mathbf{k}|$.
For $v_0 \geq 0.4$ (region III), the minority-spin (spin-down) momentum density starts to become depleted around $\abs{\mathbf{k}}=0$,  with the depletion becoming more pronounced as the strength of the interaction is increased further.
At $v_0=0.5$, the region in which the minority-spin momentum density is depleted roughly matches the region of $k$-space occupied by the unpaired majority-spin electrons.

Our results at strong interactions can be explained by considering a Bose-Fermi mixture~\cite{taylorSpinpolarizedFermiSuperfluids2007,pieriTrappedFermionsDensity2006,shinRealizationStronglyInteracting2008}.
As Table~\ref{tab:v0_as} shows, the pairing energy per particle, $\abs{\epsilon_b}$, greatly exceeds any other energy scale in the system for $v_0 > 0.3$ (region III).
In this regime, previous studies have suggested that the SIFG is best regarded as a mixture of strongly-bound composite bosons (each of which is a tightly-bound pair of fermions) and a sea of unpaired excess majority-spin fermions~\cite{taylorSpinpolarizedFermiSuperfluids2007,pieriTrappedFermionsDensity2006,shinRealizationStronglyInteracting2008}.
The unpaired majority spins occupy the lowest momentum states, while the paired majority- and minority-spin fermions, the wavefunctions of which vary on the very short length scale associated with the tightly-bound Cooper pairs, contribute a diffuse momentum density spread over a wide region of $k$-space.
The depletion of the minority-spin momentum density at small $\abs{\mathbf{k}}$ occurs because Pauli's exclusion principle prevents the paired majority spins from occupying states already occupied by unpaired majority spins.
Since a majority-spin fermion of momentum $\mathbf{k}$ pairs with a minority-spin fermion of momentum $-\mathbf{k}$, the minority-spin fermions are blocked from the same region of momentum space.
Within the Fermi surface of the unpaired majority-spin fermions, the pairs contribute very little momentum density in either spin channel.
These results are qualitatively consistent with BCS calculations in the thermodynamic limit, the results of which are illustrated in Fig.~(\ref{fig:bcs}). The left- and right-hand panels show the BCS momentum densities of both spin species in the weak and strong interaction limits.
A more detailed discussion of our BCS calculations can be found in the End Matter.

Next, we determine the pairing states and detect the existence of the FFLO phase by computing the momentum-resolved condensate fraction:
\begin{align}
  f^{\uparrow\downarrow}_{\mathbf{q}}
  &= \frac{\Omega^2}{N_\text{pair}}  \sum_\mathbf{k} \expval{\hat{c}^\dagger_{\mathbf{k}\uparrow}\hat{c}^\dagger_{\mathbf{q-k}\downarrow} \hat{c}^{\phantom{\dagger}}_{\mathbf{q-k}\downarrow} \hat{c}^{\phantom{\dagger}}_{\mathbf{k}\uparrow}}
  \notag \\
  &= \frac{\Omega}{N_\text{pair}} \int d\mathbf{r}\ e^{i\mathbf{q \cdot r}} \bar{\rho}_{\uparrow\downarrow}^{(2)}(\mathbf{r}) ,
  \label{eq:pair_momentum_density}
\end{align}
where $N_\text{pair} = \min(N^\uparrow,N^\downarrow)$ is the number of pairs in the system and $\bar{\rho}_{\uparrow\downarrow}^{(2)}(\mathbf{r})$ is the pairing part of the translationally averaged two-body density matrix (TBDM). Like the momentum density, this may be accumulated during the Monte Carlo simulation~\cite{needsVariationalDiffusionQuantum2020}.
For a conventional $s$-wave superfluid, only $f^{\uparrow\downarrow}_{\mathbf{q} = \mathbf{0}}$ is nonzero, indicating a condensation of zero-momentum Cooper pairs.
In contrast, phases with finite-momentum Cooper pairing, such as the FFLO phase, will have nonzero values of $f^{\uparrow\downarrow}_\mathbf{q}$ at the corresponding finite values of $\mathbf{q}$.
Hence, the momentum-resolved condensate fraction provides evidence for the existence of FFLO phases.

\begin{figure*}[ht!]
  \centering
  \includegraphics[width=\textwidth]{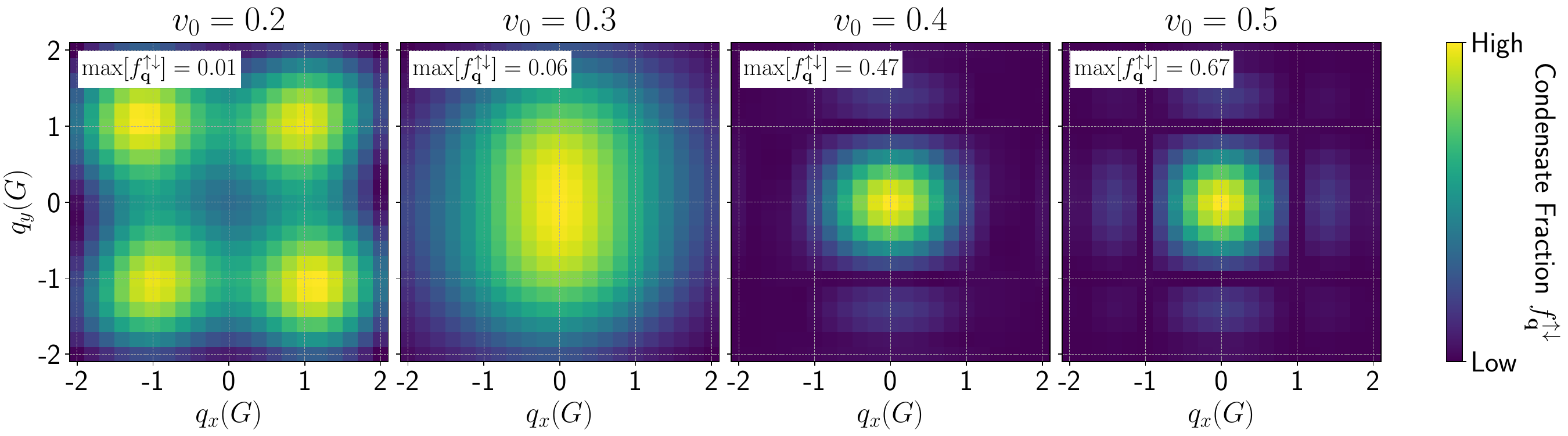}
  \caption{
    Condensate fraction of the $(N^\uparrow, N^\downarrow) = (25, 9)$ SIFG for a range of different interaction strengths across the BCS-BEC crossover.
    The condensate fraction in each subfigure has been rescaled to aid comparison, with the maximum condensate fraction shown in the top-left corner of each subfigure.
  }
  \label{fig:all-pair_momentum_density-25_9}
\end{figure*}

The condensate fraction of the $(25, 9)$ Fermi gas in momentum space for different strengths of interaction can be seen in Fig.~(\ref{fig:all-pair_momentum_density-25_9}).
At weak interaction strengths, $v_0 \lesssim 0.2$ (region I), the condensate fraction exhibits four clear peaks at $\abs{q_x} = G$ and $\abs{q_y} = G$, indicating the existence of finite-momentum Cooper pairs forming at the Fermi surface, connected by the four-fold rotational symmetry of the square simulation box.
This optimal pairing is indicated by the purple curly line in the left-hand panel in Fig.~(\ref{fig:bcs}).
All of the peaks are at least 6 standard deviations above zero, providing strong evidence for the presence of the FFLO phase.
On the other hand, at strong interactions with $v_0 \gtrsim 0.4$ (region III), the condensate fraction is dominated by the $\mathbf{q}=\mathbf{0}$ peak, with negligible signal elsewhere.
This indicates that the system has a condensate of zero-momentum Cooper pairs, making it a polarised $s$-wave superfluid.
In the intermediate regime where $v_0 \approx 0.3$ (region II), Fig.~(\ref{fig:all-pair_momentum_density-25_9}) does not allow an unambiguous identification of the phase to be made.
To investigate further, we need to look at other observables.

The \emph{exact} ground state of any finite system with a translationally invariant Hamiltonian and periodic boundary conditions must be translationally invariant.
Systems with broken translational symmetry, such as crystals, appear in nature via the mechanism of spontaneous symmetry breaking.
At weak interactions, \emph{e.g.}, $v_0 \lesssim 0.2$, we do indeed see a translationally invariant ground state.
However, if we compute the one-particle densities separately for different spin channels at intermediate interaction strengths, $v_0 \approx 0.3$, we observe a breaking of the translational symmetry. The Hamiltonian remains translationally invariant, so the role of the infinitesimal external perturbation that causes spontaneous symmetry breaking in real materials is being played by the optimization noise.

\begin{figure*}[ht!]
  \centering
  \includegraphics[width=\textwidth]{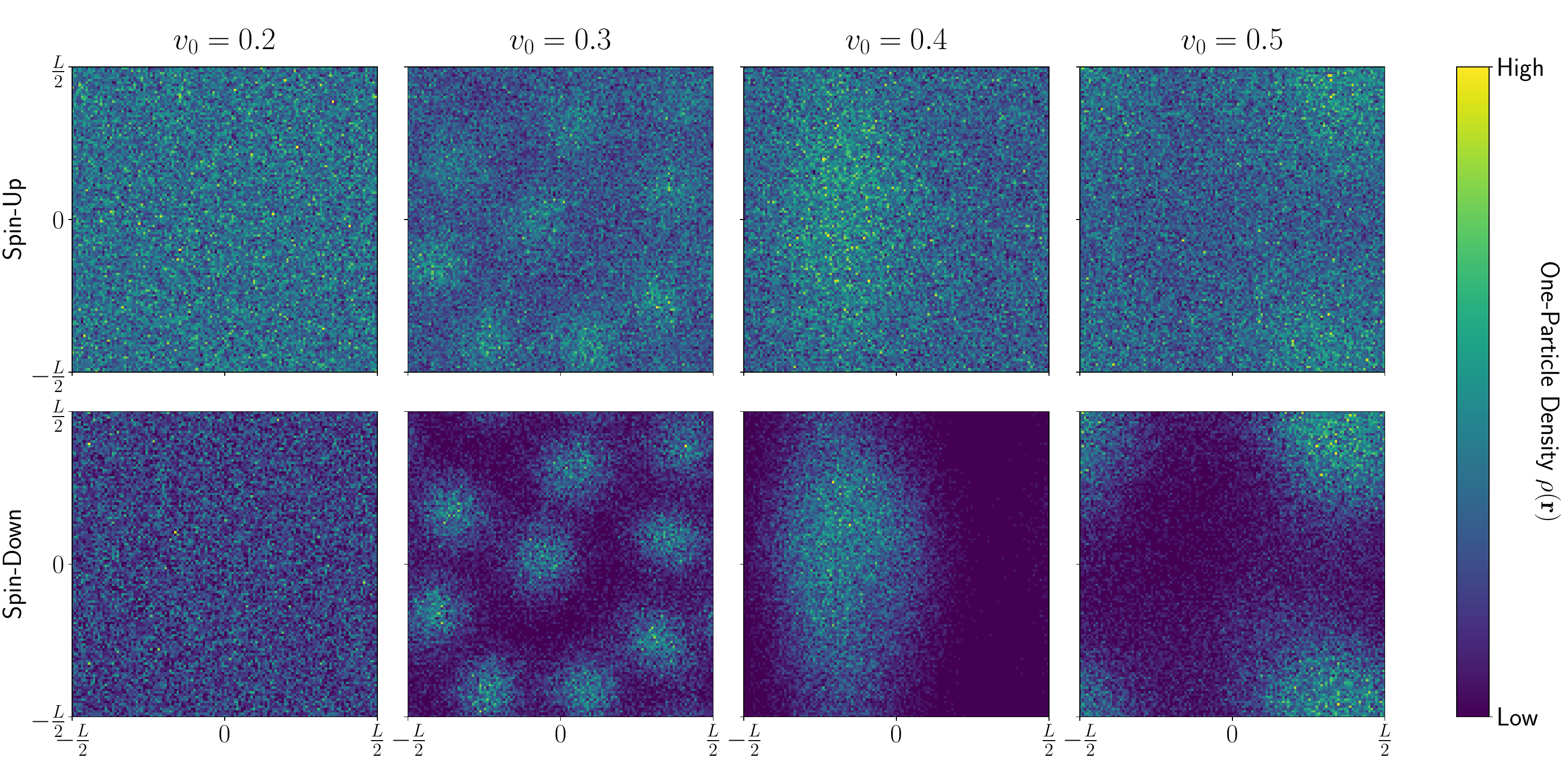}
  \caption{
    One-particle density of the $(N^{\uparrow},N^{\downarrow}) = (25, 9)$ SIFG for a range of different interaction strengths across the BCS-BEC crossover.
    The plots on the top row are the densities of spin-up particles and the plots on the bottom row are the densities of spin-down particles.
  }
  \label{fig:all-one_density-25_9}
\end{figure*}

The nature of the symmetry-broken state is shown in Fig.~(\ref{fig:all-one_density-25_9}), where the minority (spin-down) fermions are seen to be localized on a triangular lattice, disrupted by the square shape of the periodic simulation cell.
The number density of majority (spin-up) fermions increases close to the localized minority spins and is roughly uniform elsewhere.
The system has formed a crystal of Cooper pairs submerged in a sea of unpaired spin-up fermions.
This is very surprising, as we are seeing crystal formation from a system of fermions with \emph{purely attractive interactions}. Translational symmetry breaking in the uniform electron gas, as seen in Wigner crystallization, requires repulsive interactions.
To the best of our knowledge, the crystalline phase we observe has not been seen previously in the 2D SIFG, perhaps because it only exists in a small range of interaction strengths between $\eta\sim 0$ and $\eta\sim -1$.

The formation of a crystal could indicate that there exists an effective repulsion between the strongly bound pairs of fermions that prevents their collapse into a Bose liquid.
Indeed, if we increase the strength of the interaction further and push the system into the strongly interacting BEC limit, the dimers condense into a liquid and the system becomes phase separated in real space, as shown in Fig.~(\ref{fig:all-one_density-25_9}).
On the other hand, in the weakly interacting regime with $\eta > 0$, the attraction is not sufficiently strong to support the formation of bosonic dimers of fermions and the translation symmetry is not broken.
It is only in the intermediate crossover region that the balance between the attraction and the effective repulsion leads to the formation of the crystalline phase.

%% file: sections/discussion.tex
This work presented a study of the 2D two-component SIFG with short-range attractive interactions between opposite-spin fermions.
We chose two system sizes, $(N^\uparrow,N^\downarrow)=(13,5)$ (see Supplementary Material) and $(25,9)$, varied the strength of the interaction to drive the system to different scattering lengths, and investigated the ground-state phase diagram.
In the weakly interacting BCS limit, labeled (I), where $\eta = \ln(k_F a_s) \gtrsim 0.018$, we observe an FFLO phase with four-fold rotational symmetry originating from the symmetries of the square simulation cell in real space and the filled shells of occupied wave vectors in $k$-space.
In the strongly interacting BEC limit, labeled (III), where $\eta = \ln(k_F a_s) \lesssim -1.1$, the system is a standard $s$-wave superfluid but becomes phase separated in real space, with the pairs clustered into islands surrounded by unpaired majority-spin fermions.
In the far-BEC regime, where the binding energy per particle, $\abs{\epsilon_b}$, exceeds any other energy scale of the system, the momentum density of the minority spins becomes depleted within the region of $k$-space occupied by the unpaired majority-spin fermions, which act like a nearly non-interacting Fermi liquid.
This is qualitatively consistent with our mean-field BCS calculations described in the End Matter.

In the regime of intermediate interaction strength, labeled (II), the one-particle spin densities in real space exhibit translational symmetry breaking.
The strongly bound Cooper pairs localize and order to form a crystal, while the unpaired majority spins make up a nearly uniform background.
To the best of our knowledge, this $T=0$ phase of the SIFG has not previously been observed or predicted
\footnote{
  Although the system simulated is too small to allow us to reliably mitigate finite-size errors by twist averaging and/or size-extrapolation techniques, the form of the phase diagram remains the same when we reduce the (already small) system size or change the neural network architecture.
  This gives us confidence that the results are qualitatively robust.
}.

The observation of a crystal in a 2D Fermi gas is suprising as it suggests the emergence of effective repulsions between the bosonic dimers in a \emph{purely attractive} Fermi gas.
However, previous work based on diagrammatic perturbation theory has indeed shown that, in a strongly interacting attractive Fermi gas, there may indeed exist an effective repulsive interaction between the tightly bound pairs, mediated by the unpaired fermions~\cite{petrovSuperfluidTransitionQuasitwodimensional2003,bertainaBCSBECCrossoverTwoDimensional2011,heQuantumFluctuationsBCSBEC2015}.
On either side of the intermediate regime within which we observe spontaneous symmetry breaking, it seems that the attractive interaction is either too weak or too strong to allow the formation of a crystal.
Hence, the emergence of the crystalline phase requires a fine balance between the bare attraction and the effective repulsion.

%
%
The body of prior work using numerical methods to study the spin-imbalanced attractive Fermi gas in 2D is very limited and we hope that the new directions suggested by these results will inspire further study using theoretical, computational and experimental methods.

%% file: sections/acknowledgments.tex
We gratefully acknowledge the \href{www.gauss-centre.eu}{Gauss Centre for Supercomputing e.V.} for funding this project by providing computing time through the John von Neumann Institute for Computing (NIC) on the GCS Supercomputer JUWELS at J\"ulich Supercomputing Centre.
The authors acknowledge the use of resources provided by the Isambard-AI National AI Research Resource (AIRR).
Isambard-AI is operated by the University of Bristol and is funded by the UK Government’s Department for Science, Innovation and Technology (DSIT) via UK Research and Innovation; and the Science and Technology Facilities Council [ST/AIRR/I-A-I/1023].
J.K. acknowledges support from the Deutsche Forschungsgemeinschaft (DFG, German Research Foundation) under Germany’s Excellence Strategy–EXC–2111–390814868, DFG grants No. KN1254/1-2, KN1254/2-1, and TRR 360 - 492547816, as well as the Munich Quantum Valley, which is supported by the Bavarian state government with funds from the Hightech Agenda Bayern Plus.
H.S.’s PhD work was supported by the Aker Scholarship.
G.C.’s PhD work was supported by the UK Engineering and Physical Sciences Research Council (EP/T51780X/1).
A.P.F.’s PhD work is supported by the UK Engineering and Physical Sciences Research Council (EP/W524323/1).
W.T.L.'s PhD work was supported by the Imperial College President's Scholarship.

%
%

%% file: sections/end_matter.tex
\paragraph{Appendix I}\label{sec:v0_as}
In this section, we list the set of interaction strength parameter, $v_0$, and the corresponding values of $\eta=\ln(k_F a_s)$ in Table~(\ref{tab:v0_as}).
\begin{table}[ht!]
  \centering
  \begin{tabular}{lrlc}
    \toprule
    $v_0$ & $\eta$ \phantom{--} & $\abs{\epsilon_b}/\epsilon_F$ & Region \\
    \midrule
    0.10 &  4.367 & 4.061$\times 10^{-4}$ & \multirow{4}{*}{(I)} \\
    0.15 &  1.956 & 5.047$\times 10^{-2}$ & \\
    0.20 &  0.747 & 5.661$\times 10^{-1}$ & \\
    0.25 &  0.018 & 2.431 & \\
    \midrule
    0.30 & -0.470 & 6.457 & \multirow{2}{*}{(II)} \\
    0.35 & -0.821 & 13.04 & \\
    \midrule
    0.40 & -1.087 & 22.18 & \multirow{3}{*}{(III)} \\
    0.45 & -1.298 & 33.82 & \\
    0.50 & -1.465 & 47.19 & \\
    \bottomrule
  \end{tabular}
  \caption{
    Values of $v_0$ and the corresponding values of $\eta=\ln(k_F a_s)$, as well as the binding energy per particle $\abs{\epsilon_b}/\epsilon_F = 8e^{-2(\gamma + \eta)}$ at a fixed $\mu=24k_F$.
    $\gamma \approx 0.577$ is the Euler-Mascheroni constant.
  }
  \label{tab:v0_as}
\end{table}
The Fermi energy $\epsilon_F$ is the average Fermi energy, defined as
\begin{equation}
	\epsilon_F = \frac{\hbar^2 k_F^2}{2m} = \frac{\pi \hbar^2}{m\Omega}(N^\uparrow + N^\downarrow) ,
\end{equation}
and $\Omega$ is the area of the 2D simulation cell.
The binding energy is defined as
\begin{equation}
	\abs{\epsilon_b} = 8e^{-2(\gamma + \eta)} \epsilon_F .
\end{equation}
The rows in Table~(\ref{tab:v0_as}) are divided into three groups, as indicated by (I), (II) and (III), characterized by their ground-state qualitative behavior, \emph{e.g.}, the phase of the ground state.

\begin{figure}[ht!]
  \centering
  \begin{subfigure}[b]{0.49\columnwidth}
    \includegraphics[width=\columnwidth]{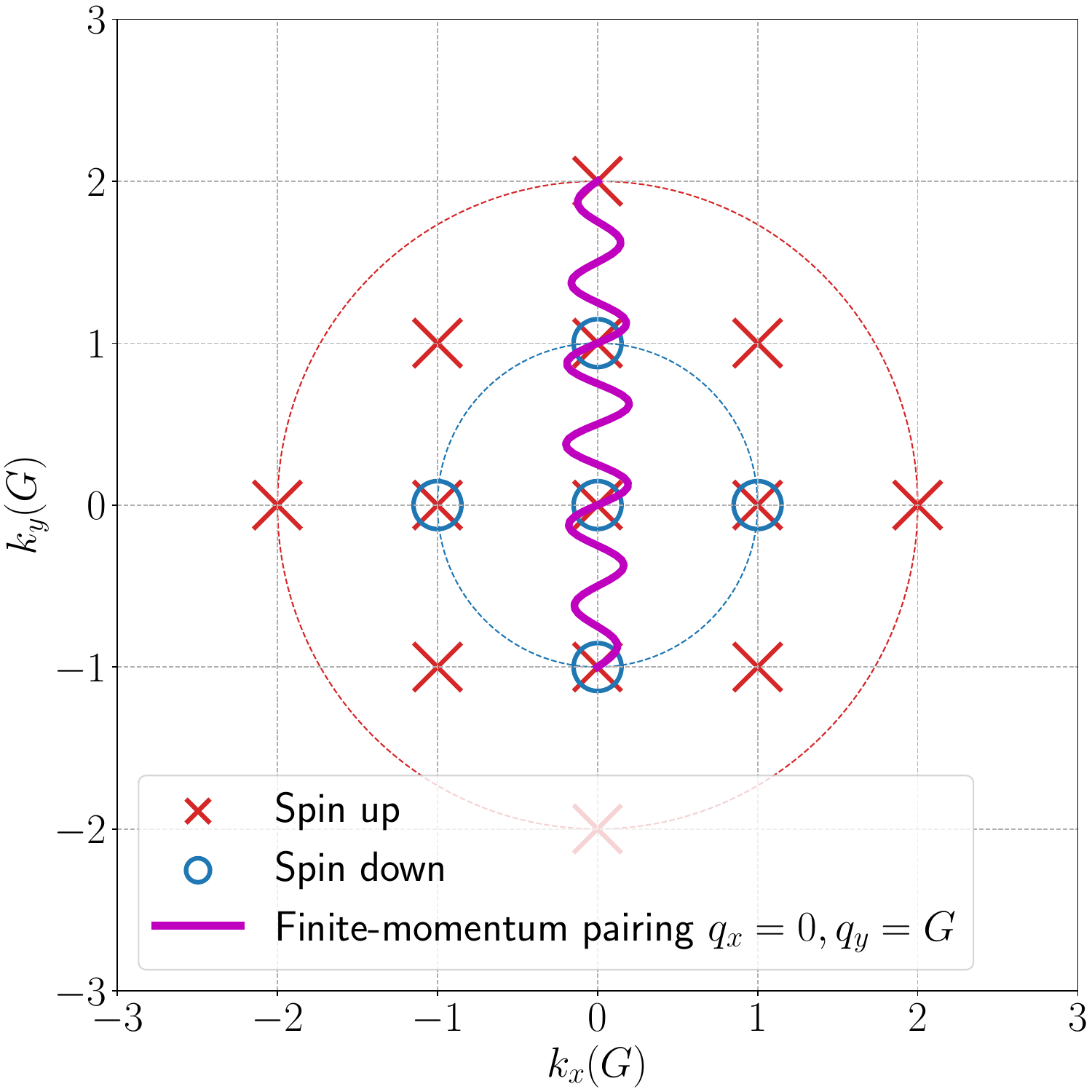}
    \caption{$(N^\uparrow, N^\downarrow) = (13, 5)$}
  \end{subfigure}
  \hfill
  \begin{subfigure}[b]{0.49\columnwidth}
    \includegraphics[width=\columnwidth]{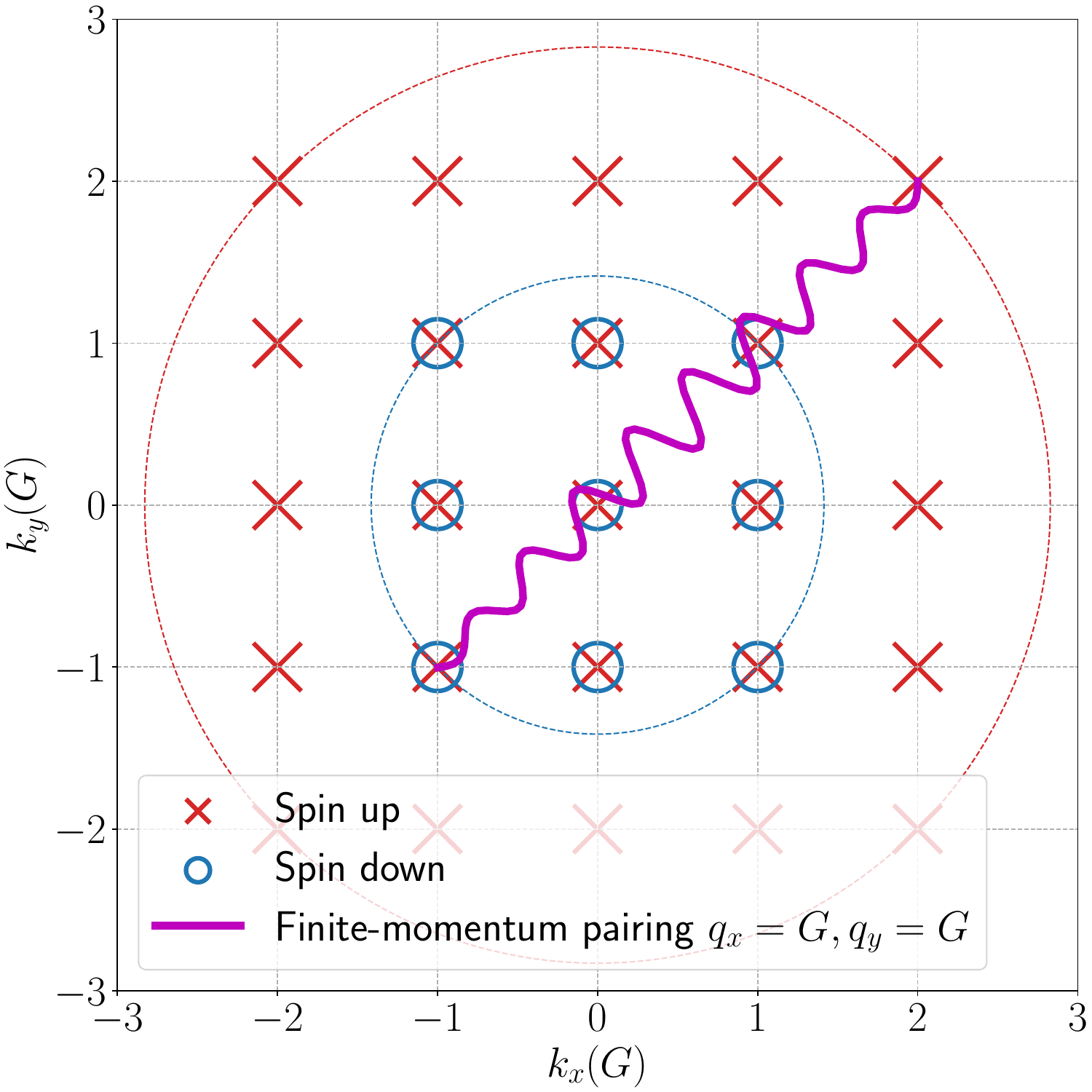}
    \caption{$(N^\uparrow, N^\downarrow) = (25, 9)$}
  \end{subfigure}
  \caption{
    Occupied momentum states of the SIFG in our simulations.
    The red crosses and the blue circles represent spin-up (majority) and spin-down (minority) particles, respectively.
    The big red and blue dashed circles represent the ``Fermi surfaces''.
    The yellow curly lines denote one of the optimal finite-momentum pairing in each case.
  }
  \label{fig:discrete_fermi_surfaces}
\end{figure}

We simulated two different system sizes: $(N^\uparrow, N^\downarrow) = (13, 5)$ and $(N^\uparrow, N^\downarrow) = (25, 9)$.
The corresponding Fermi surfaces (occupied momentum states) in the non-interacting limit for both systems can be seen in Fig.~(\ref{fig:discrete_fermi_surfaces}).

\paragraph{Appendix II}\label{sec:bcs}
In this section, we present the mean-field BCS calculations of the SIFG in the strongly interacting limit.
For a two-component spin-imbalanced system with unequal checmical potentials, $\mu_\uparrow$ and $\mu_\downarrow$, the single particle energies are given by
\begin{equation}
  \begin{aligned}
    \xi_{\mathbf{k+q}/2, \uparrow} &= \epsilon_{\mathbf{k+q}/2} - \mu_\uparrow \\
    \xi_{\mathbf{-k+q}/2, \downarrow} &= \epsilon_{\mathbf{-k+q}/2} - \mu_\downarrow
  \end{aligned}
\end{equation}
where $\epsilon_{\mathbf{k}} = \frac{\hbar^2 k^2}{2m}$ is the single particle dispersion.
From BCS theory, the quasiparticle energies are given by
\begin{equation}
  \begin{aligned}
    E_{\mathbf{kq}\uparrow}
    &= \sqrt{\xi^{+2}_{\mathbf{kq}} + \Delta_\mathbf{q}^2} + \xi^{-}_{\mathbf{kq}} \\
    E_{\mathbf{kq}\downarrow}
    &= \sqrt{\xi^{+2}_{\mathbf{kq}} + \Delta_\mathbf{q}^2} - \xi^{-}_{\mathbf{kq}}
  \end{aligned} .
\end{equation}
where $\xi^{\pm}_{\mathbf{kq}}$ are defined as
\begin{equation}
	\begin{aligned}
	  \xi^{\pm}_{\mathbf{kq}} = \frac{1}{2} (\xi_{\mathbf{k}+\mathbf{q}/2\uparrow} \pm \xi_{\mathbf{-k}+\mathbf{q}/2\downarrow})
	\end{aligned} .
\end{equation}
The spin-resolved momentum densities at $T=0$ from BCS theory is given by
\begin{multline}
	n_{\mathbf{kq}\sigma} = \frac{1}{2} \left(
	  1 - \frac{\xi^{+}_{\mathbf{kq}}}{\sqrt{\xi^{+2}_{\mathbf{kq}} + \Delta_\mathbf{q}^2}}
	\right) \\
	\times \left[
	  1 - \Theta(-E_{\mathbf{kq}\uparrow}) - \Theta(-E_{\mathbf{kq}\downarrow})
	\right] + \Theta(-E_{\mathbf{kq}\sigma})
	\label{eq:bcs_spin_resolved_md}
\end{multline}
where $\Theta(x)$ is the Heaviside step function.

\begin{figure}[ht!]
  \centering
  \begin{subfigure}[b]{0.49\columnwidth}
    \includegraphics[width=\columnwidth]{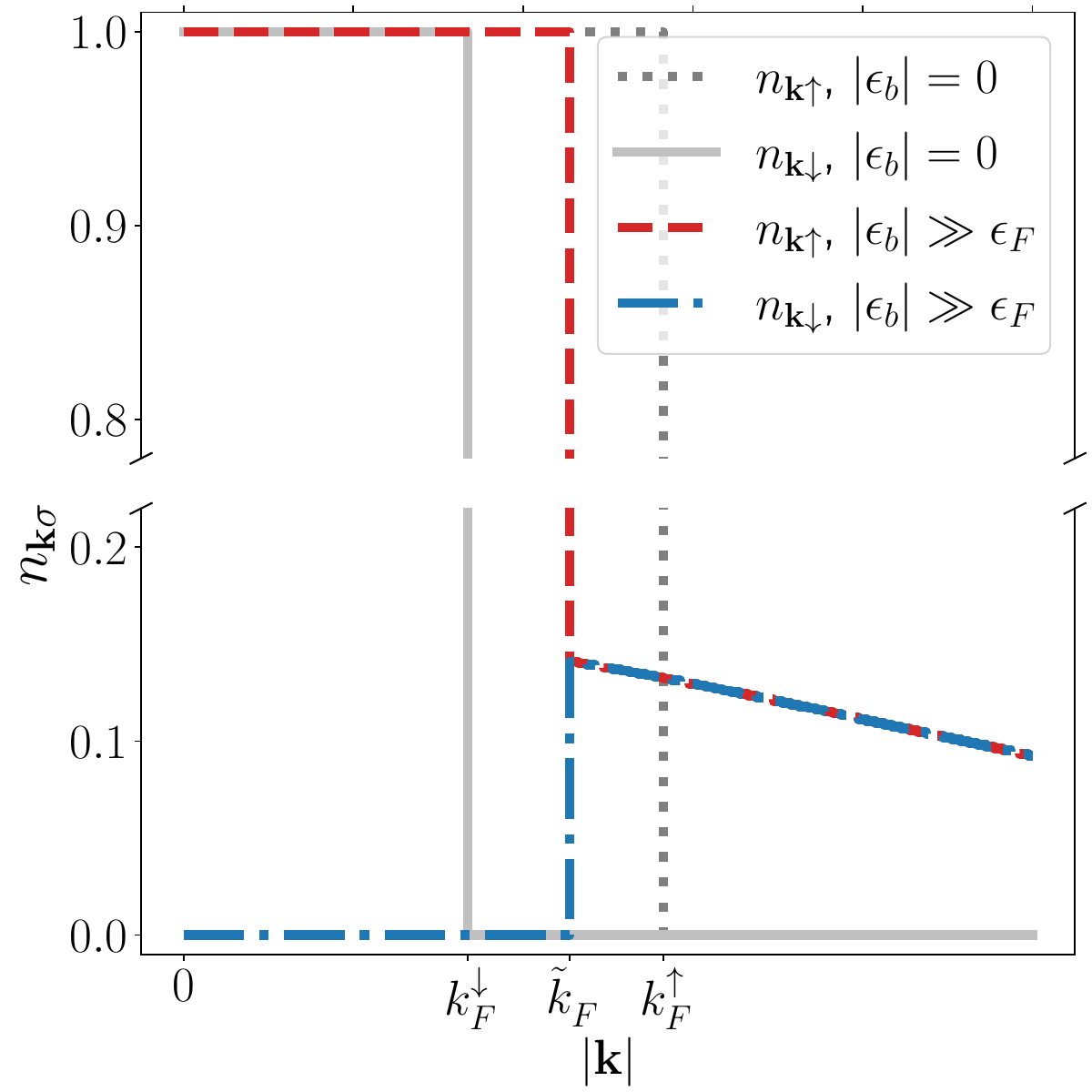}
    \caption{BCS spin-resolved momentum densities with zero/strong interactions.}
  \end{subfigure}
  \hfill
  \begin{subfigure}[b]{0.49\columnwidth}
    \includegraphics[width=\columnwidth]{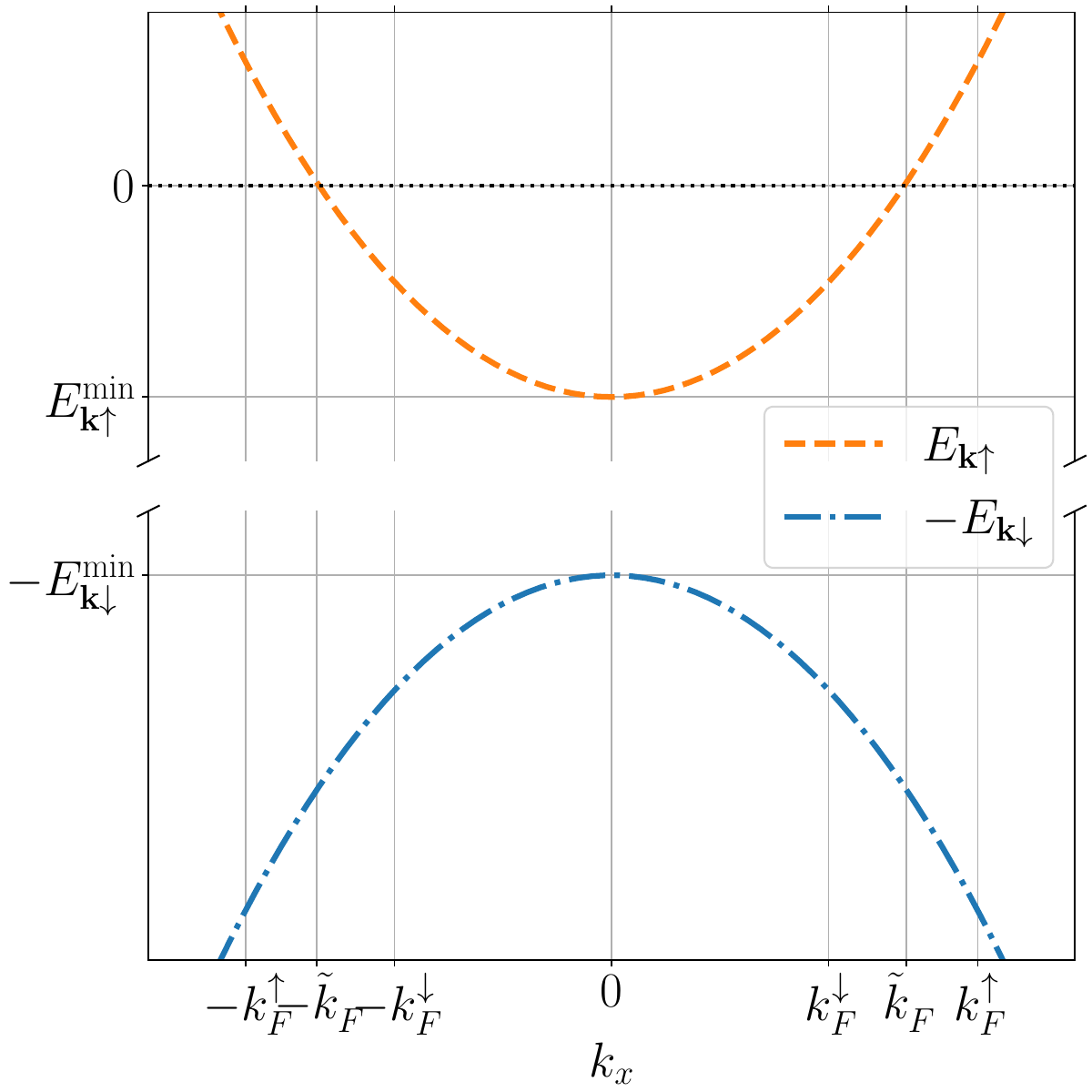}
    \caption{BCS band structure of the strongly interacting SIFG.}
  \end{subfigure}
  \caption{The spin-resolved momentum density and the band structure of the SIFG from the mean-field BCS calculations.}
  \label{fig:bcs_md_and_bands}
\end{figure}

In the strongly interacting limit, where the binding energy is much greater than the average Fermi energy, $\abs{\epsilon_b} \gg \epsilon_F$, the most energetically favorable pairing state is $\abs{\mathbf{q}}=0$, \emph{i.e.}, zero-momentum pairing state.
To get the results in Fig.~(\ref{fig:bcs}), we set $\mathbf{q}=0$ in Eq.~(\ref{eq:bcs_spin_resolved_md}) while keeping the gap parameter, $\Delta$, fixed, and vary the chemical potentials such that the following condition is satisfied for each spin channel,
\begin{equation}
	\sum_\mathbf{k} n_{\mathbf{k}\sigma} = N^\sigma
	\qquad
	\sigma \in \{\uparrow,\downarrow\} .
\end{equation}
In the regime where $\abs{\epsilon_b} \gg \epsilon_F$, the majority and minority spin chemical potentials become positive and negative, respectively, \emph{i.e.}, $\mu_\uparrow > 0, \mu_\downarrow < 0$.
This gives rise to the ``hole'' in the minority-spin momentum density and a shrinked majority Fermi surface at
\begin{equation}
  \tilde{k}_F = \sqrt{k_F^{\uparrow 2} - k_F^{\downarrow 2}} \propto \sqrt{N^\uparrow - N^\downarrow} ,
  \label{eq:tilde_k_F}
\end{equation}
in Fig.~(\ref{fig:bcs}), \emph{i.e.}, the new Fermi surface is formed by the unpaired majority-spin particles.
The quasiparticle energies and the spin-resolved momentum densities at zero and at a very strong interaction are plotted in Fig.~(\ref{fig:bcs_md_and_bands}).

%% file: sections/appendix.tex
\section{Experimental Setup}
\label{appendix}

In this section, we report the FermiNet setup and hyperparameters used in this work.

\subsection{FermiNet}\label{appendix:ferminet}
The periodic version of the FermiNet implemented by Cassella \emph{et al.}~\cite{cassellaDiscoveringQuantumPhase2023}, which can be found in the FermiNet repository \cite{ferminet_github}, was used as the basis for the AGPs code.
The small modifications required to support AGPs wave functions were made using the JAX Python library \cite{GoogleJax2024}.
The exact form of the FermiNet geminal we use in Eq.~(\ref{eq:agpsferminet_unpaired}) is given by~\cite{louNeuralWaveFunctions2024}.
For simplicity, the APGs FermiNet ``geminal'', $\varphi^{k}(\mathbf{r}^\uparrow_i, \mathbf{r}^{\downarrow}_j)$, has been written as a function of the coordinates of two particles only. In reality, it also depends in an exchange-symmetric manner on the coordinates of the other particles~\cite{louNeuralWaveFunctions2024}: $\varphi^{k}(\mathbf{r}^\uparrow_i, \mathbf{r}^{\downarrow}_j) \equiv \varphi^{k}(\mathbf{r}^\uparrow_i, \mathbf{r}^{\downarrow}_j; \{\mathbf{r}^\uparrow_{/i}\};\{\mathbf{r}^{\downarrow}_{/j}\})$, where $\{\mathbf{r}^{\sigma}_{/i}\}$ denotes the set of all particle coodinates of spin $\sigma$ excluding those of electron $i$.
The unpaired spin-up electrons occupy FermiNet ``orbitals'' of the form~\cite{pfauAbInitioSolutionManyElectron2020} $\phi^{k\uparrow}_i(\mathbf{r}^\uparrow_{j}) \equiv \phi^{k\uparrow}_i(\mathbf{r}^\uparrow_{j};\{\mathbf{r}^\uparrow_{/j}\}; \{\mathbf{r}^{\downarrow}\})$, which are invariant under permutations of the coordinates in $\{\mathbf{r}^{\uparrow}_{/j}\}$ and $\{\mathbf{r}^{\downarrow}\}$.
The FermiNet geminals and orbitals are many-body analogues of two-particle BCS pairing orbitals and one-particle Hartree-Fock orbitals, respectively.

We use the FermiNet geminal proposed in Ref.~\cite{louNeuralWaveFunctions2024} in our work.
In particular, we use the FermiNet geminal with both one- and two-electron streams as this yields better convergence and accuracy than the FermiNet geminal with one-electron streams only:
\begin{equation}
  \varphi^{k}(\mathbf{r}^\uparrow_i, \mathbf{r}^{\downarrow}_j; \{\mathbf{r}^\uparrow_{/i}\};\{\mathbf{r}^{\downarrow}_{/j}\})
  \\
  = \left[
    \mathbf{w}^k_1 \cdot \left(\mathbf{h}^{L\uparrow}_i \odot \mathbf{h}^{L\downarrow}_j\right)
	  + \mathbf{w}^k_2 \cdot \left(\mathbf{h}^{L\uparrow\downarrow}_{ij} \odot \mathbf{h}^{L\downarrow\uparrow}_{ji}\right)
	\right] \chi^{k\uparrow}(\mathbf{r}^\uparrow_i) \chi^{k\downarrow}(\mathbf{r}^{\downarrow}_j)
\end{equation}
where $\mathbf{w}^k_1\in \mathbb{R}^{n_L}$ and $\mathbf{w}^k_2\in \mathbb{R}^{m_L}$ are vectors of variational parameters, $\mathbf{h}^{L\alpha}_i$ and $\mathbf{h}^{L\uparrow\downarrow}_{ij}$ are the output vectors from the final layer, $L$, of the neural network, and $\chi^\alpha(\mathbf{r}^{\alpha}_i)$ is an envelope function chosen to ensure that the relevant boundary conditions are satisfied.
In this work, we use the multiwave envelope proposed by Cassella~\emph{et al.} in Ref.~\cite{cassellaDiscoveringQuantumPhase2023}.
The vectors $\mathbf{h}^{L\alpha}_i$ and $\mathbf{h}^{L\uparrow\downarrow}_{ij}$ comes from the one-electron and two-electron streams of the neural network, respectively~\cite{pfauAbInitioSolutionManyElectron2020,louNeuralWaveFunctions2024}.
Similarly, the FermiNet orbital we use in Eq.~(\ref{eq:agpsferminet_unpaired}) for the extra spin-up particles is given by~\cite{pfauAbInitioSolutionManyElectron2020}
\begin{equation}
	\phi^{k\uparrow}_i(\mathbf{r}^\uparrow_j; \{\mathbf{r}^\uparrow_{/j}\}; \{\mathbf{r}^\downarrow\})
	= \left( \mathbf{w}^{k\uparrow}_i \cdot \mathbf{h}^{L\uparrow}_j \right) \chi^{k\uparrow}_i(\mathbf{r}^\uparrow_j)
\end{equation}
where $\mathbf{w}^{k\uparrow}_i\in \mathbb{R}^{n_L}$ is also a vector of variational parameters.

For $(N^\uparrow,N^\downarrow)=(13, 5)$, four NVIDIA A100 GPUs were used.
For $(N^\uparrow,N^\downarrow)=(25, 9)$, we used four nodes with a total of sixteen A100 GPUs to speed up the calculations.
A JAX implementation of the Kronecker-factored approximate curvature (KFAC) gradient descent algorithm \cite{martensOptimizingNeuralNetworks2015,kfac-jax2022github} was used for optimization.
The initial parameters of the network are initialized using Xavier (random) initialization\cite{glorotUnderstandingDifficultyTraining2010} and the positions of the particles are initialized uniformly in the simulation box.
We do not observe significant run-to-run variation in the final energy of a given system as a function of the random initialization.
The FermiNet hyperparameters are shown in Table (\ref{tab:hyperparams}) and the network sizes in Table (\ref{tab:network_hyperparameters}).
All training runs used $3 \times 10^5$ iterations to ensure convergence.
When evaluating the expectation values with an optimized wave function, $1000$ inference steps were used.
\begin{table}[ht!]
  \centering
  \begin{tabularx}{0.45\textwidth}{ c c }
    \midrule \midrule
    Parameter & Value \\
    \midrule
    Batch size & 4096 \\
    Training iterations & 3e5 \\
    Pretraining iterations & None \\
    Learning rate & $(2e4 + t)^{-1}$ \\
    Local energy clipping  & 5.0 \\
    KFAC Momentum & 0 \\
    KFAC Covariance moving average decay & 0.95 \\
    KFAC Norm constraint & 1e-3 \\
    KFAC Damping & 1e-3 \\
    MCMC Proposal std.\ dev.\ (per dimension) & 0.02 \\
    MCMC Steps between parameter updates & 10 \\
    \midrule
  \end{tabularx}
  \caption{Hyperparameters used in all simulations.}
  \label{tab:hyperparams}
\end{table}

\begin{table}[ht!]
  \centering
  \begin{tabularx}{0.42\textwidth}{ c c c }
    \midrule \midrule
    Parameter & Symbol & Value \\
    \midrule
    One-electron Stream Network Size & $n_l$ & 512 \\
    Two-electron Stream Network Size & $m_l$ & 64 \\
    Number of Network Layers & $L$ & 4 \\
    Number of Determinants & $D$ & 32 \\
    \midrule
  \end{tabularx}
  \caption{Network sizes and number of determinants used in all simulations. The corresponding mathematical symbols mentioned in the main text of the paper, where available, are also listed.}
  \label{tab:network_hyperparameters}
\end{table}

\section{Observables}\label{appendix:observables}
The one-body density matrix (OBDM) and the two-body density matrix (TBDM) are defined in terms of the fermionic field operators via:
\begin{align}
  \rho_{\alpha}^{(1)}(\mathbf{r}'_1; \mathbf{r}_1)
  &= \expval{\hat{\psi}^\dagger_\alpha(\mathbf{r}'_1) \hat{\psi}^{\phantom{\dagger}}_\alpha(\mathbf{r}_1)}
  \label{eq:general_second_quantised_obdm} \\
  \rho_{\alpha\beta}^{(2)}(\mathbf{r}'_1, \mathbf{r}'_2; \mathbf{r}_1, \mathbf{r}_2)
  &= \expval{\hat{\psi}^\dagger_\alpha(\mathbf{r}'_1) \hat{\psi}^\dagger_\beta(\mathbf{r}'_2) \hat{\psi}^{\phantom{\dagger}}_\beta(\mathbf{r}_2) \hat{\psi}^{\phantom{\dagger}}_\alpha(\mathbf{r}_1)}
  \label{eq:general_second_quantised_tbdm}
\end{align}
where $\alpha, \beta$ are the spin indices of particles 1 and 2, respectively.
The normalisation is chosen such that
\begin{align}
  \sum_\alpha \int_{\Omega} d\mathbf{r}_1\ d\mathbf{r}'_1\ \delta(\mathbf{r}'_1-\mathbf{r}_1) \rho_{\alpha}^{(1)}(\mathbf{r}'_1; \mathbf{r}_1) &= N \\
  \sum_\alpha \sum_\beta \int_{\Omega} d\mathbf{r}_1\ d\mathbf{r}'_1\ d\mathbf{r}_2\ d\mathbf{r}'_2\ \delta(\mathbf{r}'_1-\mathbf{r}_1) \delta(\mathbf{r}'_2-\mathbf{r}_2) \rho_{\alpha\beta}^{(2)}(\mathbf{r}'_1, \mathbf{r}'_2; \mathbf{r}_1, \mathbf{r}_2) &= N(N-1) .
  \label{eq:density_matrix_normalisation}
\end{align}
If the system is homogeneous, one can define the \emph{translationally averaged} OBDM and TBDM as \cite{needsVariationalDiffusionQuantum2020}
\begin{align}
  \bar{\rho}_{\alpha}^{(1)}(\mathbf{r})
  &= \frac{1}{\Omega} \int_{\Omega} d\mathbf{r}_1\ \rho_{\alpha}^{(1)}(\mathbf{r}_1 + \mathbf{r}; \mathbf{r}_1)
  \label{eq:general_translational_avg_obdm} \\
  \bar{\rho}_{\alpha\beta}^{(2)}(\mathbf{r})
  &= \frac{1}{\Omega^2} \int_{\Omega} d\mathbf{r}_1\ d\mathbf{r}_2\ \rho_{\alpha\beta}^{(2)}(\mathbf{r}_1 + \mathbf{r}, \mathbf{r}_2 + \mathbf{r}; \mathbf{r}_1, \mathbf{r}_2)
  \label{eq:general_translational_avg_tbdm}
\end{align}
where $\Omega$ is the volume of the system, and the bar at the top of the $\rho$s denotes the translational average.

\subsection{Momentum Density and the Condensate Fraction}
The Fourier transform of the translationally-averaged OBDM and TBDM in Eq.~(\ref{eq:general_translational_avg_obdm}) and Eq.~(\ref{eq:general_translational_avg_tbdm}) are known as the momentum density and the pair-momentum density, respectively.
The momentum density is given in Eq.~(\ref{eq:momentum_density}), and the condensate fraction, $f^{\uparrow\downarrow}_{\mathbf{q}}$, defined in Eq.~(\ref{eq:pair_momentum_density}) is related to the pair-momentum density,
\begin{equation}
  g^{\uparrow\downarrow}_{\mathbf{q}} = \frac{1}{\Omega} \int d\mathbf{r}\ e^{i\mathbf{q \cdot r}} \bar{\rho}_{\uparrow\downarrow}^{(2)}(\mathbf{r}) ,
\end{equation}
by,
\begin{equation}
  f^{\uparrow\downarrow}_{\mathbf{q}}
  = \frac{\Omega^2}{N_\text{pair}} \sum_\mathbf{k} \expval{\hat{c}^\dagger_{\mathbf{k}\uparrow}\hat{c}^\dagger_{\mathbf{q-k}\downarrow} \hat{c}^{\phantom{\dagger}}_{\mathbf{q-k}\downarrow} \hat{c}^{\phantom{\dagger}}_{\mathbf{k}\uparrow}}
  = \frac{\Omega^2}{N_\text{pair}} g^{\uparrow\downarrow}_{\mathbf{q}} .
\end{equation}
To obtain the momentum density and the condensate fraction in our calculations, we first estimate the OBDM and the TBDM across the whole simulation cell with QMC accumulation, followed by a discrete Fourier transform of the two quantities.
The details of the QMC accumulation can be found in the next section, as well as the CASINO manual in Ref.~\cite{needsVariationalDiffusionQuantum2020}.

\subsection{QMC Accumulation}
The OBDM and TBDM in first quantized notation can be written as~\cite{needsVariationalDiffusionQuantum2020}
\begin{align}
  \rho_{\alpha}^{(1)}(\mathbf{r}_1; \mathbf{r}'_1)
  &= N_\alpha \frac{\int |\Psi(\mathbf{R})|^2 \frac{\Psi(\mathbf{r}'_1)}{\Psi(\mathbf{r}_1)}d\mathbf{r}_2 \dots d\mathbf{r}_N}{\int |\Psi(\mathbf{R})|^2 d\mathbf{R}} ,
  \label{eq:obdm_first}
  \\
  \rho_{\alpha\beta}^{(2)}(\mathbf{r}_1, \mathbf{r}_2; \mathbf{r}'_1, \mathbf{r}'_2)
  &= N_\alpha (N_\beta - \delta_{\alpha\beta}) \frac{\int |\Psi(\mathbf{R})|^2 \frac{\Psi(\mathbf{r}'_1, \mathbf{r}'_2)}{\Psi(\mathbf{r}_1,\mathbf{r}_2)}d\mathbf{r}_3 \dots d\mathbf{r}_N}{\int |\Psi(\mathbf{R})|^2 d\mathbf{R}} ,
  \label{eq:tbdm_first}
\end{align}
where $\alpha$ and $\beta$ denote the spin species.
To reduce the effect of the one-body contribution to the TBDM, we use an improved estimator of the TBDM in Eq.~(\ref{eq:modified_tbdm}) that removes the one-body contribution explicitly~\cite{needsVariationalDiffusionQuantum2020}:
\begin{equation}
  \rho_{\alpha\beta}^{(2)}(\mathbf{r}_1, \mathbf{r}_2; \mathbf{r}'_1, \mathbf{r}'_2) = N_\alpha (N_\beta - \delta_{\alpha\beta}) \frac{\int |\Psi(\mathbf{R})|^2
  \left[
    \frac{\Psi(\mathbf{r}'_1, \mathbf{r}'_2)}{\Psi(\mathbf{r}_1,\mathbf{r}_1)} - \frac{\Psi(\mathbf{r}'_1, \mathbf{r}_2)}{\Psi(\mathbf{r}_1,\mathbf{r}_2)} \frac{\Psi(\mathbf{r}_1, \mathbf{r}'_2)}{\Psi(\mathbf{r}_1,\mathbf{r}_2)}
  \right] d\mathbf{r}_3 \dots d\mathbf{r}_N}{\int |\Psi(\mathbf{R})|^2 d\mathbf{R}}.
  \label{eq:modified_tbdm}
\end{equation}
All of these quantities can then be estimated using Monte Carlo sampling.

\section{All Data}\label{appendix:data}
In this section, we present all the results of our simulations, including the momentum densities, the condensate fraction and the one-particle real-space density.

\subsection{Momentum Density}
\begin{figure}[ht!]
  \centering
  \begin{subfigure}[b]{0.32\columnwidth}
    \centering
    \includegraphics[width=\columnwidth]{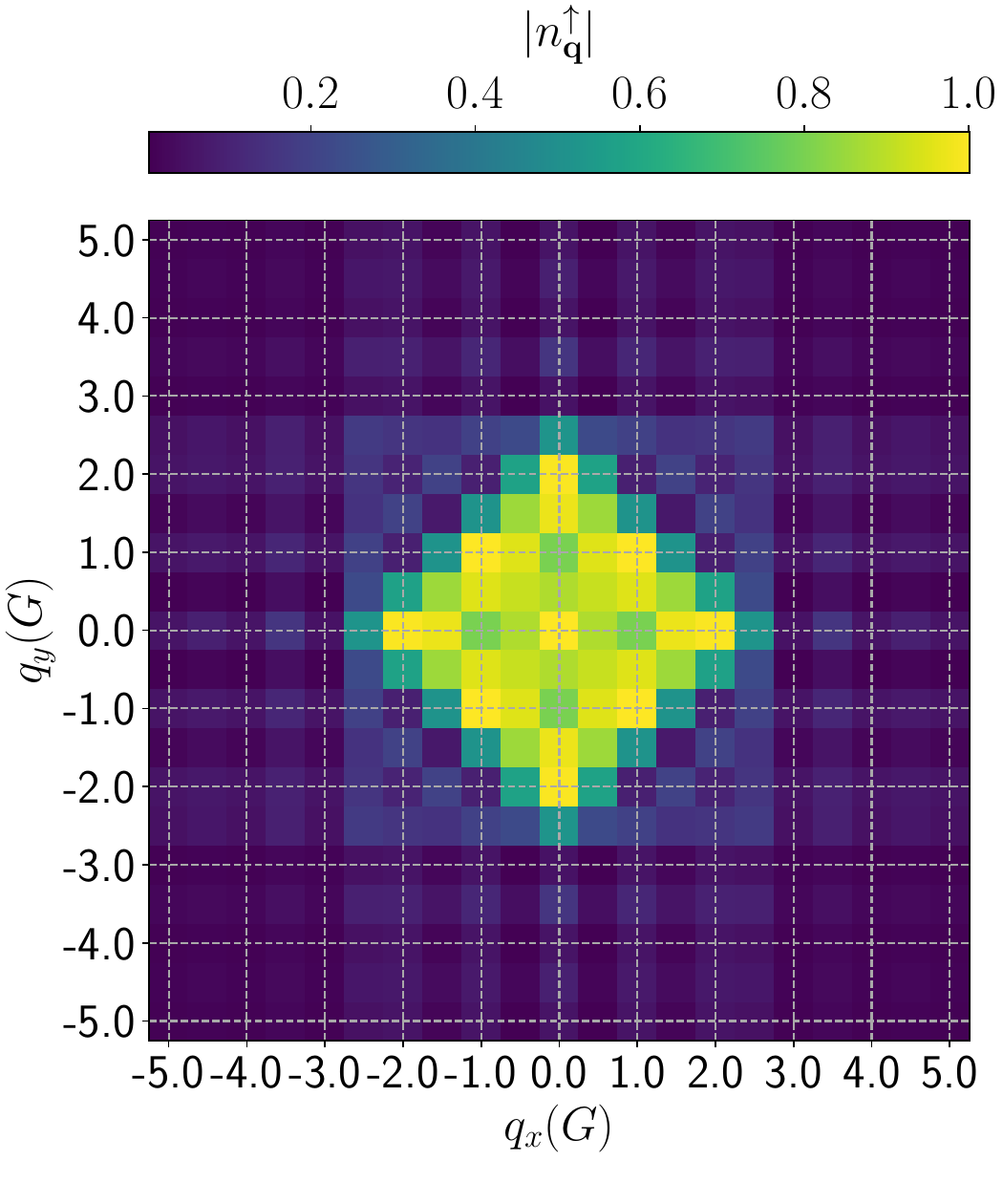}
    \caption{$v_0=0.1$, spin-up}
  \end{subfigure}
  \hfill
  \begin{subfigure}[b]{0.32\columnwidth}
    \centering
    \includegraphics[width=\columnwidth]{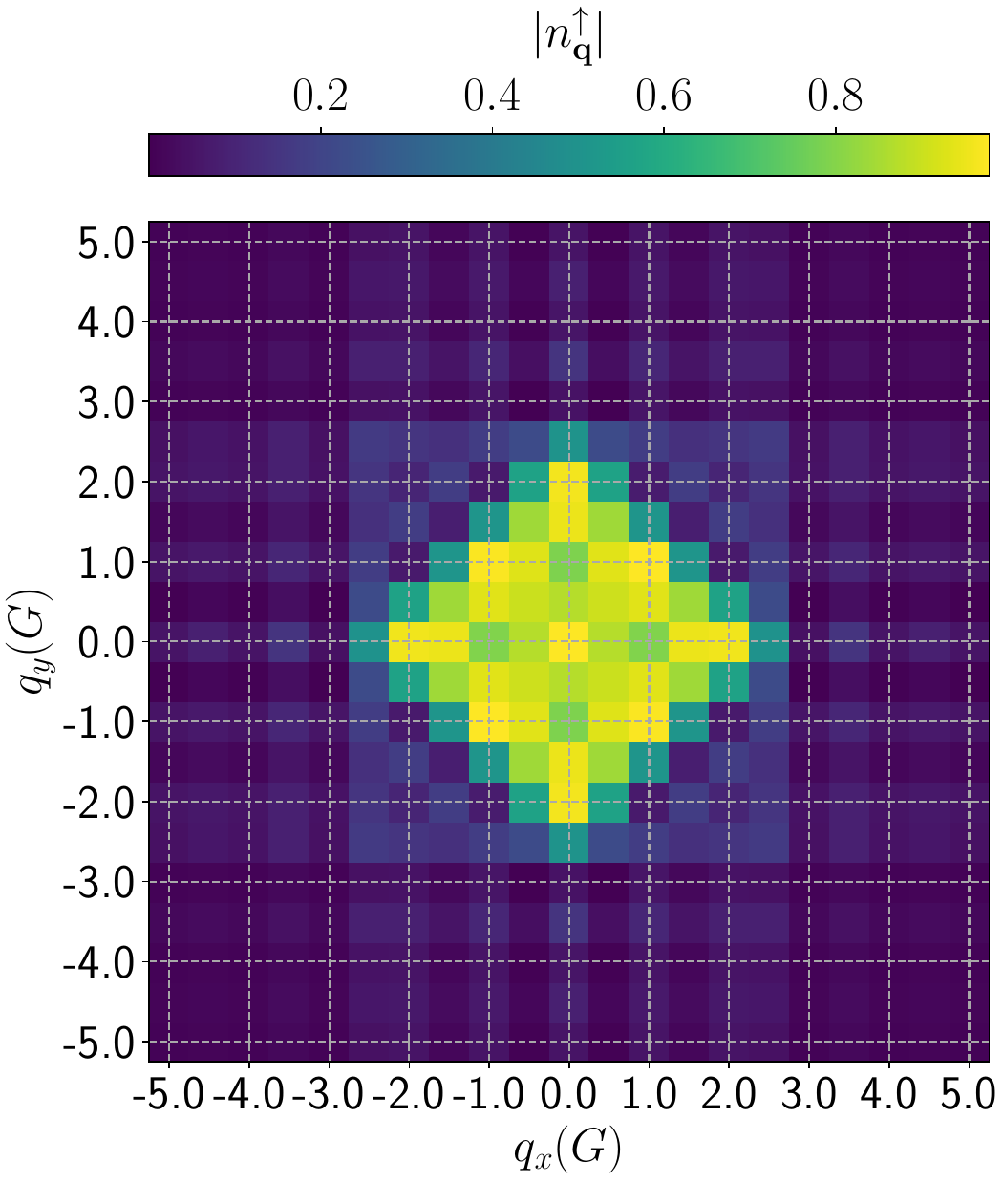}
    \caption{$v_0=0.15$, spin-up}
  \end{subfigure}
  \hfill
  \begin{subfigure}[b]{0.32\columnwidth}
    \centering
    \includegraphics[width=\columnwidth]{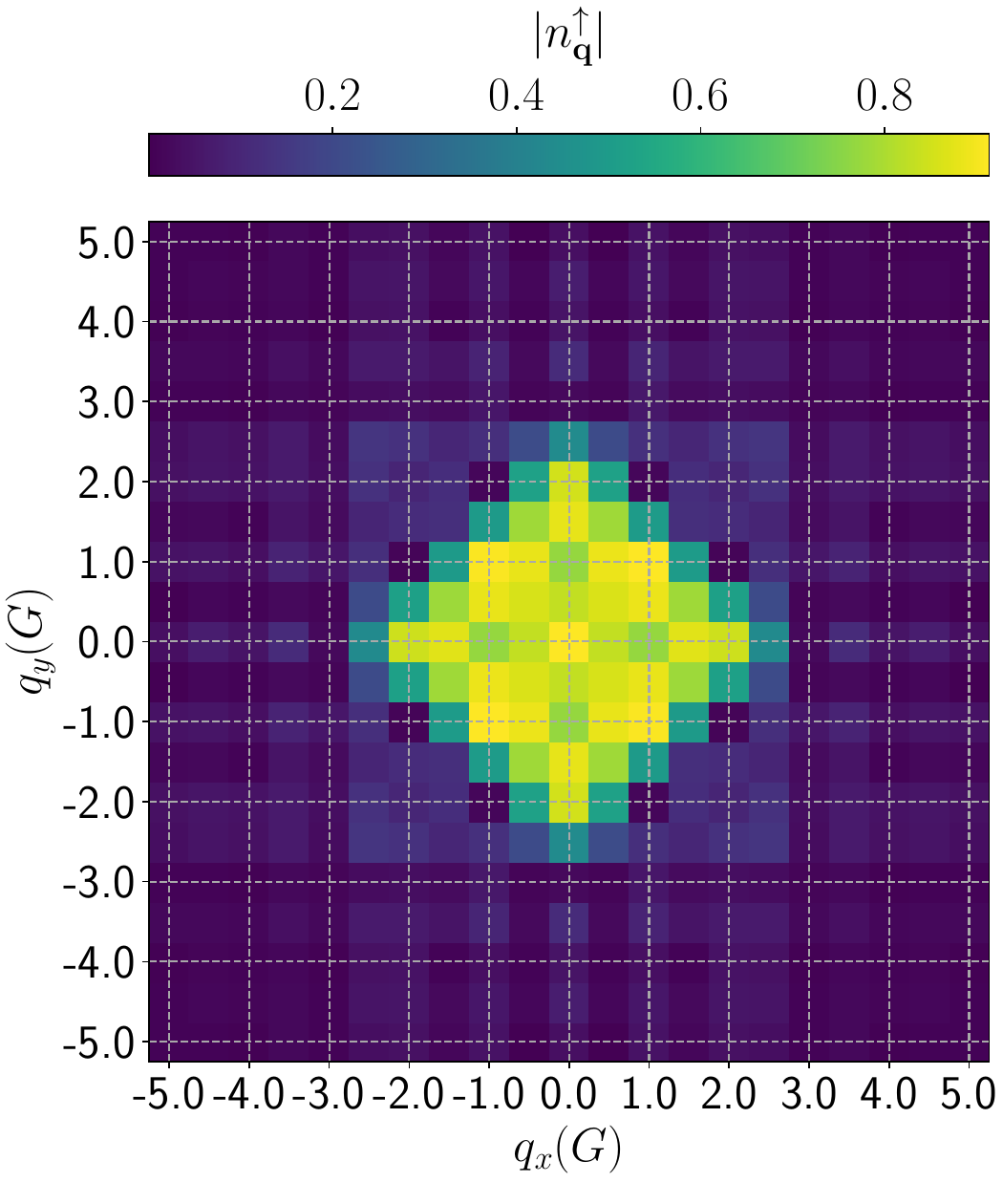}
    \caption{$v_0=0.2$, spin-up}
  \end{subfigure}
  \hfill
  \begin{subfigure}[b]{0.32\columnwidth}
    \centering
    \includegraphics[width=\columnwidth]{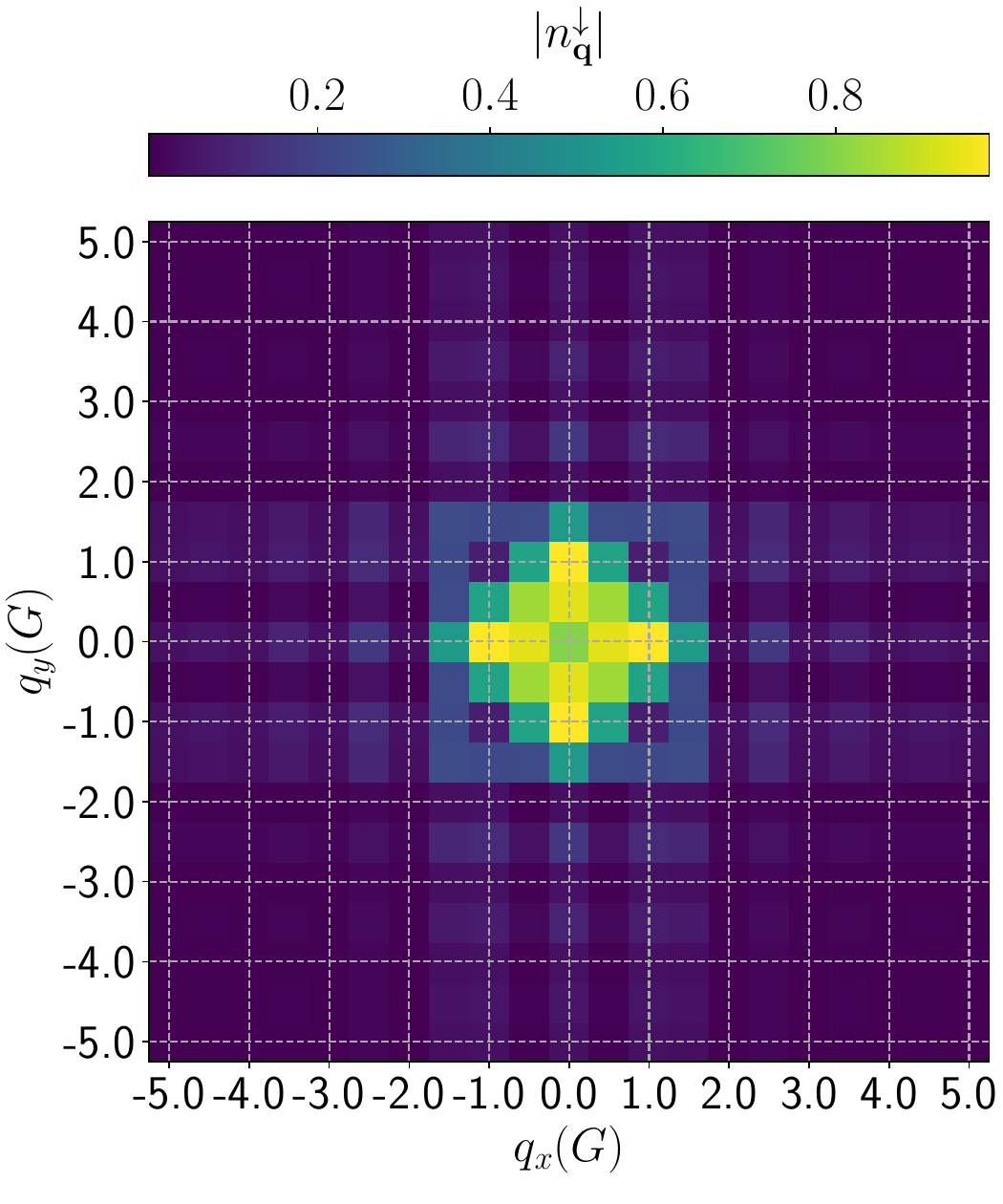}
    \caption{$v_0=0.1$, spin-down}
  \end{subfigure}
  \hfill
  \begin{subfigure}[b]{0.32\columnwidth}
    \centering
    \includegraphics[width=\columnwidth]{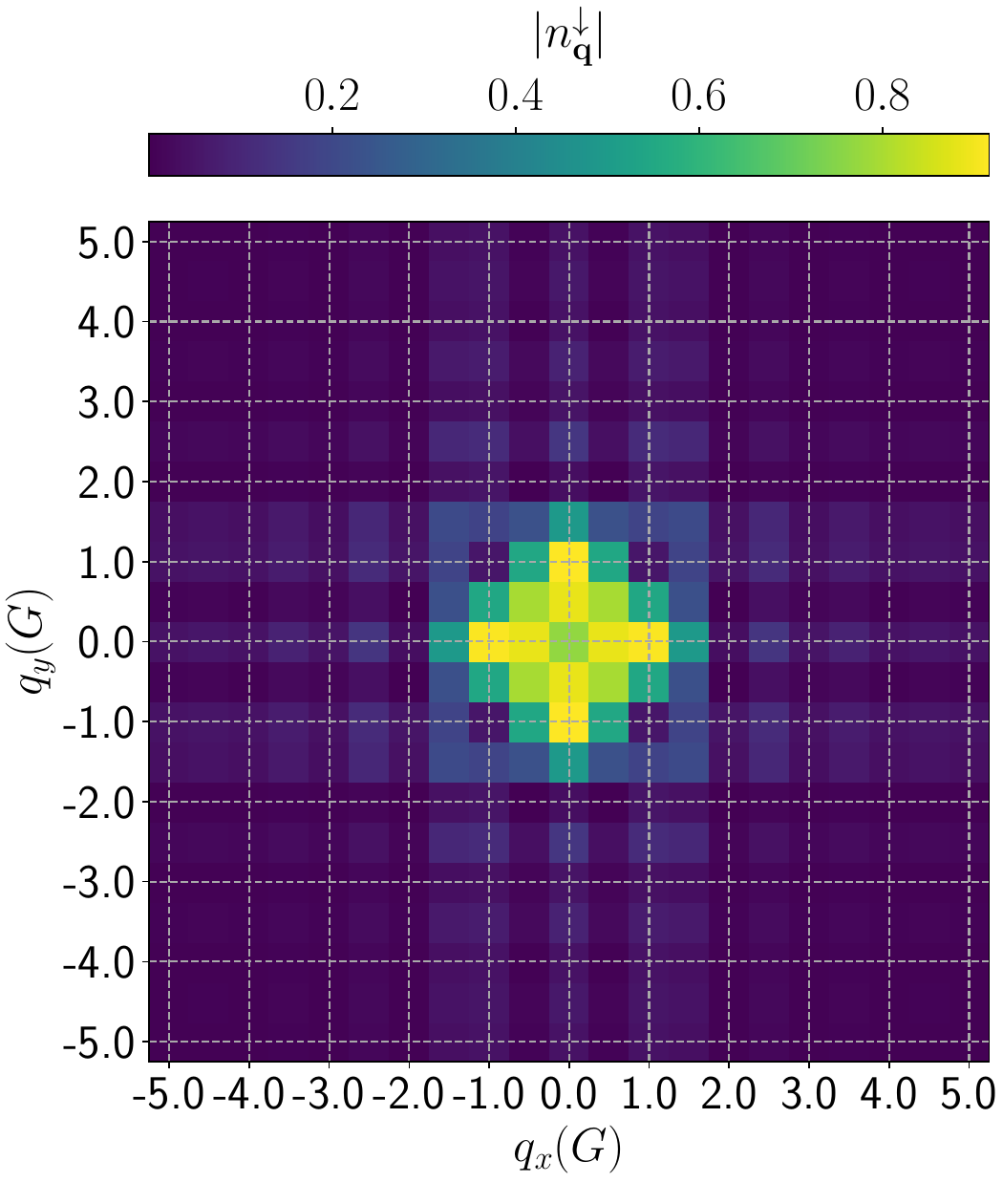}
    \caption{$v_0=0.15$, spin-down}
  \end{subfigure}
  \hfill
  \begin{subfigure}[b]{0.32\columnwidth}
    \centering
    \includegraphics[width=\columnwidth]{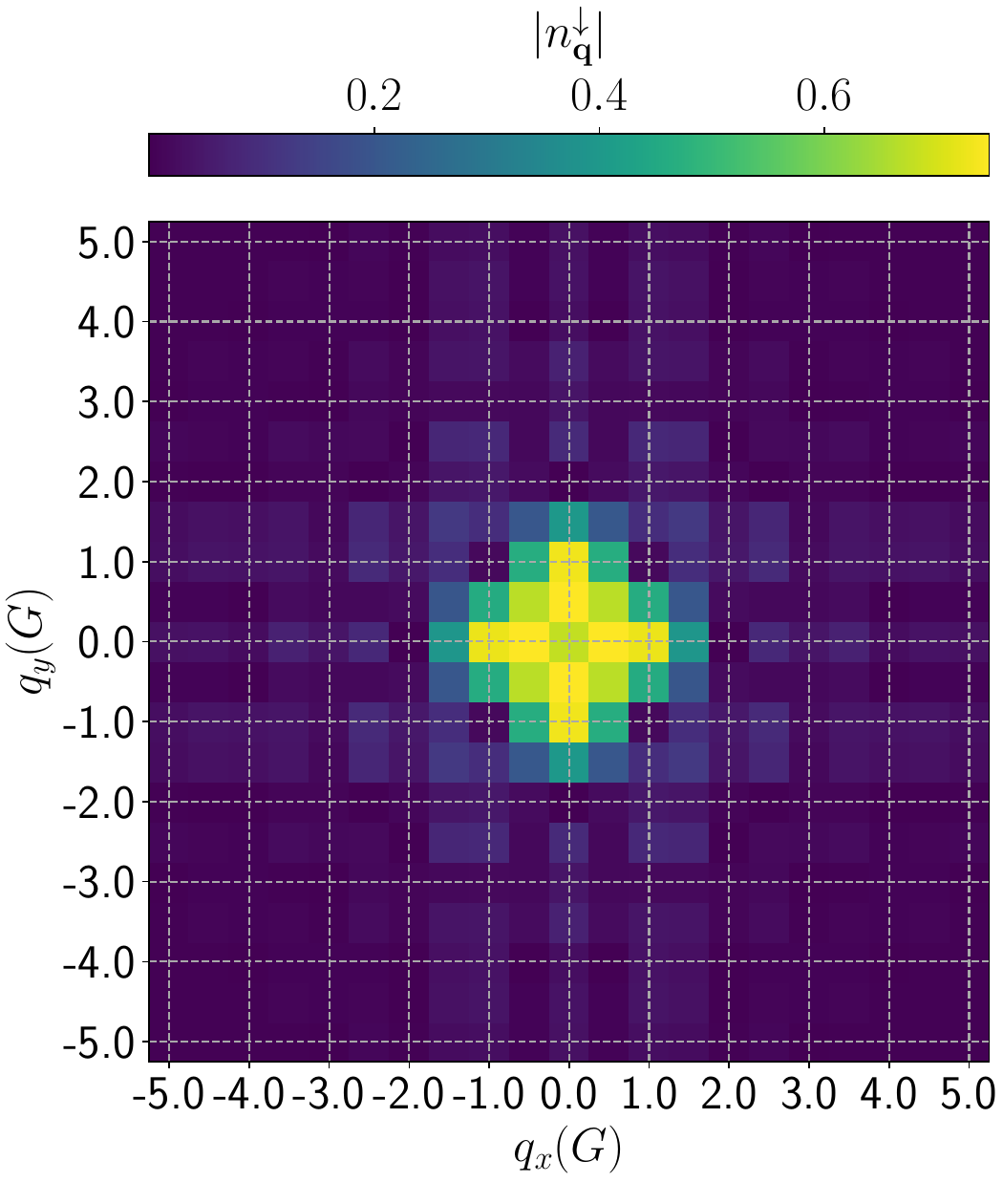}
    \caption{$v_0=0.2$, spin-down}
  \end{subfigure}
  \caption{Momentum density of the $(N^\uparrow, N^\downarrow) = (13, 5)$ SIFG for $v_0=0.1, 0.15, 0.2$.}
  \label{fig:13_5-weak-md}
\end{figure}

\begin{figure}[ht!]
  \centering
  \begin{subfigure}[b]{0.32\columnwidth}
    \centering
    \includegraphics[width=\columnwidth]{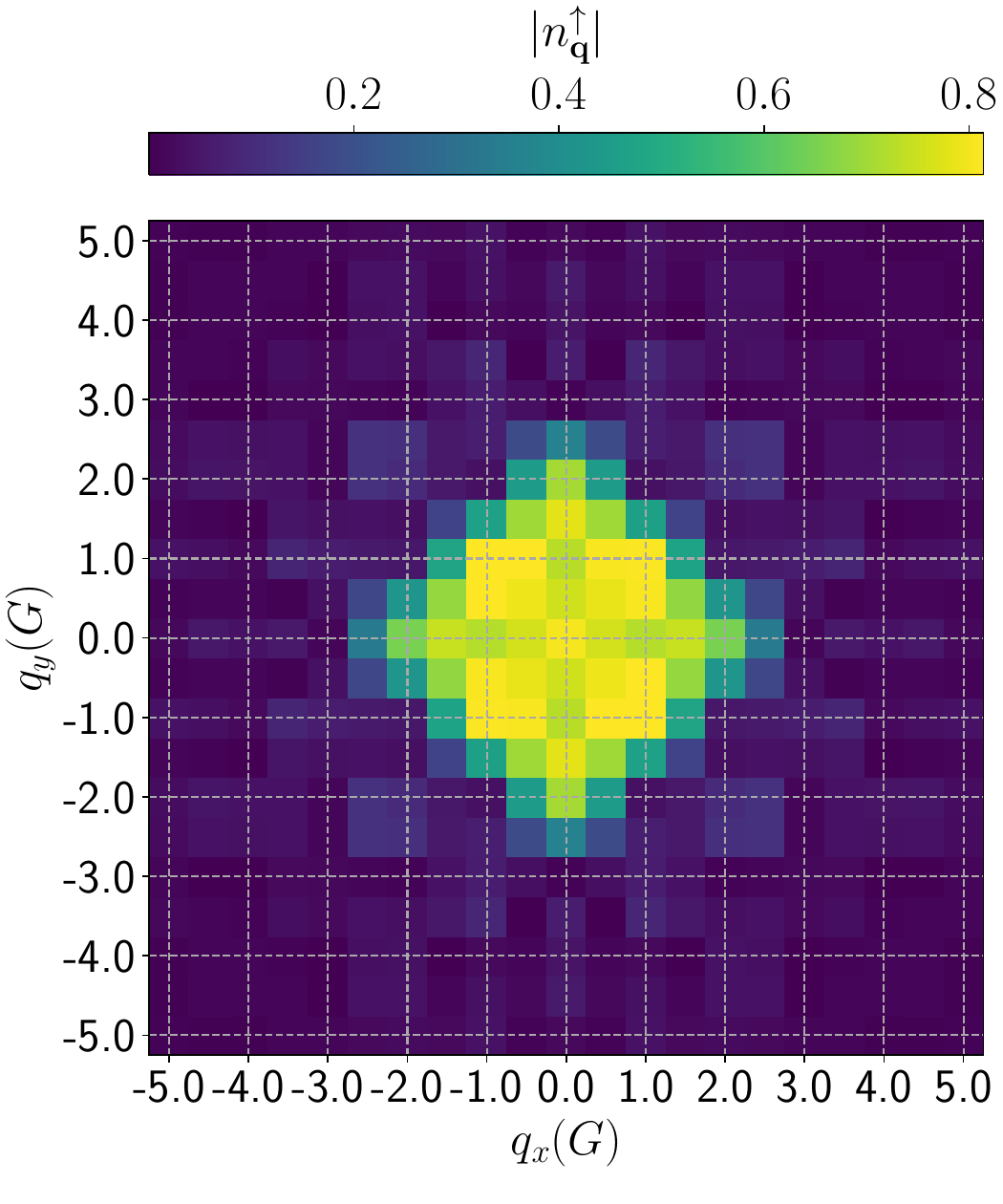}
    \caption{$v_0=0.25$, spin-up}
  \end{subfigure}
  \hfill
  \begin{subfigure}[b]{0.32\columnwidth}
    \centering
    \includegraphics[width=\columnwidth]{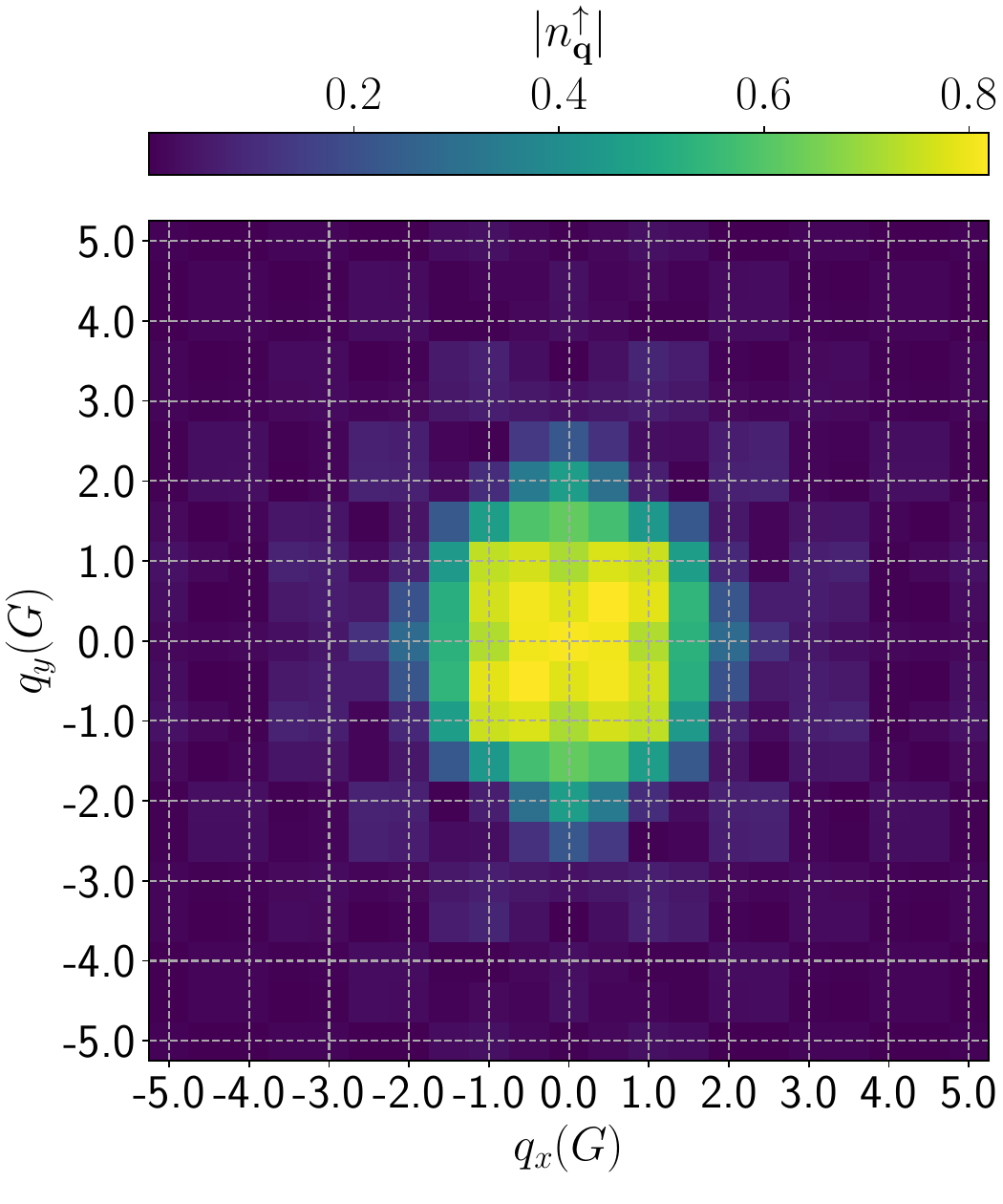}
    \caption{$v_0=0.3$, spin-up}
  \end{subfigure}
  \begin{subfigure}[b]{0.32\columnwidth}
    \centering
    \includegraphics[width=\columnwidth]{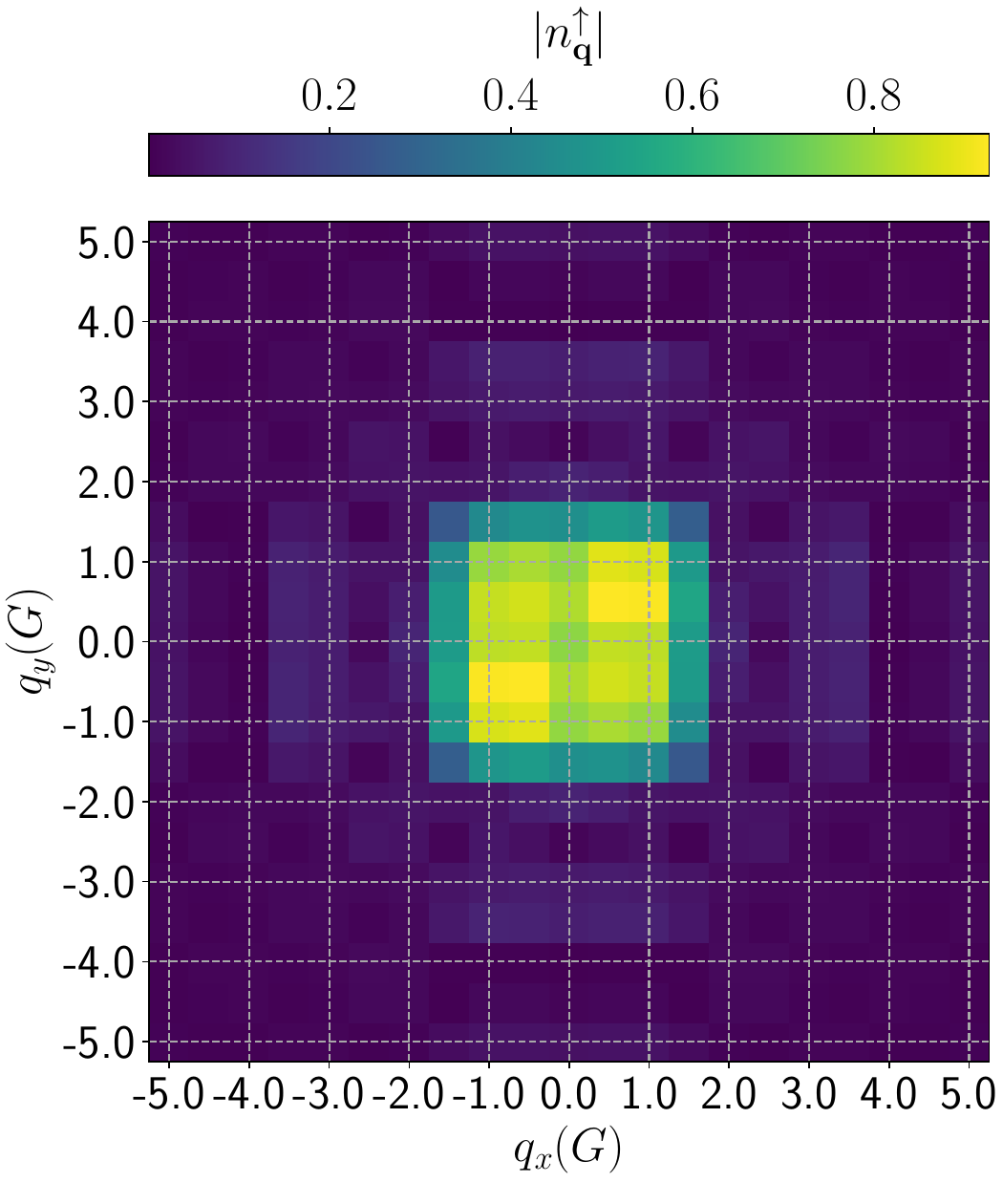}
    \caption{$v_0=0.35$, spin-up}
  \end{subfigure}
  \hfill
  \begin{subfigure}[b]{0.32\columnwidth}
    \centering
    \includegraphics[width=\columnwidth]{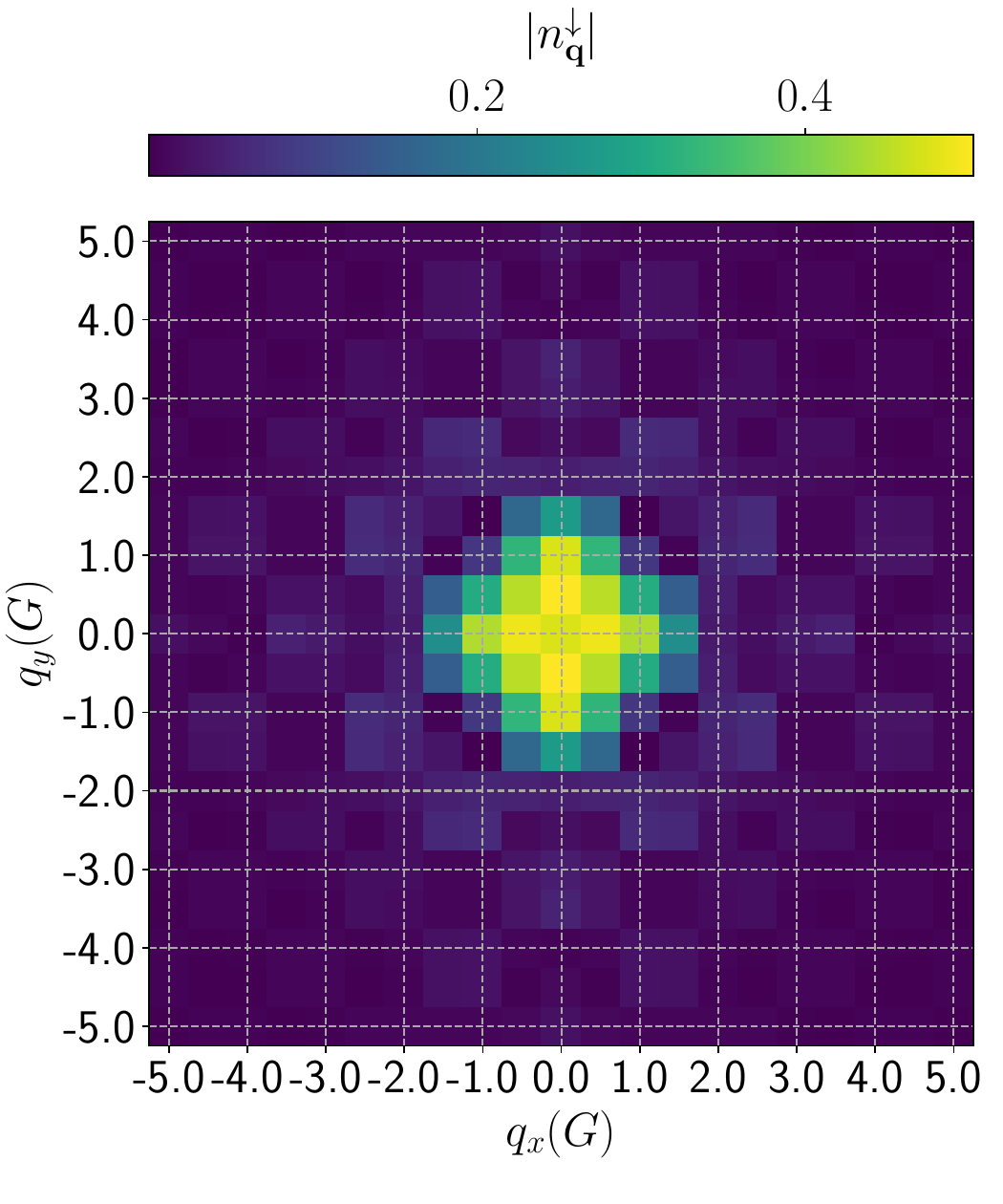}
    \caption{$v_0=0.25$, spin-down}
  \end{subfigure}
  \hfill
  \begin{subfigure}[b]{0.32\columnwidth}
    \centering
    \includegraphics[width=\columnwidth]{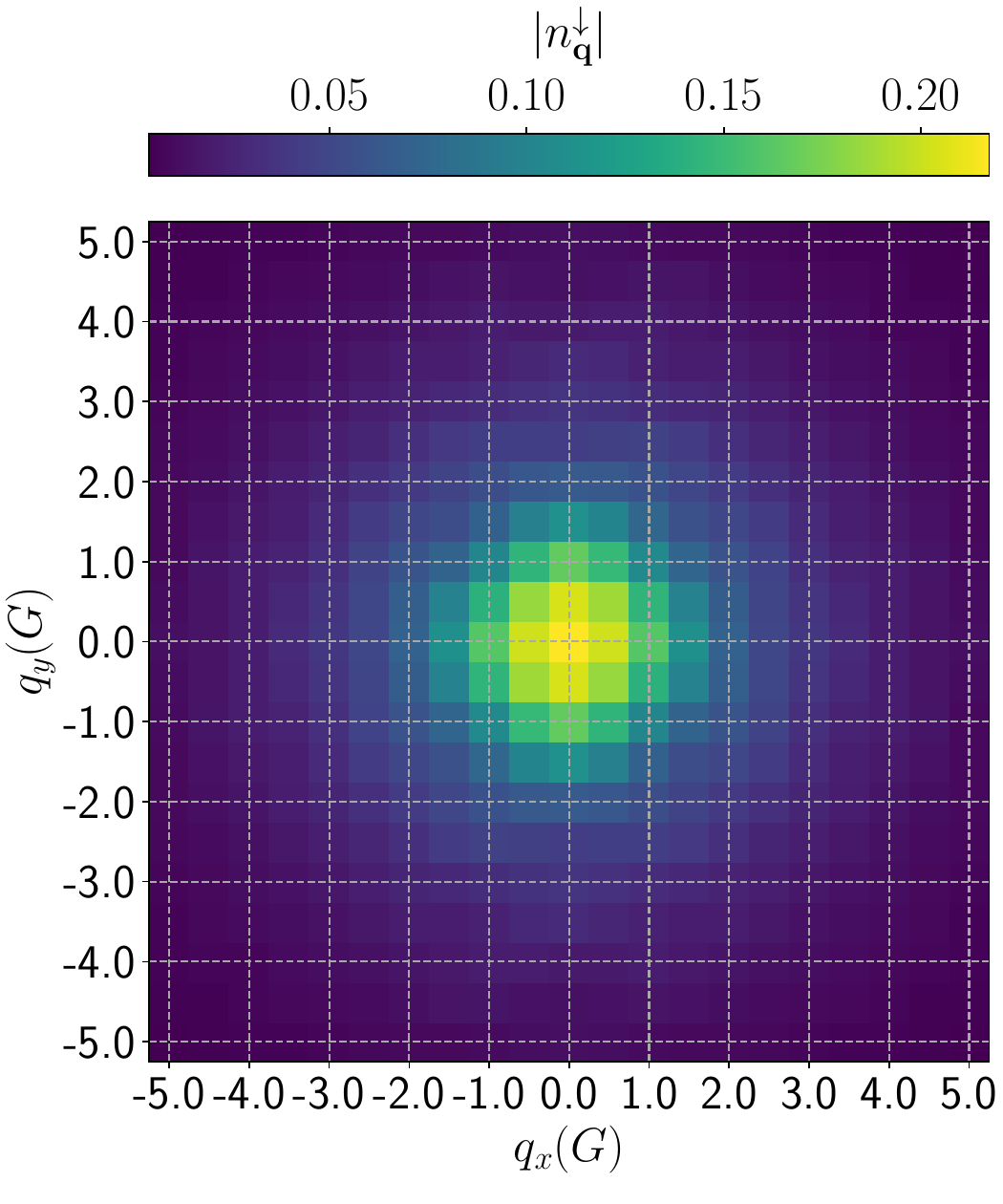}
    \caption{$v_0=0.3$, spin-down}
  \end{subfigure}
  \hfill
  \begin{subfigure}[b]{0.32\columnwidth}
    \centering
    \includegraphics[width=\columnwidth]{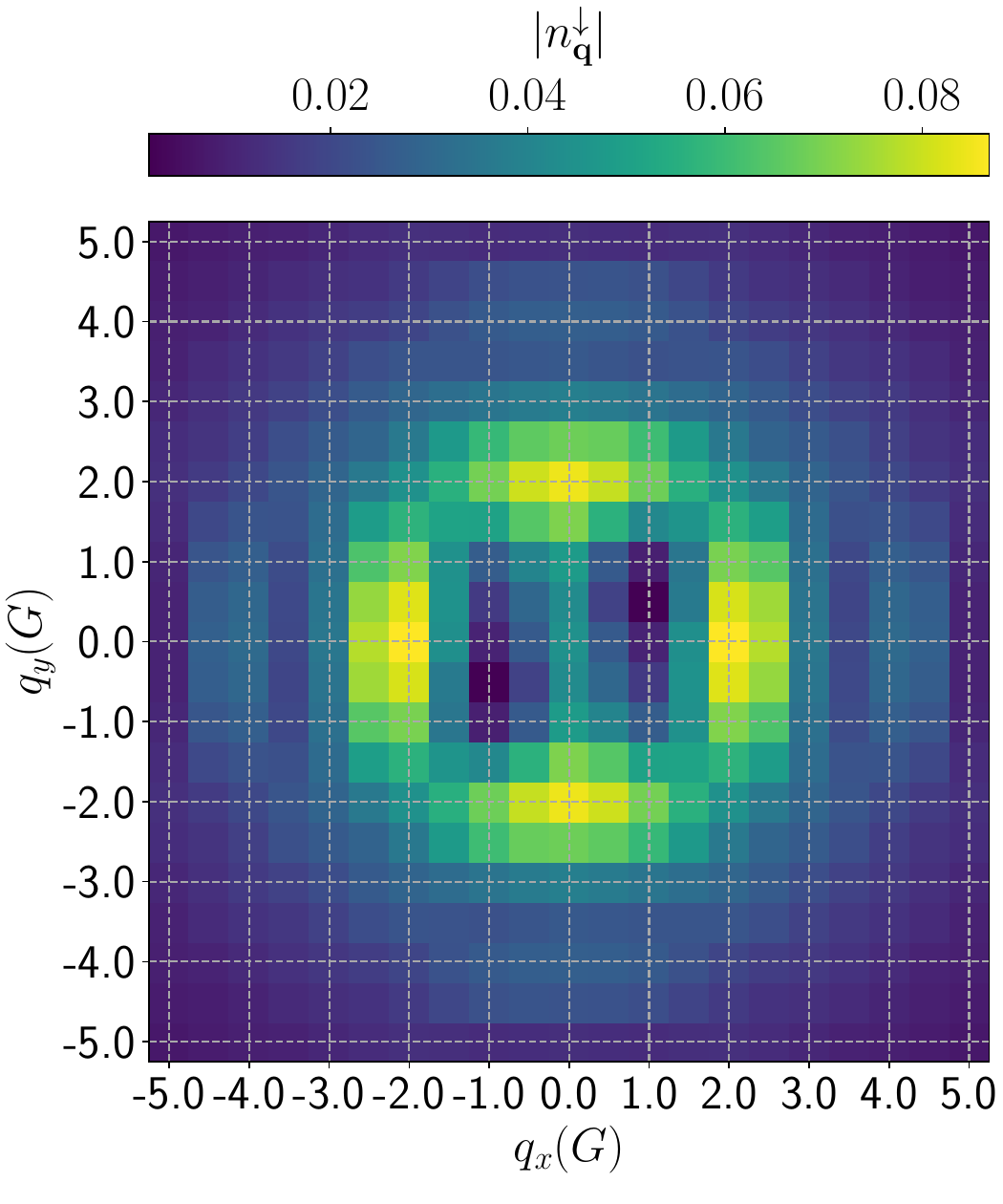}
    \caption{$v_0=0.35$, spin-down}
  \end{subfigure}
  \caption{Momentum density of the $(N^\uparrow, N^\downarrow) = (13, 5)$ SIFG for $v_0=0.25, 0.3, 0.35$.}
  \label{fig:13_5-mid-md}
\end{figure}

\begin{figure}[ht!]
  \centering
  \begin{subfigure}[b]{0.32\columnwidth}
    \centering
    \includegraphics[width=\columnwidth]{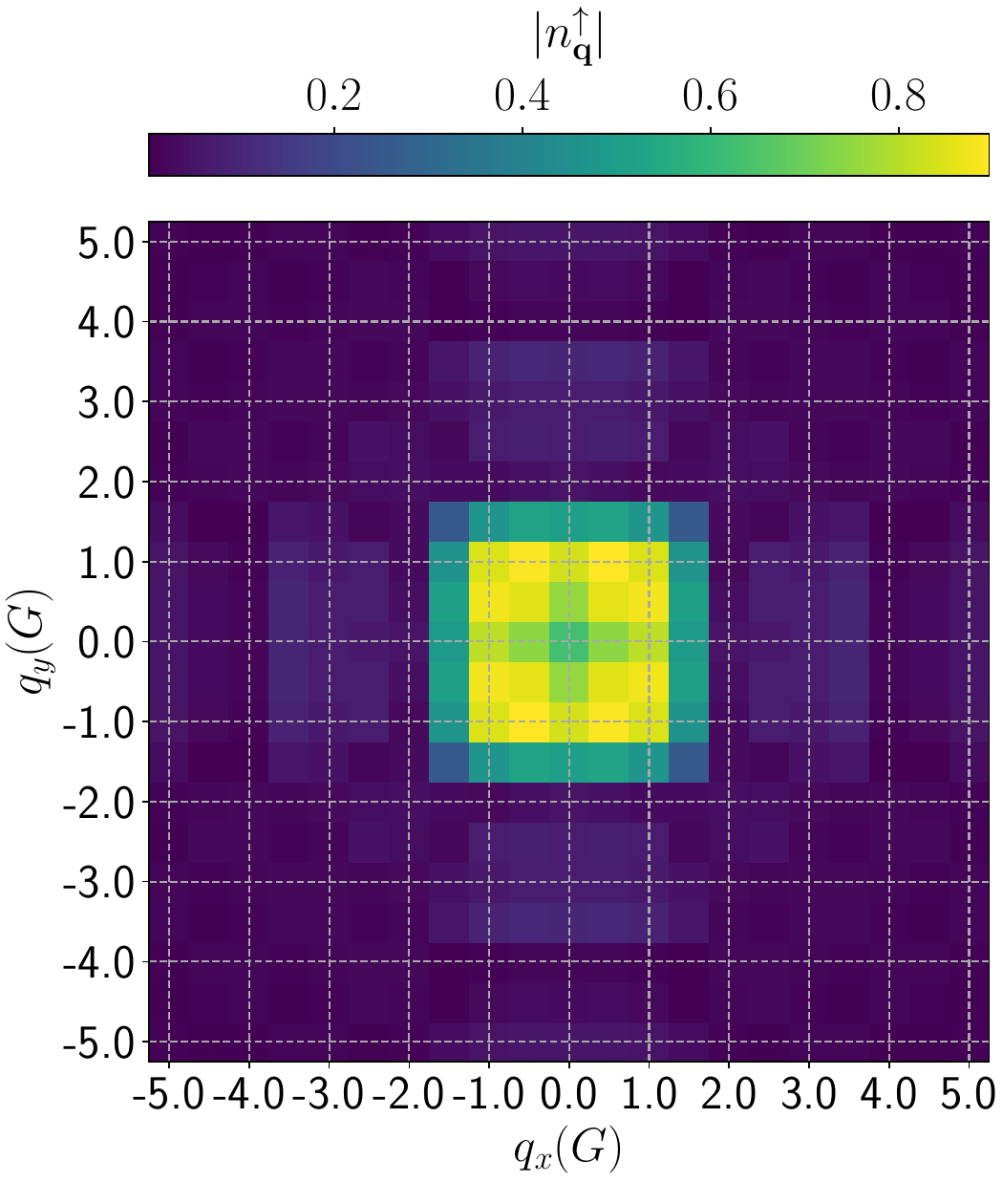}
    \caption{$v_0=0.4$, spin-up}
  \end{subfigure}
  \hfill
  \begin{subfigure}[b]{0.32\columnwidth}
    \centering
    \includegraphics[width=\columnwidth]{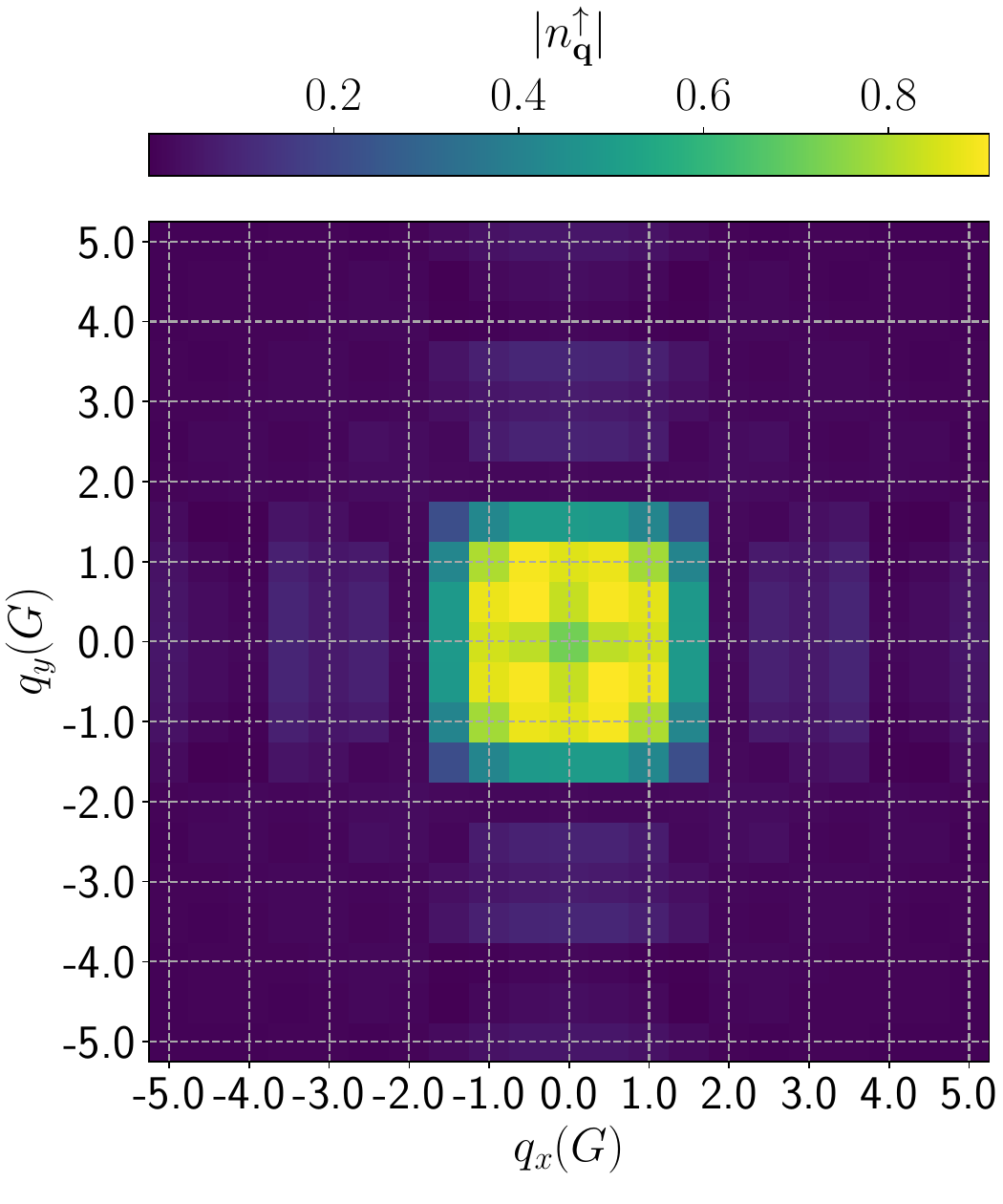}
    \caption{$v_0=0.45$, spin-up}
  \end{subfigure}
  \hfill
  \begin{subfigure}[b]{0.32\columnwidth}
    \centering
    \includegraphics[width=\columnwidth]{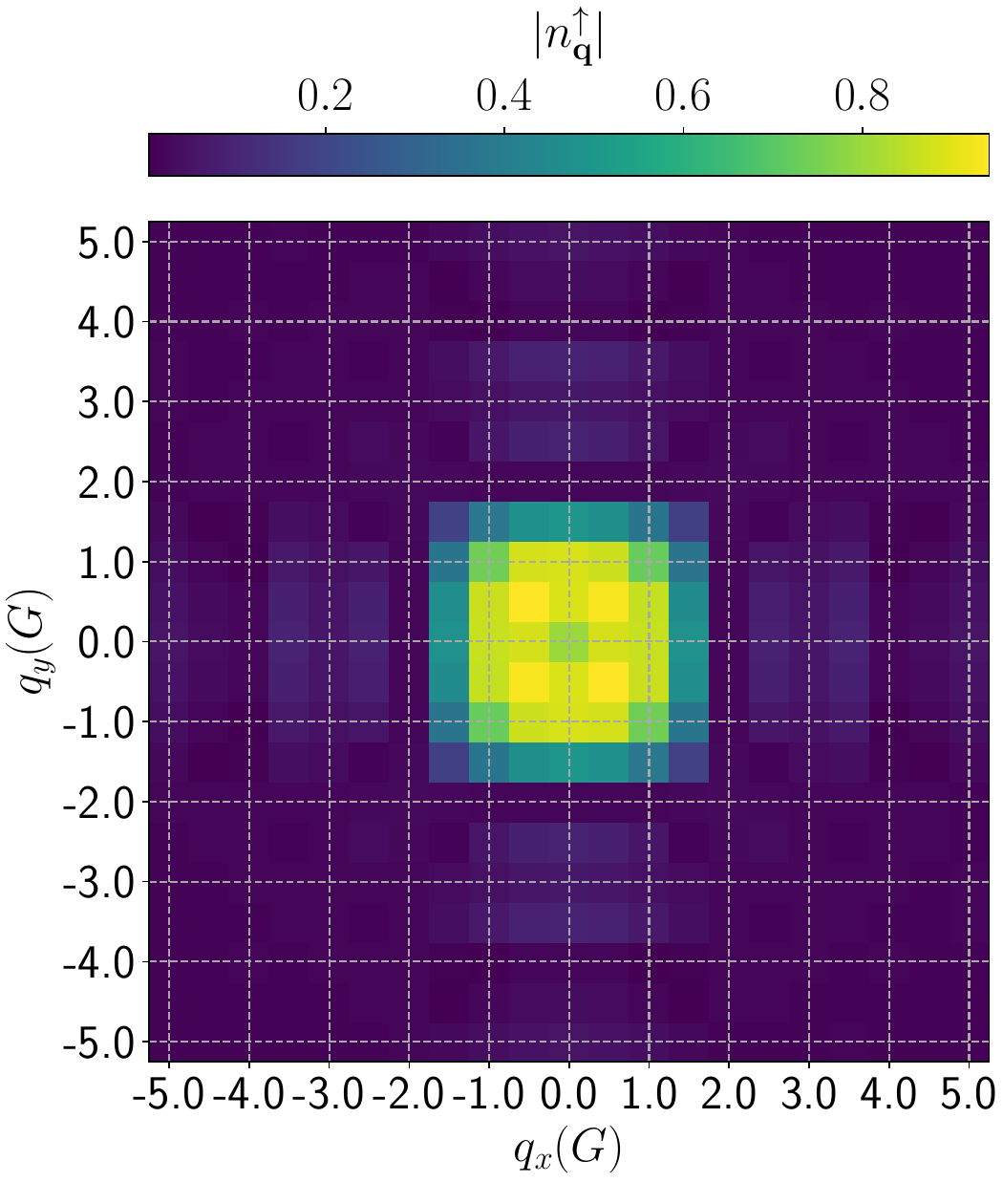}
    \caption{$v_0=0.5$, spin-up}
  \end{subfigure}
  \hfill
  \begin{subfigure}[b]{0.32\columnwidth}
    \centering
    \includegraphics[width=\columnwidth]{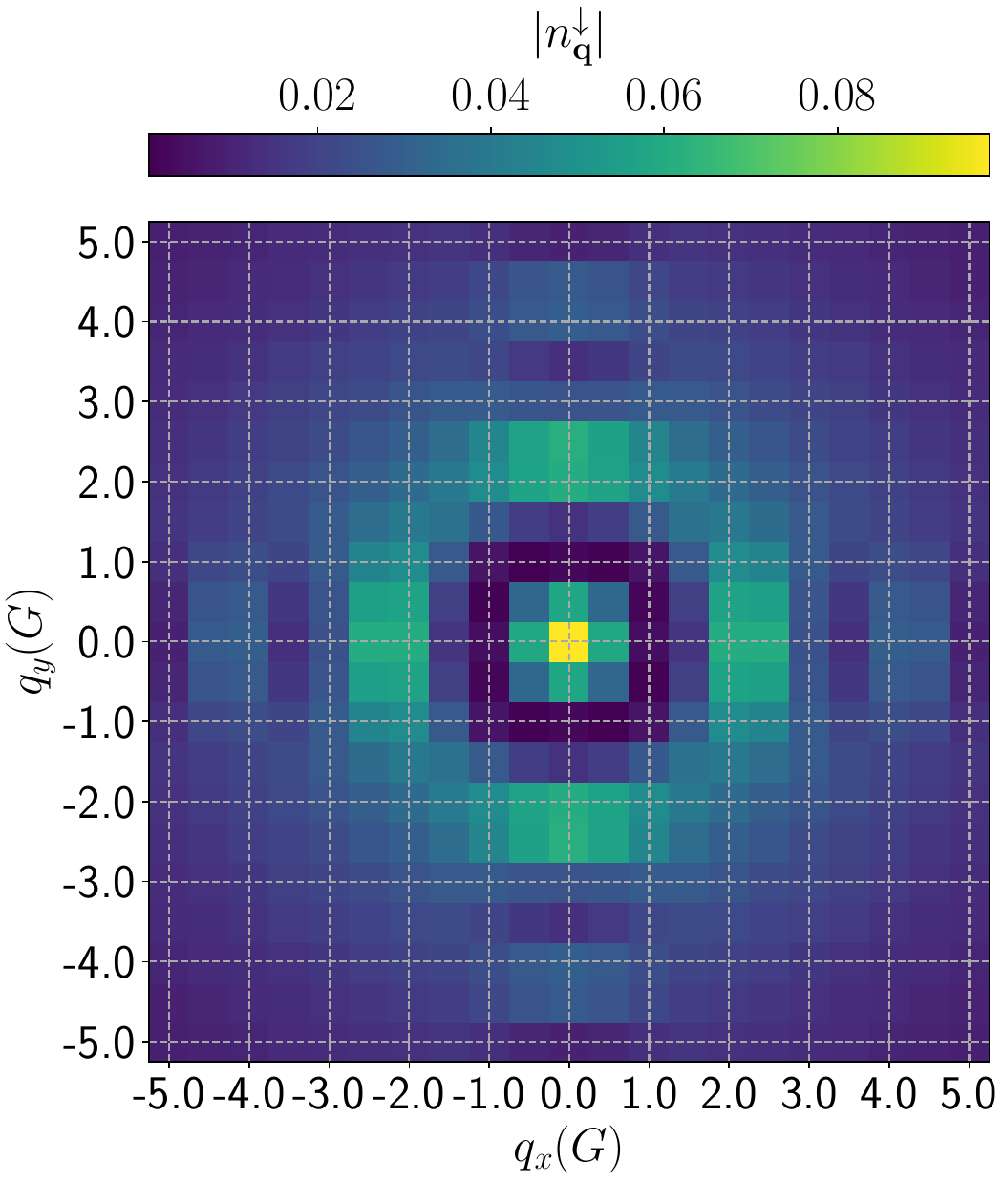}
    \caption{$v_0=0.4$, spin-down}
  \end{subfigure}
  \hfill
  \begin{subfigure}[b]{0.32\columnwidth}
    \centering
    \includegraphics[width=\columnwidth]{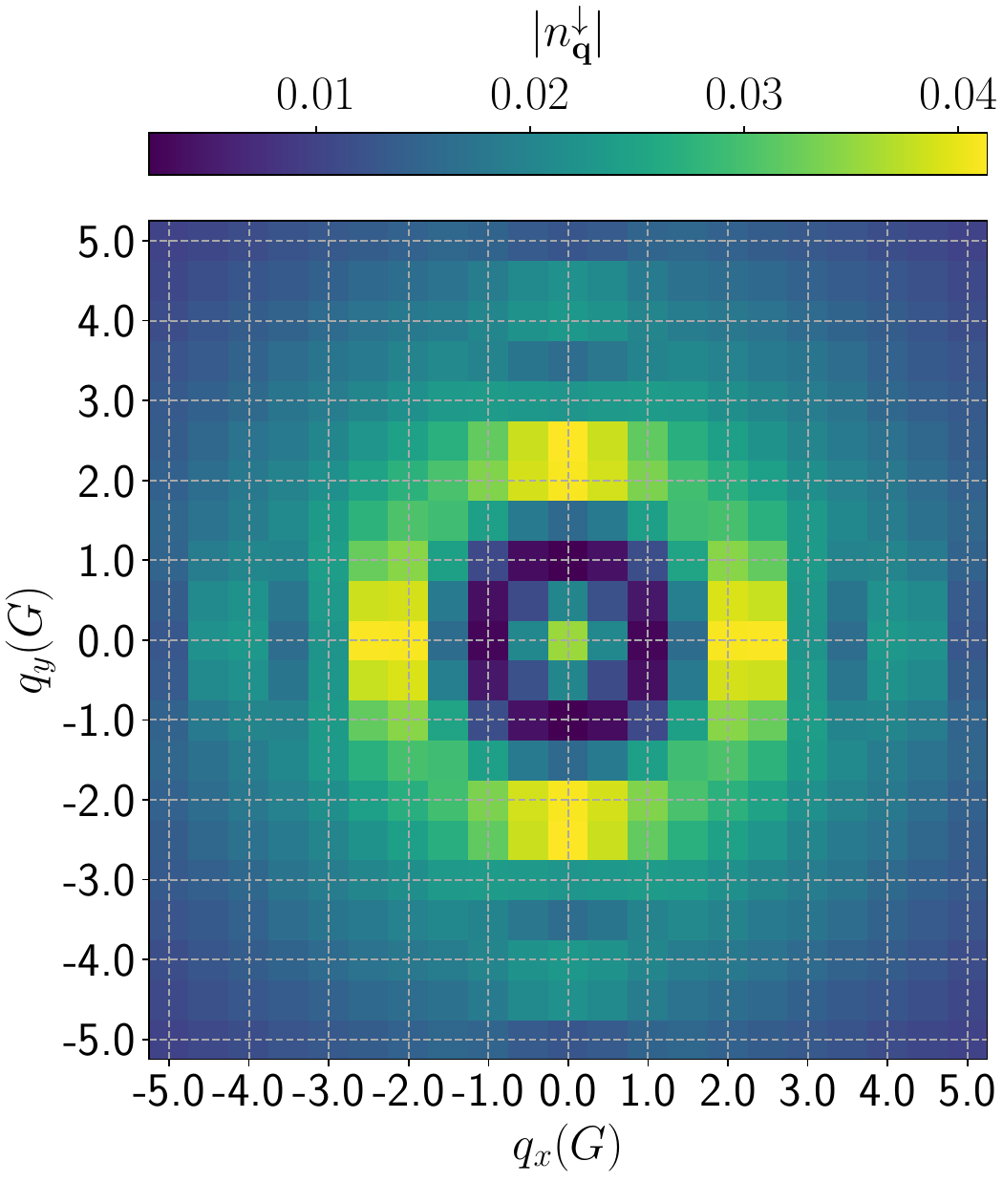}
    \caption{$v_0=0.45$, spin-down}
  \end{subfigure}
  \hfill
  \begin{subfigure}[b]{0.32\columnwidth}
    \centering
    \includegraphics[width=\columnwidth]{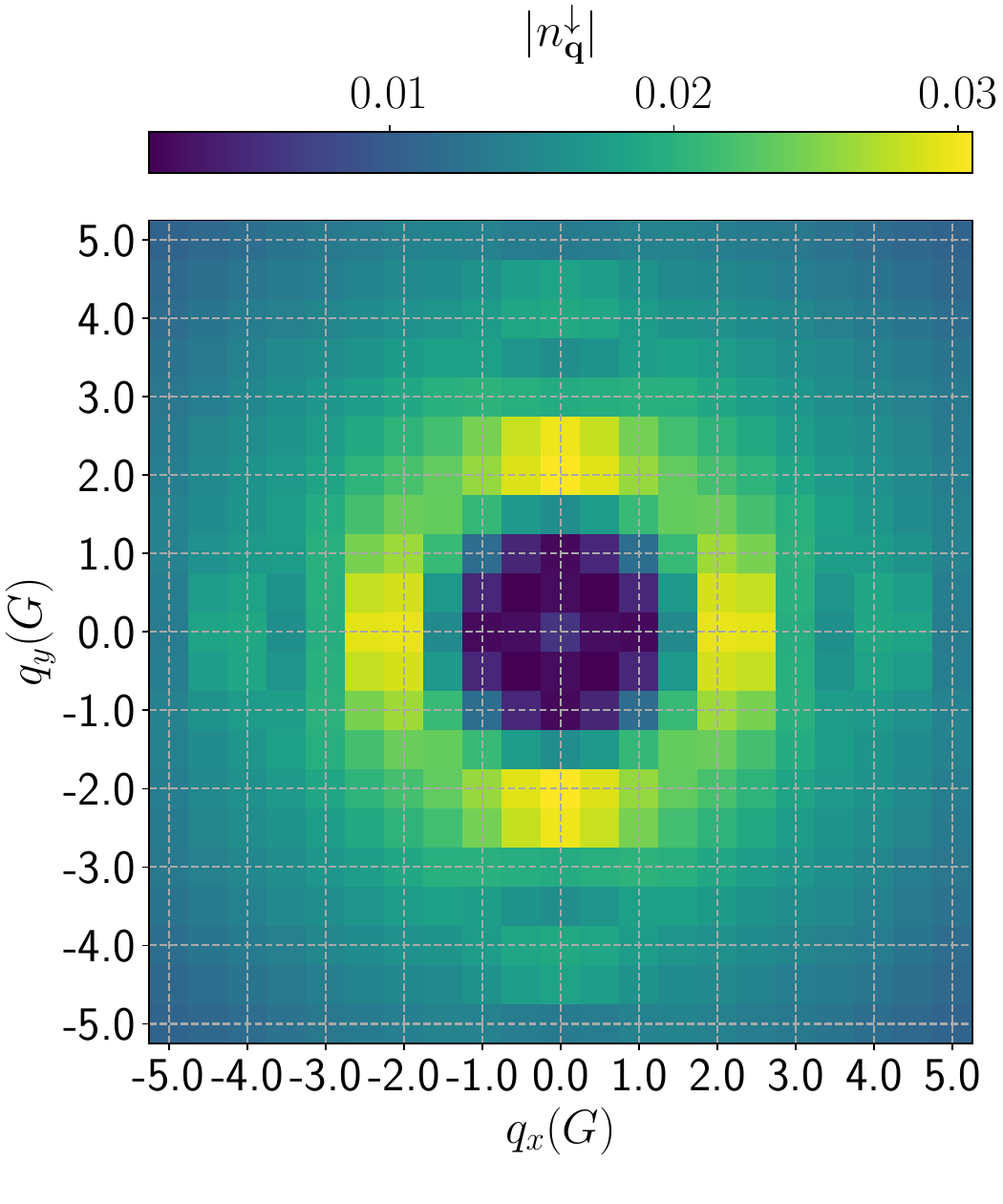}
    \caption{$v_0=0.5$, spin-down}
  \end{subfigure}
  \caption{Momentum density of the $(N^\uparrow, N^\downarrow) = (13, 5)$ SIFG for $v_0=0.4, 0.45, 0.5$.}
  \label{fig:13_5-strong-md}
\end{figure}

\begin{figure}[ht!]
  \centering
  \begin{subfigure}[b]{0.32\columnwidth}
    \centering
    \includegraphics[width=\columnwidth]{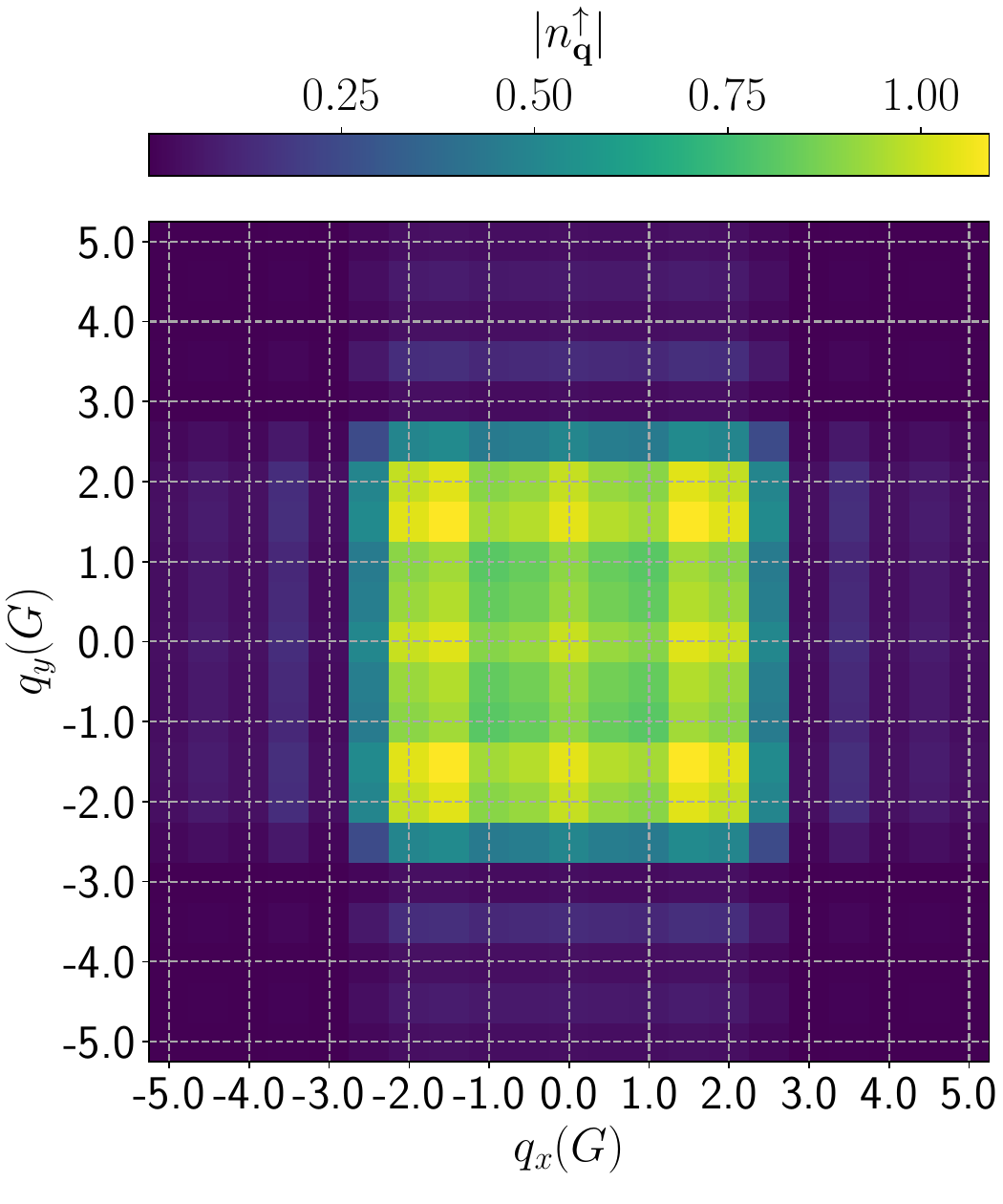}
    \caption{$v_0=0.1$, spin-up}
  \end{subfigure}
  \hfill
  \begin{subfigure}[b]{0.32\columnwidth}
    \centering
    \includegraphics[width=\columnwidth]{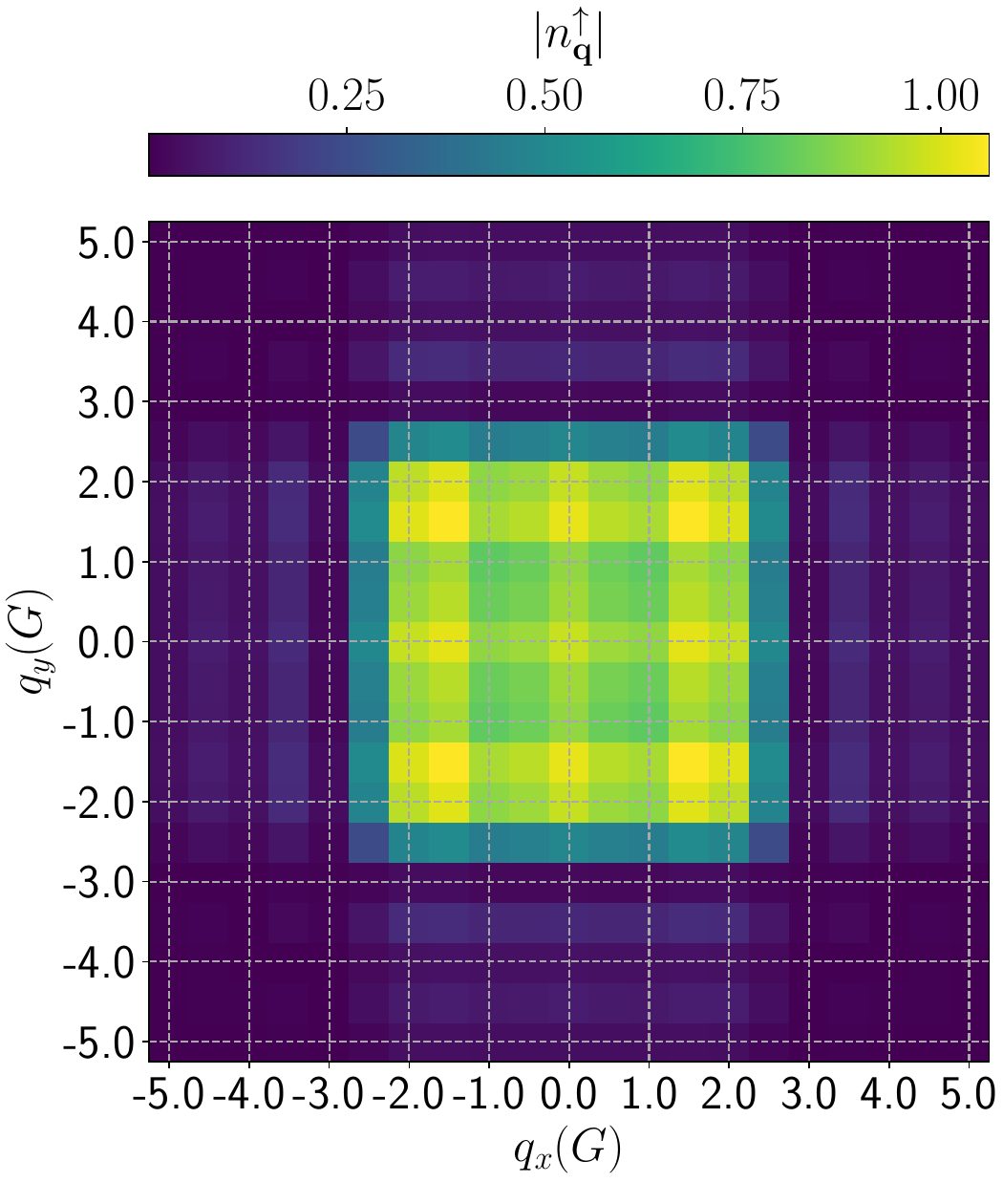}
    \caption{$v_0=0.15$, spin-up}
  \end{subfigure}
  \hfill
  \begin{subfigure}[b]{0.32\columnwidth}
    \centering
    \includegraphics[width=\columnwidth]{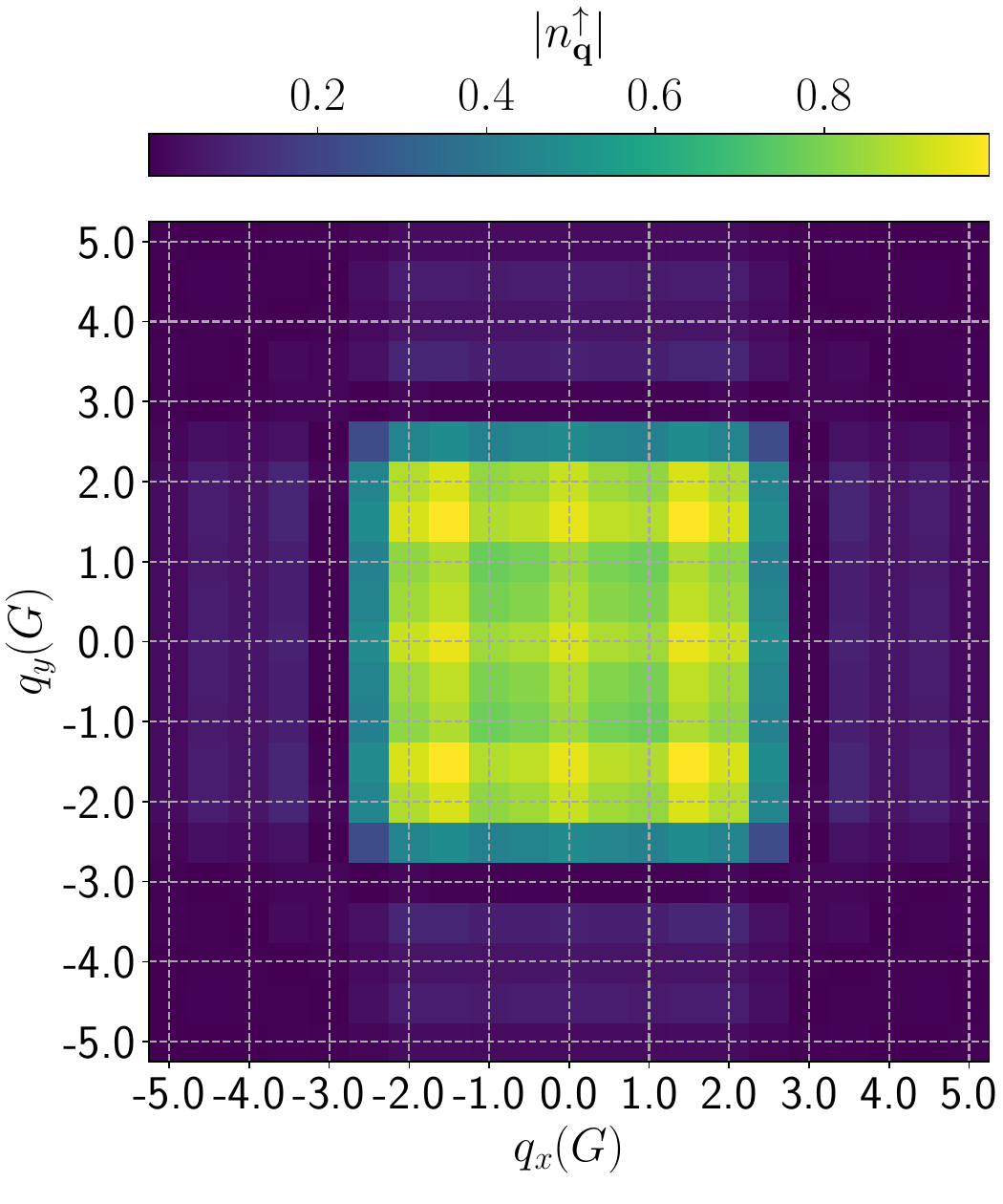}
    \caption{$v_0=0.2$, spin-up}
  \end{subfigure}
  \hfill
  \begin{subfigure}[b]{0.32\columnwidth}
    \centering
    \includegraphics[width=\columnwidth]{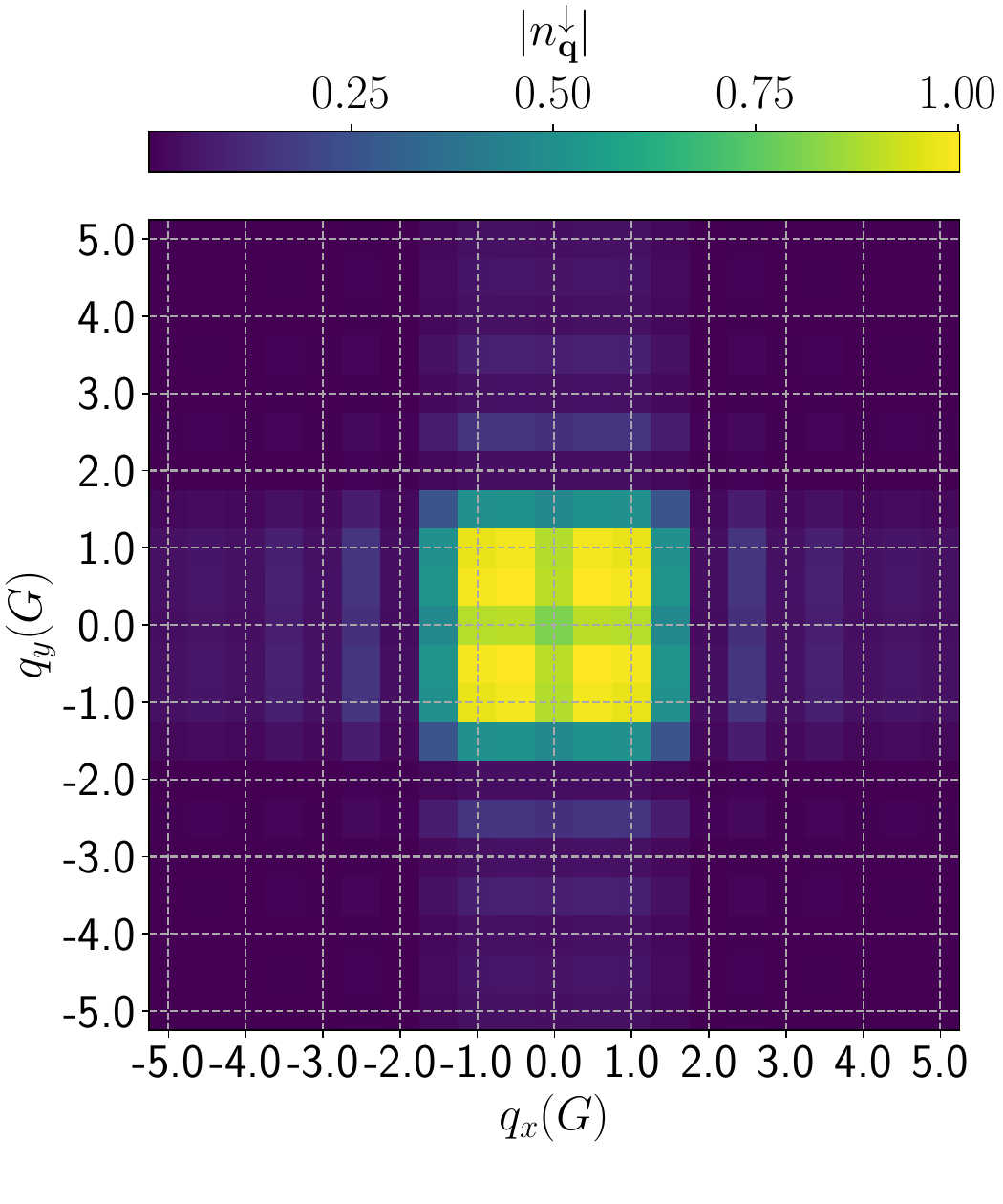}
    \caption{$v_0=0.1$, spin-down}
  \end{subfigure}
  \hfill
  \begin{subfigure}[b]{0.32\columnwidth}
    \centering
    \includegraphics[width=\columnwidth]{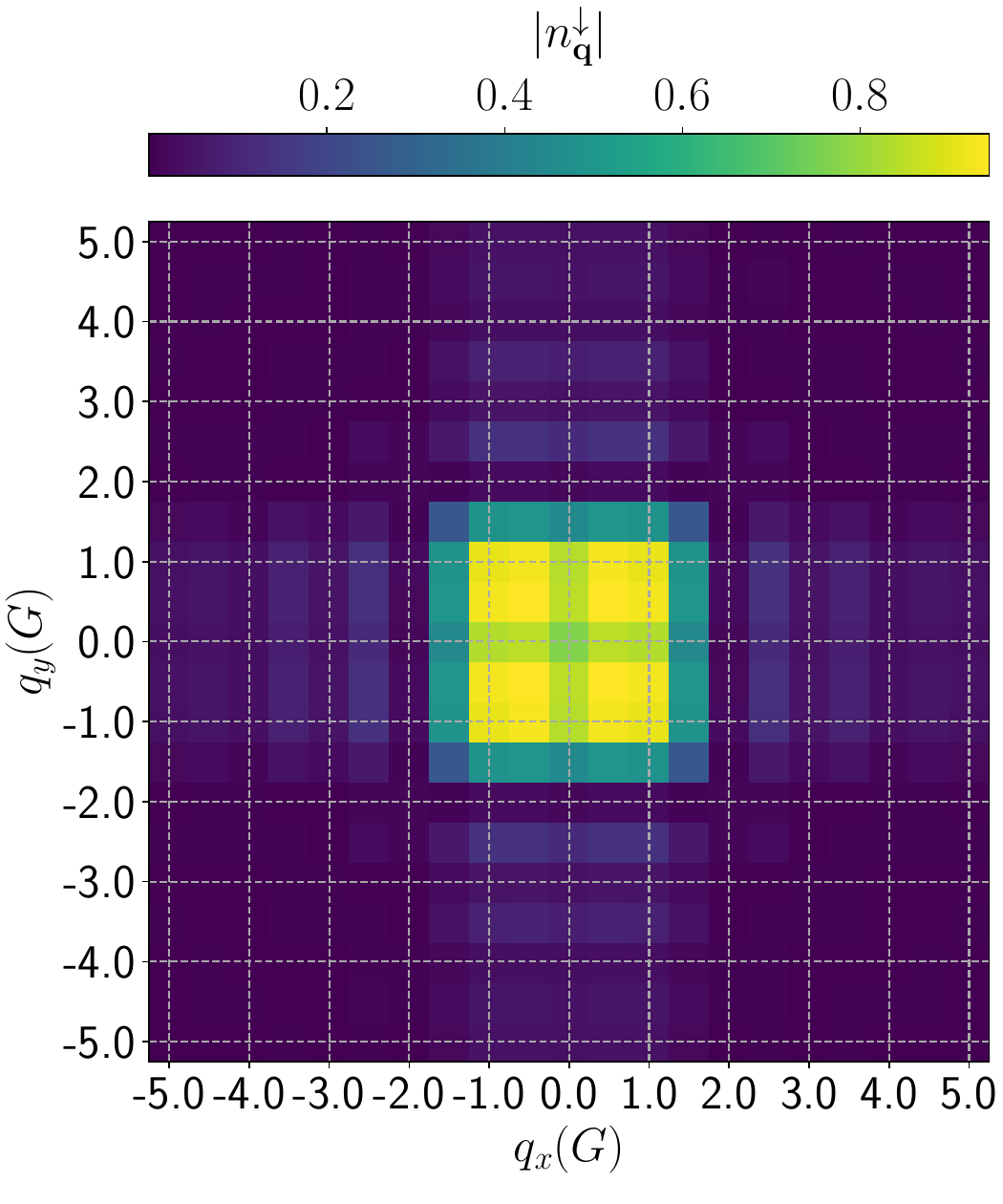}
    \caption{$v_0=0.15$, spin-down}
  \end{subfigure}
  \hfill
  \begin{subfigure}[b]{0.32\columnwidth}
    \centering
    \includegraphics[width=\columnwidth]{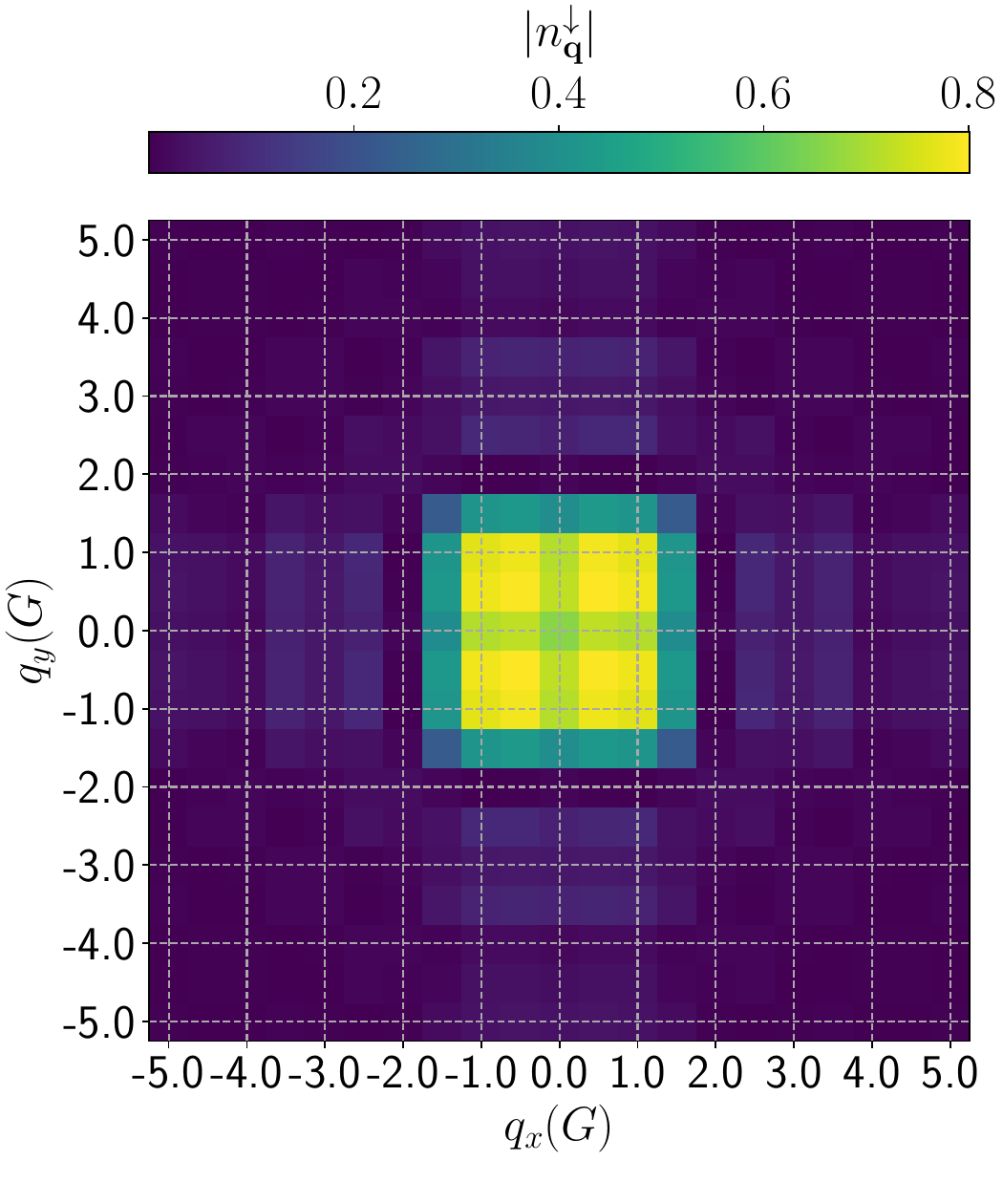}
    \caption{$v_0=0.2$, spin-down}
  \end{subfigure}
  \caption{Momentum density of the $(N^\uparrow, N^\downarrow) = (25, 9)$ SIFG for $v_0=0.1, 0.15, 0.2$.}
  \label{fig:25_9-weak-md}
\end{figure}

\begin{figure}[ht!]
  \centering
  \begin{subfigure}[b]{0.32\columnwidth}
    \centering
    \includegraphics[width=\columnwidth]{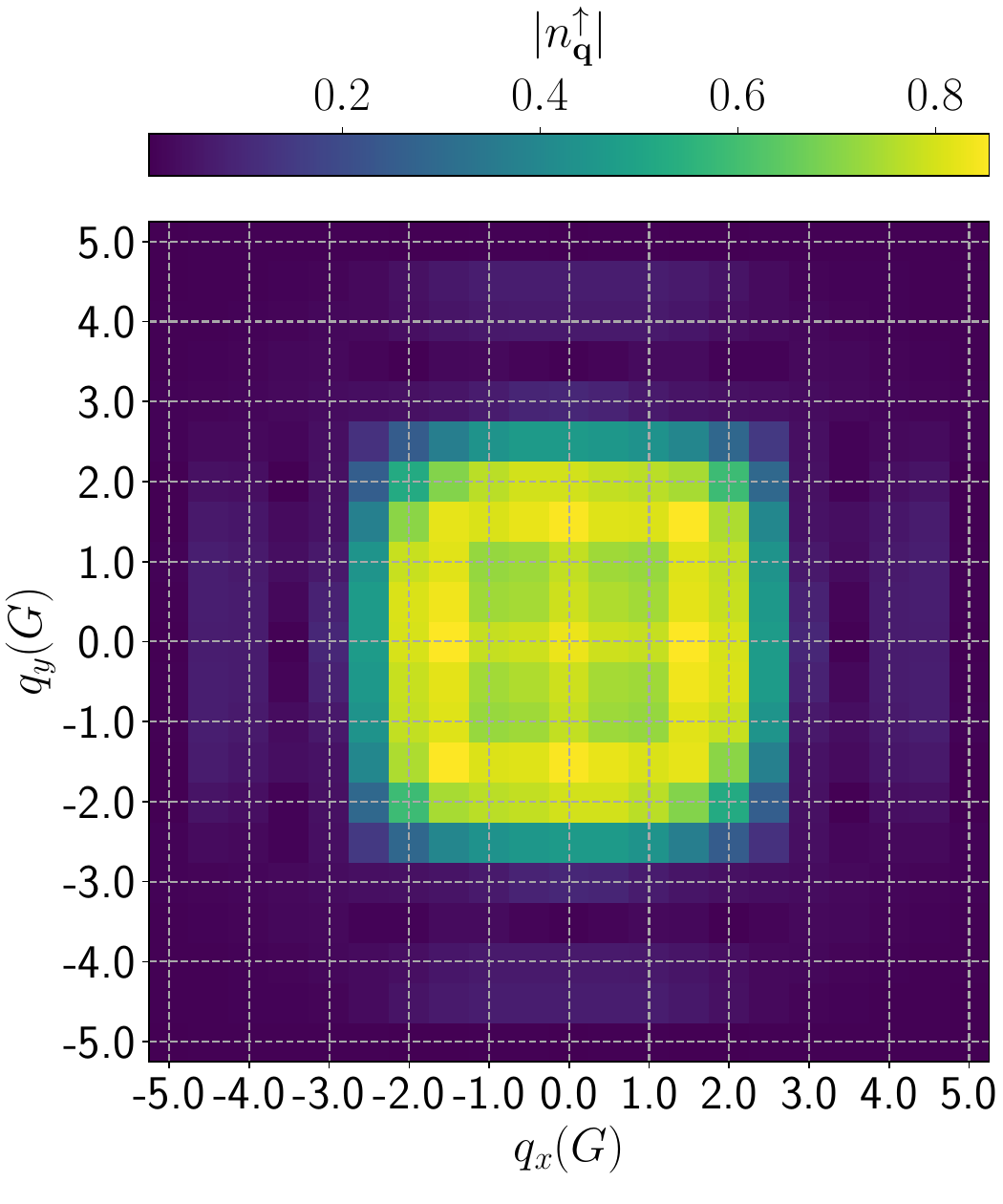}
    \caption{$v_0=0.25$, spin-up}
  \end{subfigure}
  \hfill
  \begin{subfigure}[b]{0.32\columnwidth}
    \centering
    \includegraphics[width=\columnwidth]{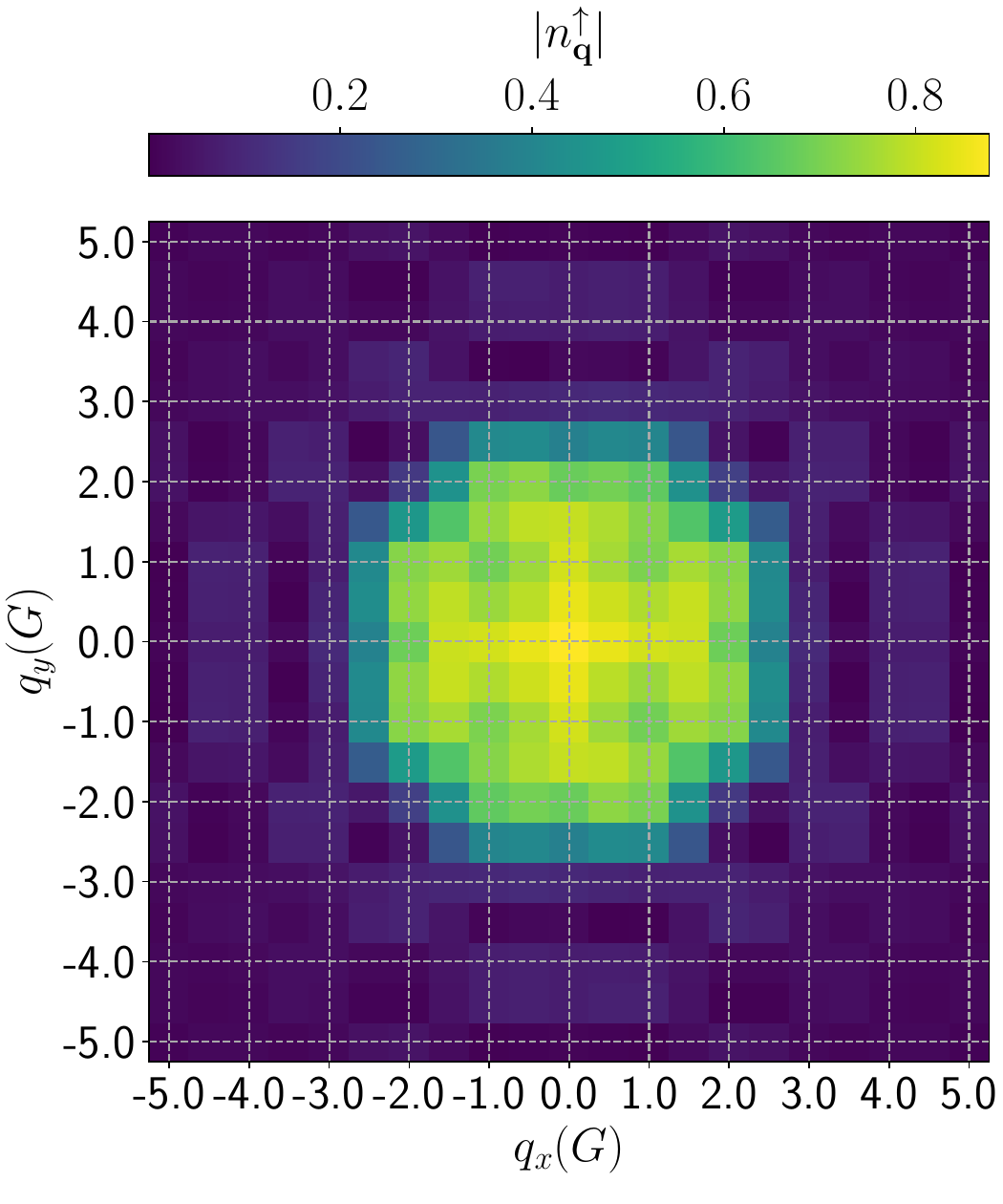}
    \caption{$v_0=0.3$, spin-up}
  \end{subfigure}
  \begin{subfigure}[b]{0.32\columnwidth}
    \centering
    \includegraphics[width=\columnwidth]{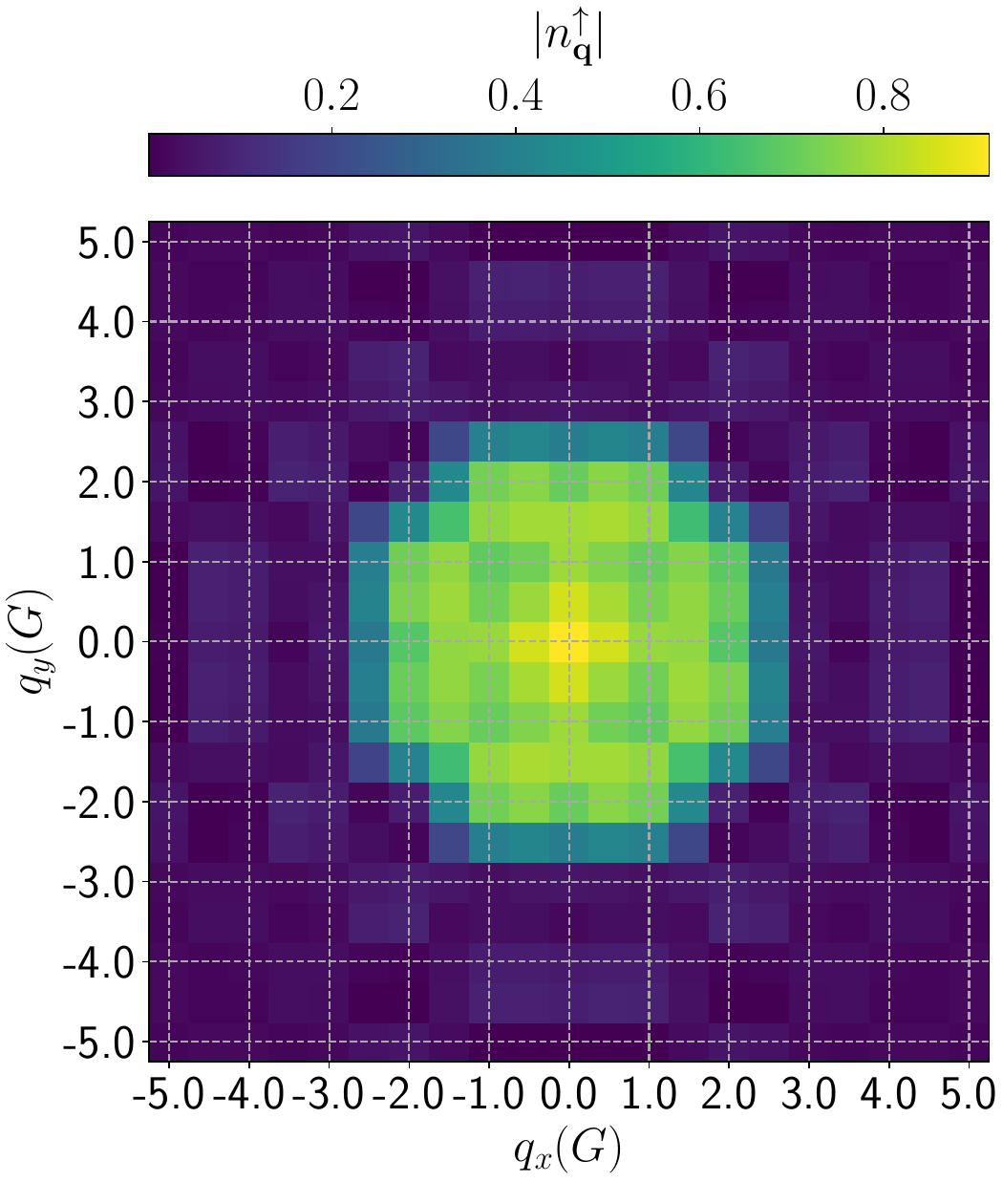}
    \caption{$v_0=0.35$, spin-up}
  \end{subfigure}
  \hfill
  \begin{subfigure}[b]{0.32\columnwidth}
    \centering
    \includegraphics[width=\columnwidth]{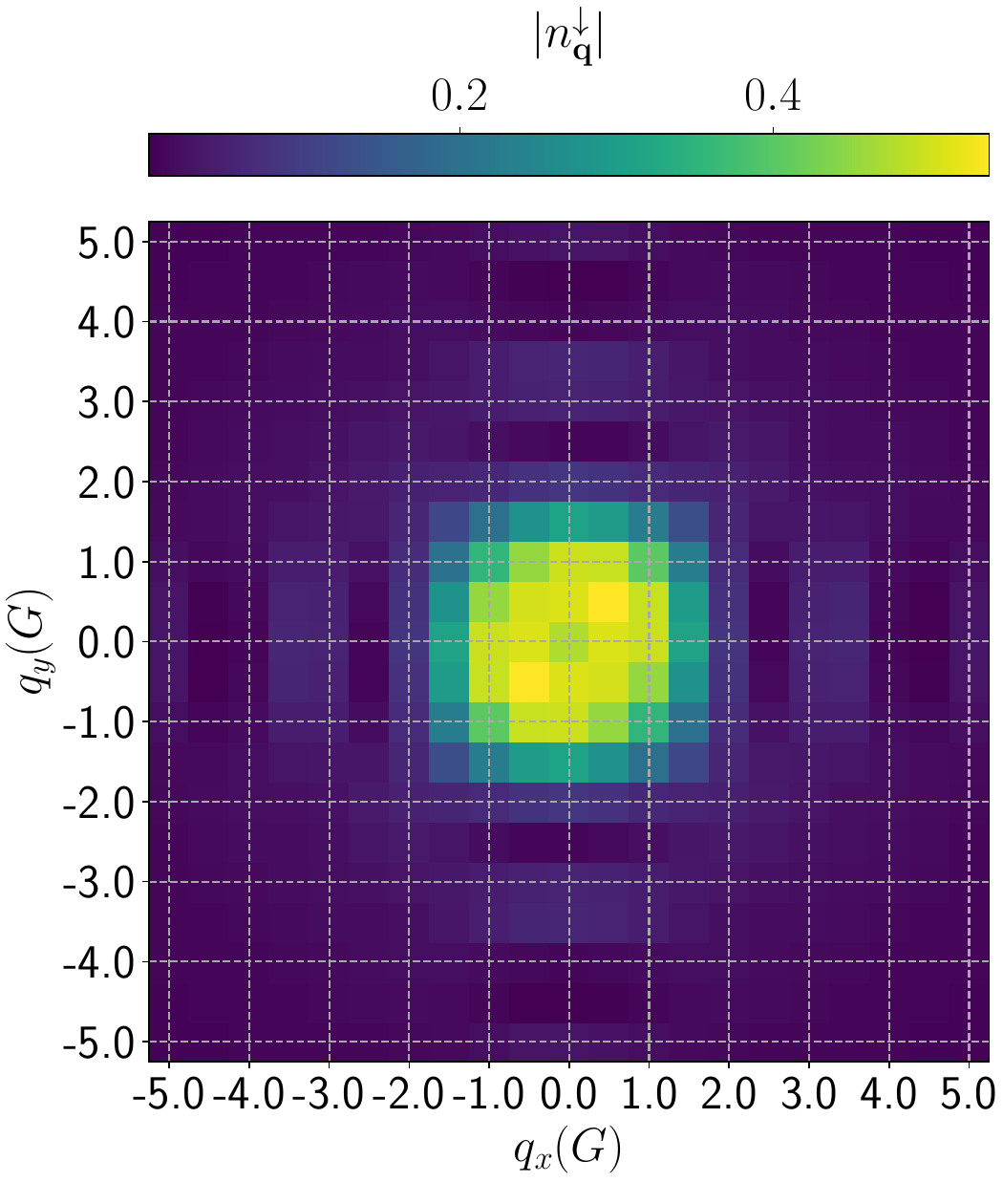}
    \caption{$v_0=0.25$, spin-down}
  \end{subfigure}
  \hfill
  \begin{subfigure}[b]{0.32\columnwidth}
    \centering
    \includegraphics[width=\columnwidth]{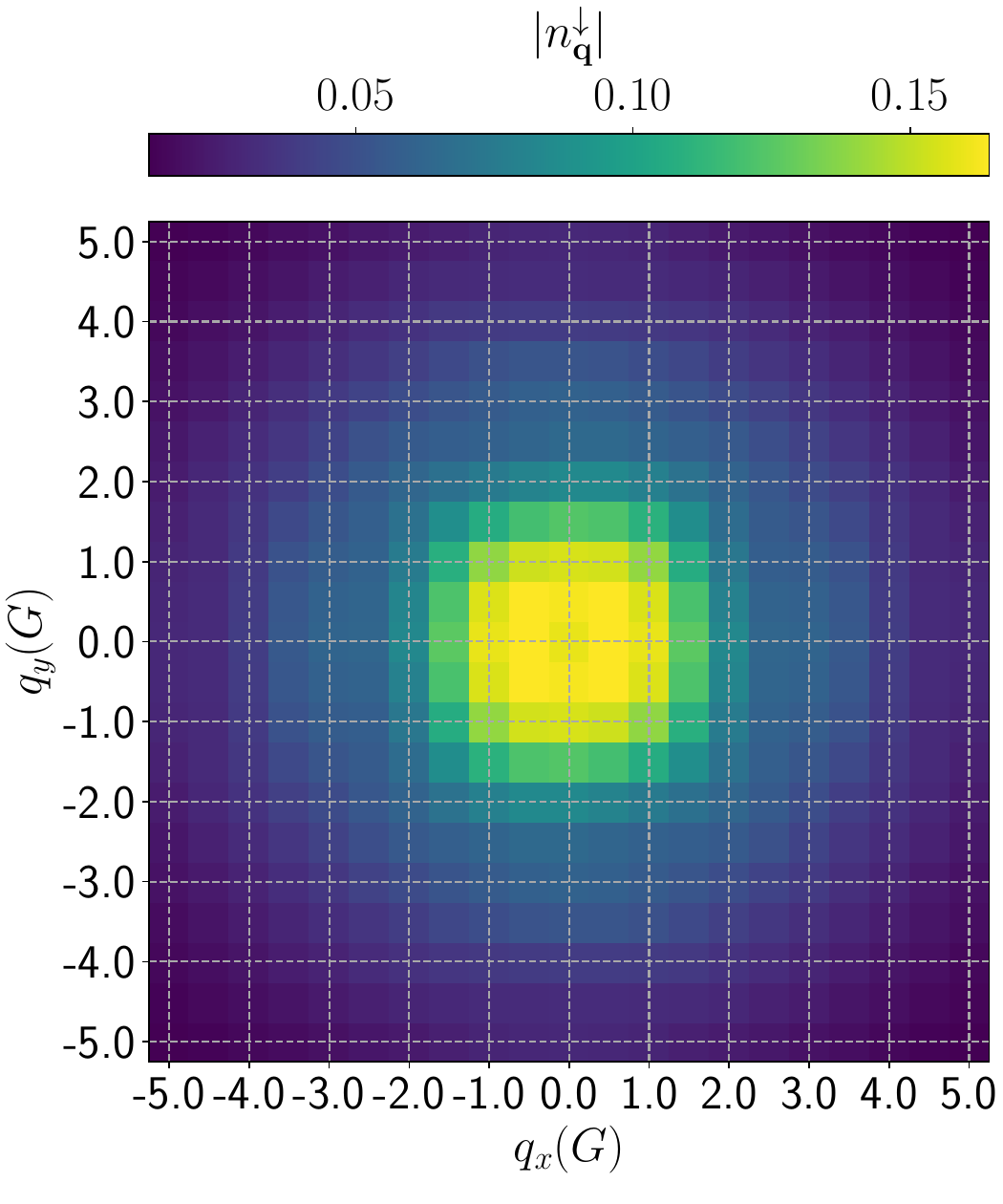}
    \caption{$v_0=0.3$, spin-down}
  \end{subfigure}
  \hfill
  \begin{subfigure}[b]{0.32\columnwidth}
    \centering
    \includegraphics[width=\columnwidth]{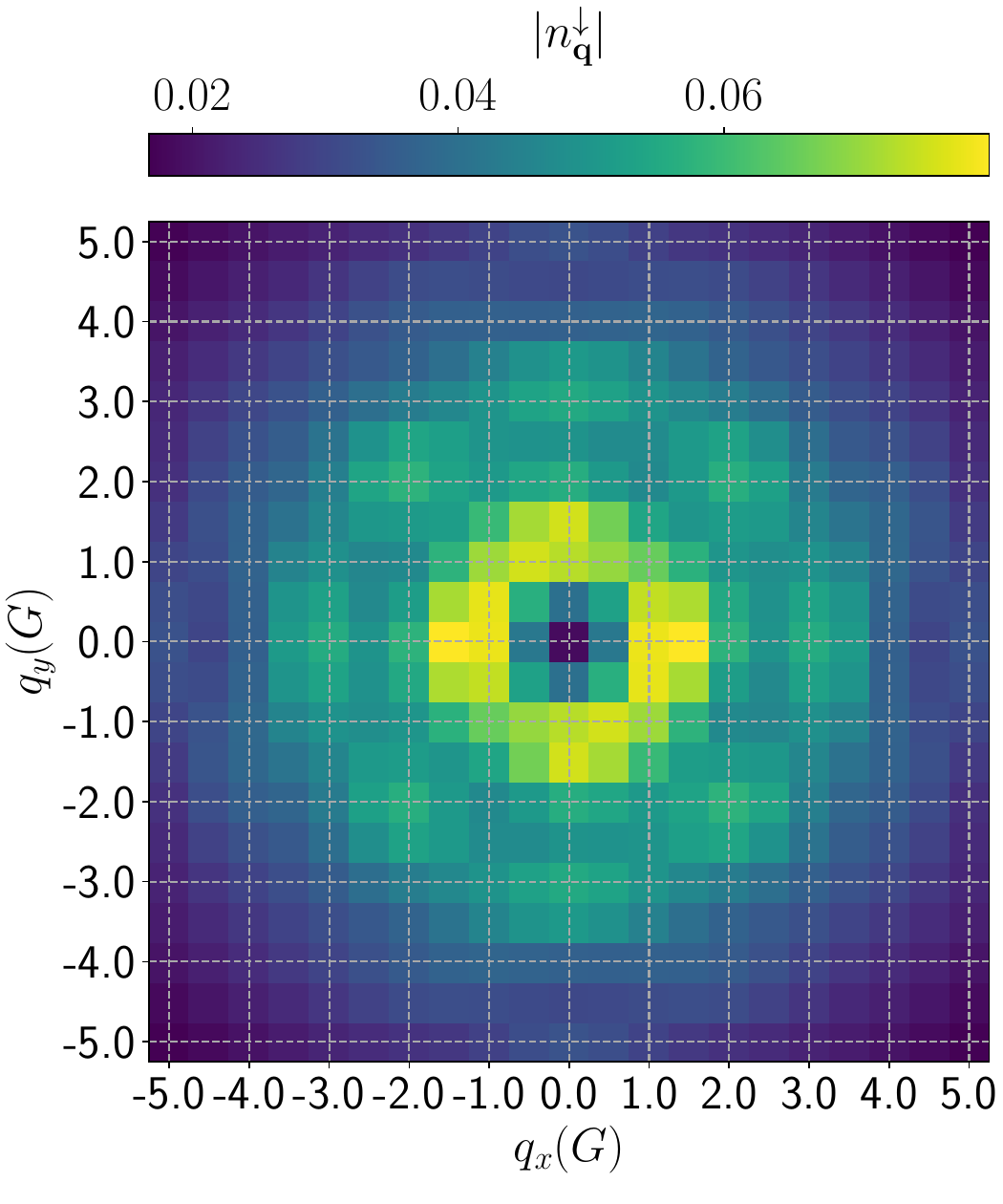}
    \caption{$v_0=0.35$, spin-down}
  \end{subfigure}
  \caption{Momentum density of the $(N^\uparrow, N^\downarrow) = (25, 9)$ SIFG for $v_0=0.25, 0.3, 0.35$.}
  \label{fig:25_9-mid-md}
\end{figure}

\begin{figure}[ht!]
  \centering
  \begin{subfigure}[b]{0.32\columnwidth}
    \centering
    \includegraphics[width=\columnwidth]{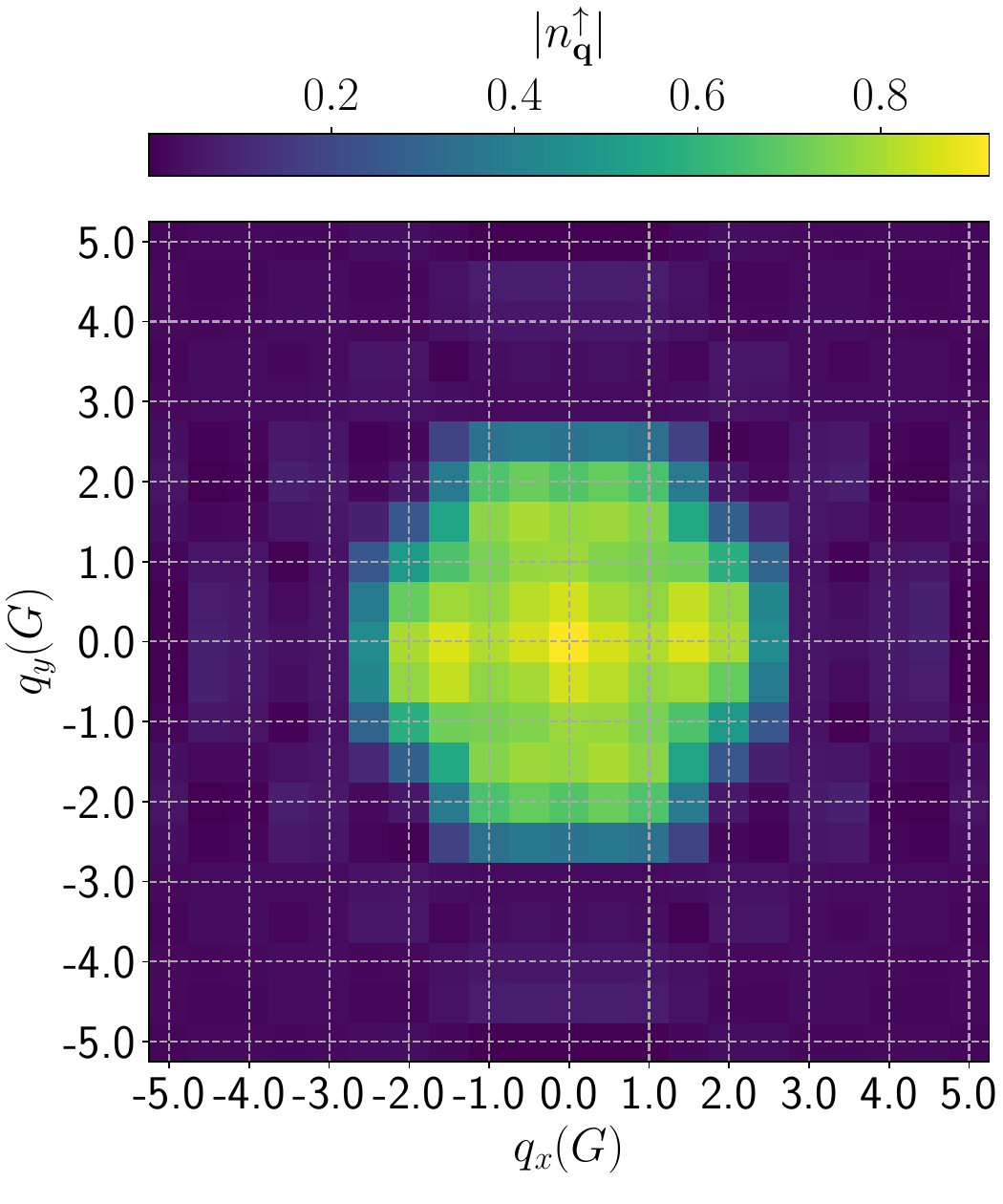}
    \caption{$v_0=0.4$, spin-up}
  \end{subfigure}
  \hfill
  \begin{subfigure}[b]{0.32\columnwidth}
    \centering
    \includegraphics[width=\columnwidth]{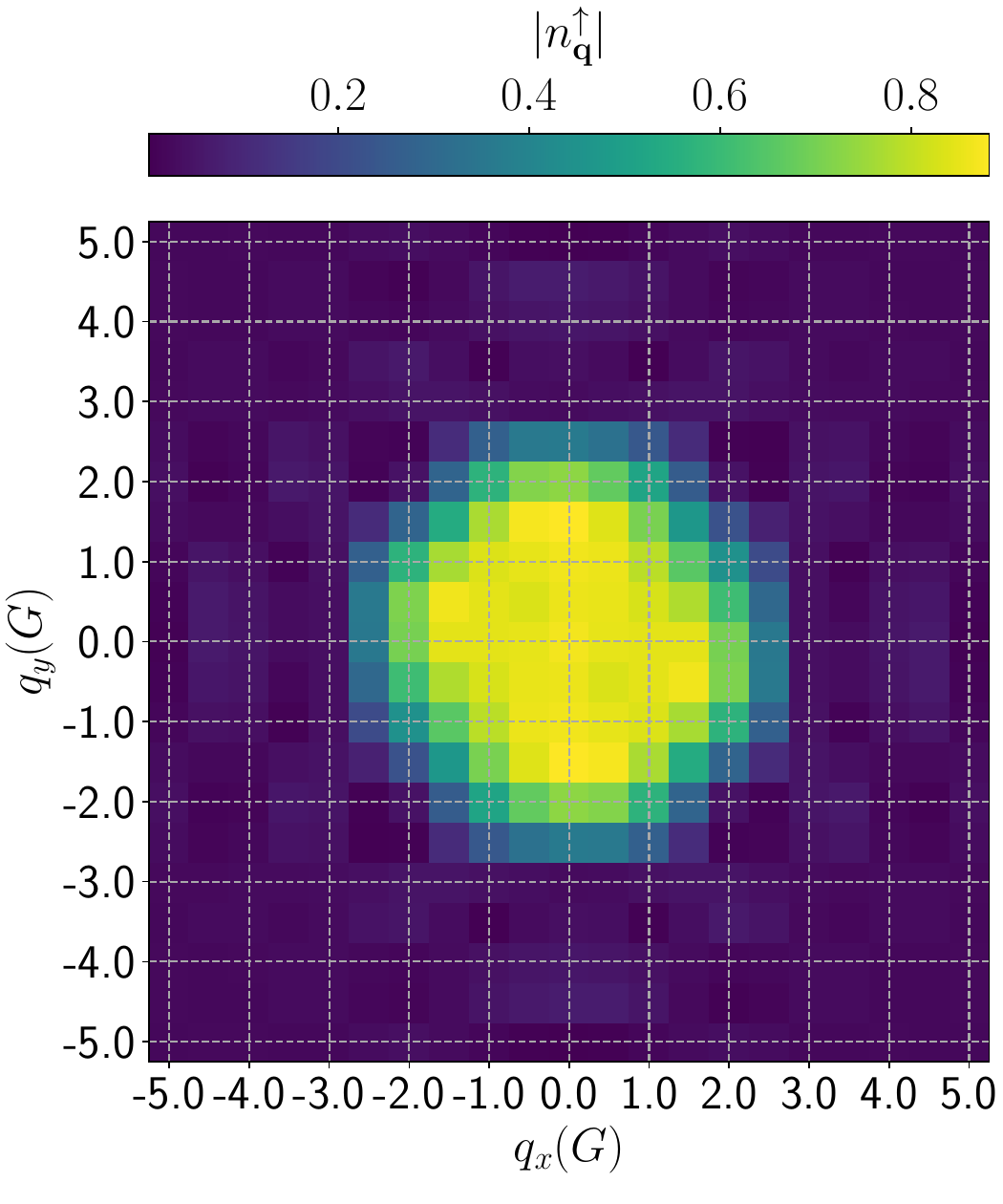}
    \caption{$v_0=0.45$, spin-up}
  \end{subfigure}
  \hfill
  \begin{subfigure}[b]{0.32\columnwidth}
    \centering
    \includegraphics[width=\columnwidth]{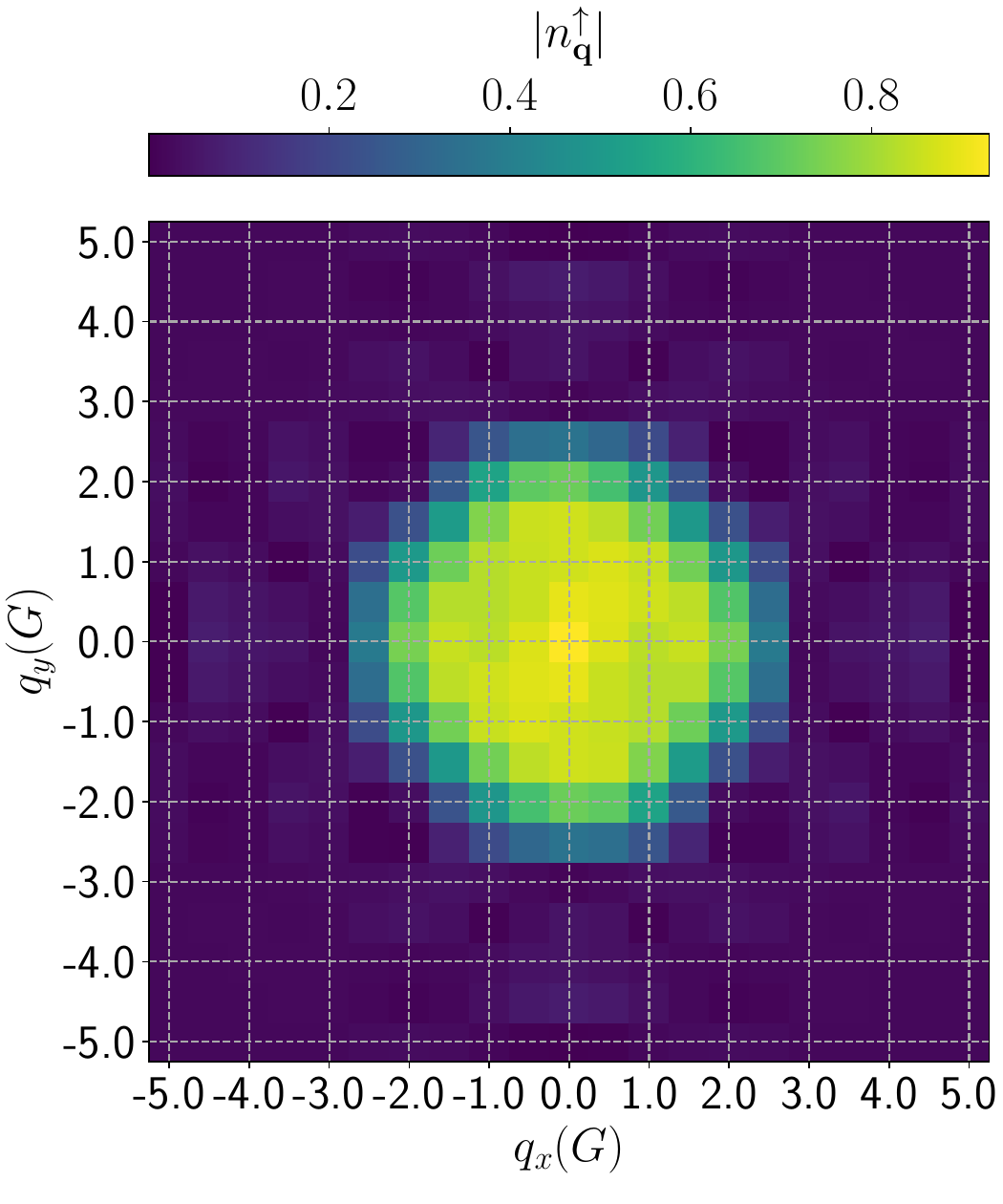}
    \caption{$v_0=0.5$, spin-up}
  \end{subfigure}
  \hfill
  \begin{subfigure}[b]{0.32\columnwidth}
    \centering
    \includegraphics[width=\columnwidth]{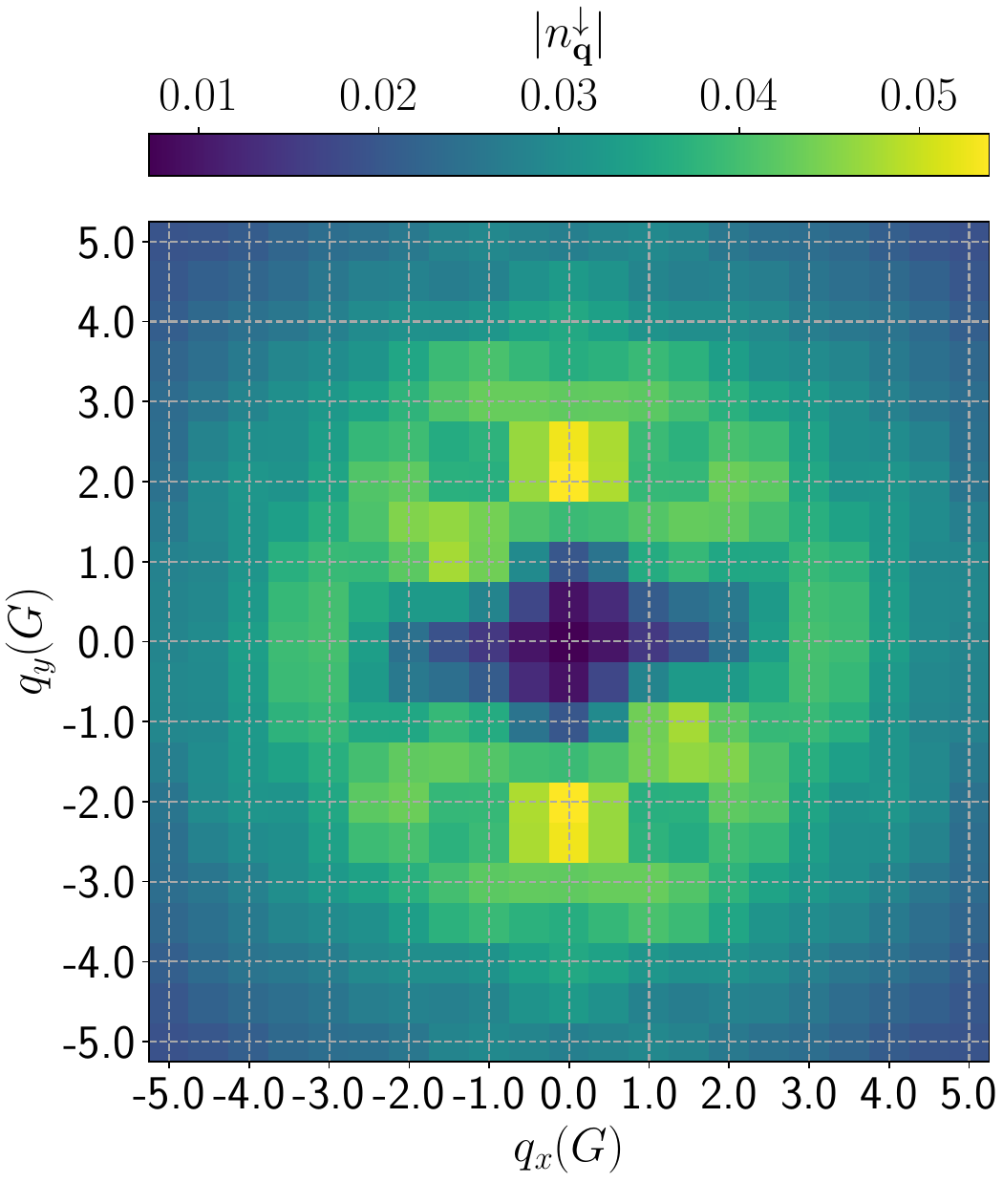}
    \caption{$v_0=0.4$, spin-down}
  \end{subfigure}
  \hfill
  \begin{subfigure}[b]{0.32\columnwidth}
    \centering
    \includegraphics[width=\columnwidth]{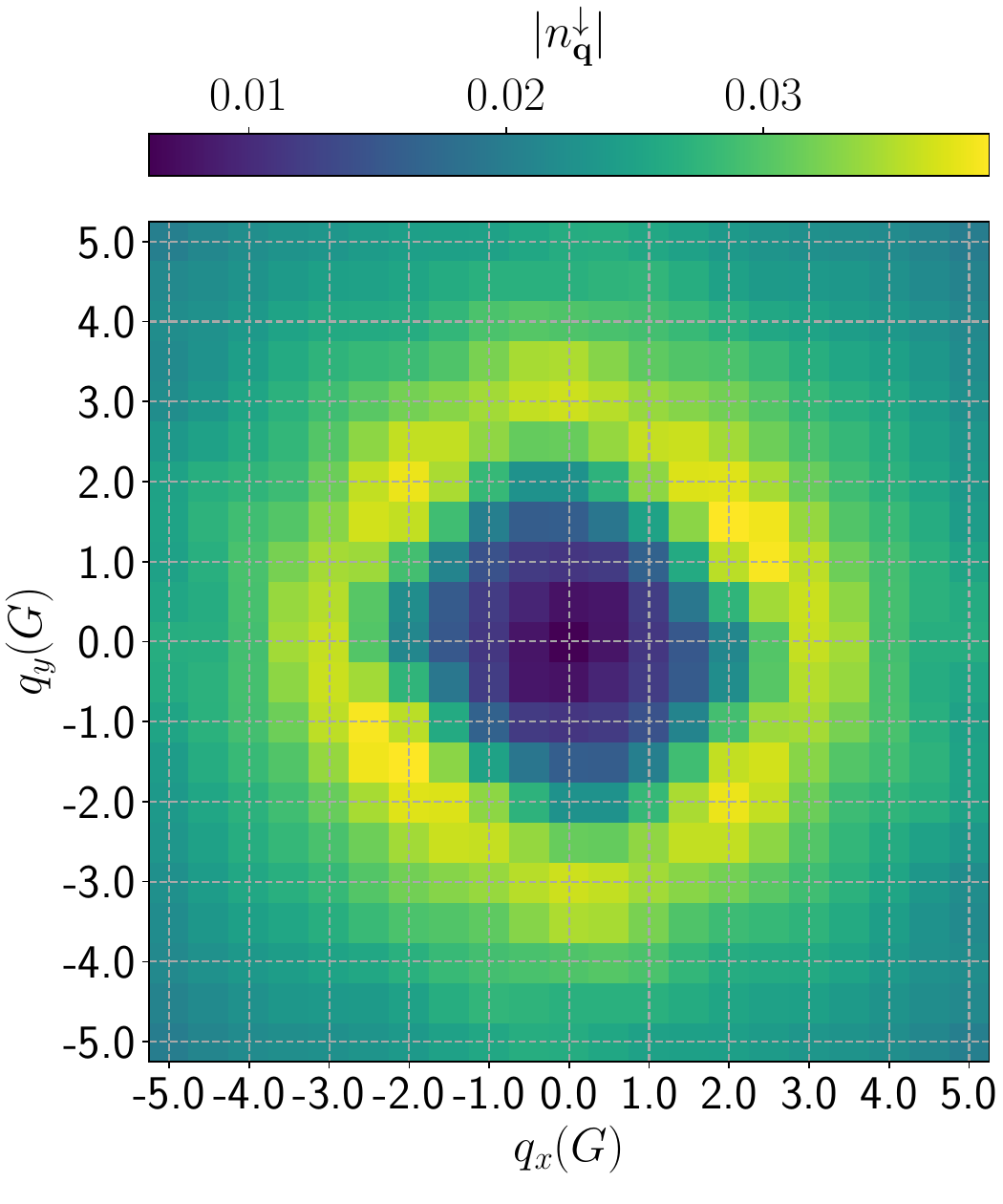}
    \caption{$v_0=0.45$, spin-down}
  \end{subfigure}
  \hfill
  \begin{subfigure}[b]{0.32\columnwidth}
    \centering
    \includegraphics[width=\columnwidth]{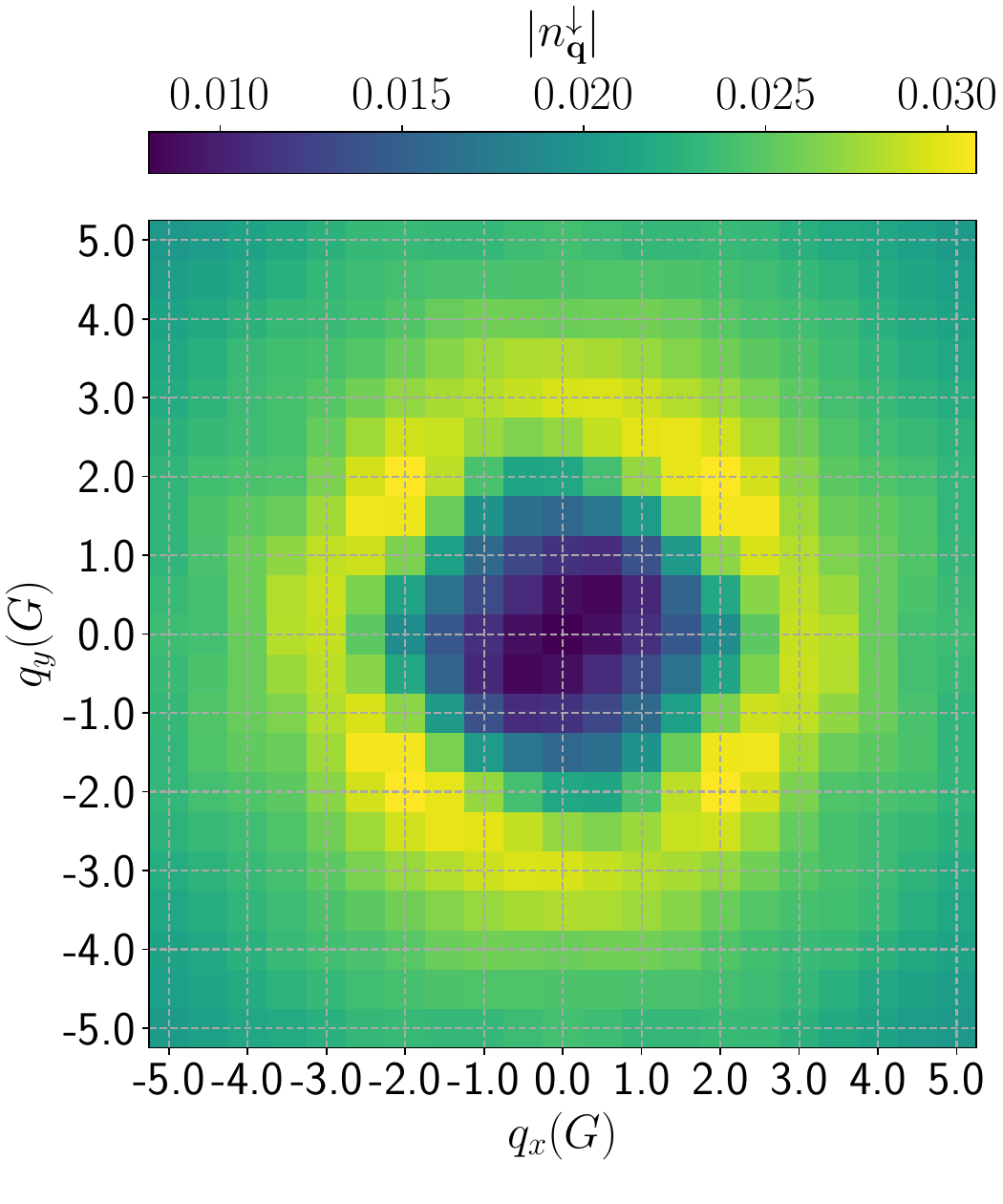}
    \caption{$v_0=0.5$, spin-down}
  \end{subfigure}
  \caption{Momentum density of the $(N^\uparrow, N^\downarrow) = (25, 9)$ SIFG for $v_0=0.4, 0.45, 0.5$.}
  \label{fig:25_9-strong-md}
\end{figure}

In this subsection, we present the momentum density of each spin species of the 2D spin-imbalanced Fermi gas with $(N^\uparrow, N^\downarrow) = (13, 5)$ and $(N^\uparrow, N^\downarrow) = (25, 9)$, respectively, across the BCS-BEC crossover.
Results of the $(N^\uparrow, N^\downarrow) = (13, 5)$ system are shown in \cref{fig:13_5-weak-md,fig:13_5-mid-md,fig:13_5-strong-md} and the results of the $(N^\uparrow, N^\downarrow) = (25, 9)$ system are shown in \cref{fig:25_9-weak-md,fig:25_9-mid-md,fig:25_9-strong-md}.
In each figure, the top and bottom rows represent the momentum density of the spin-up and spin-down particles, respectively.
The strength of interactions, $v_0$, are mentioned in the caption of each subfigure.

\subsection{Condensate Fraction}
\begin{figure}[ht!]
  \centering
  \begin{subfigure}[b]{0.32\columnwidth}
    \centering
    \includegraphics[width=\columnwidth]{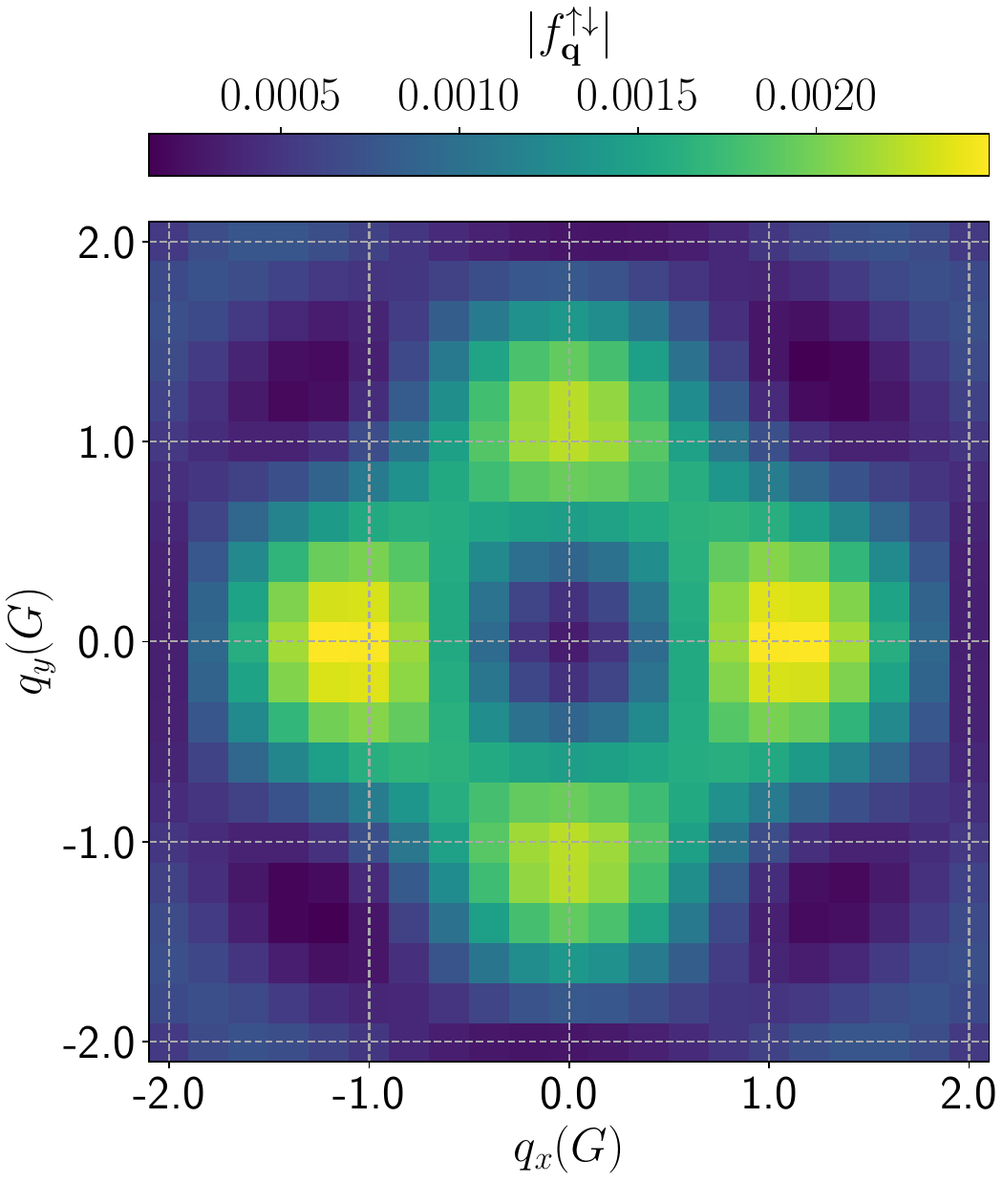}
    \caption{$v_0=0.1$}
  \end{subfigure}
  \hfill
  \begin{subfigure}[b]{0.32\columnwidth}
    \centering
    \includegraphics[width=\columnwidth]{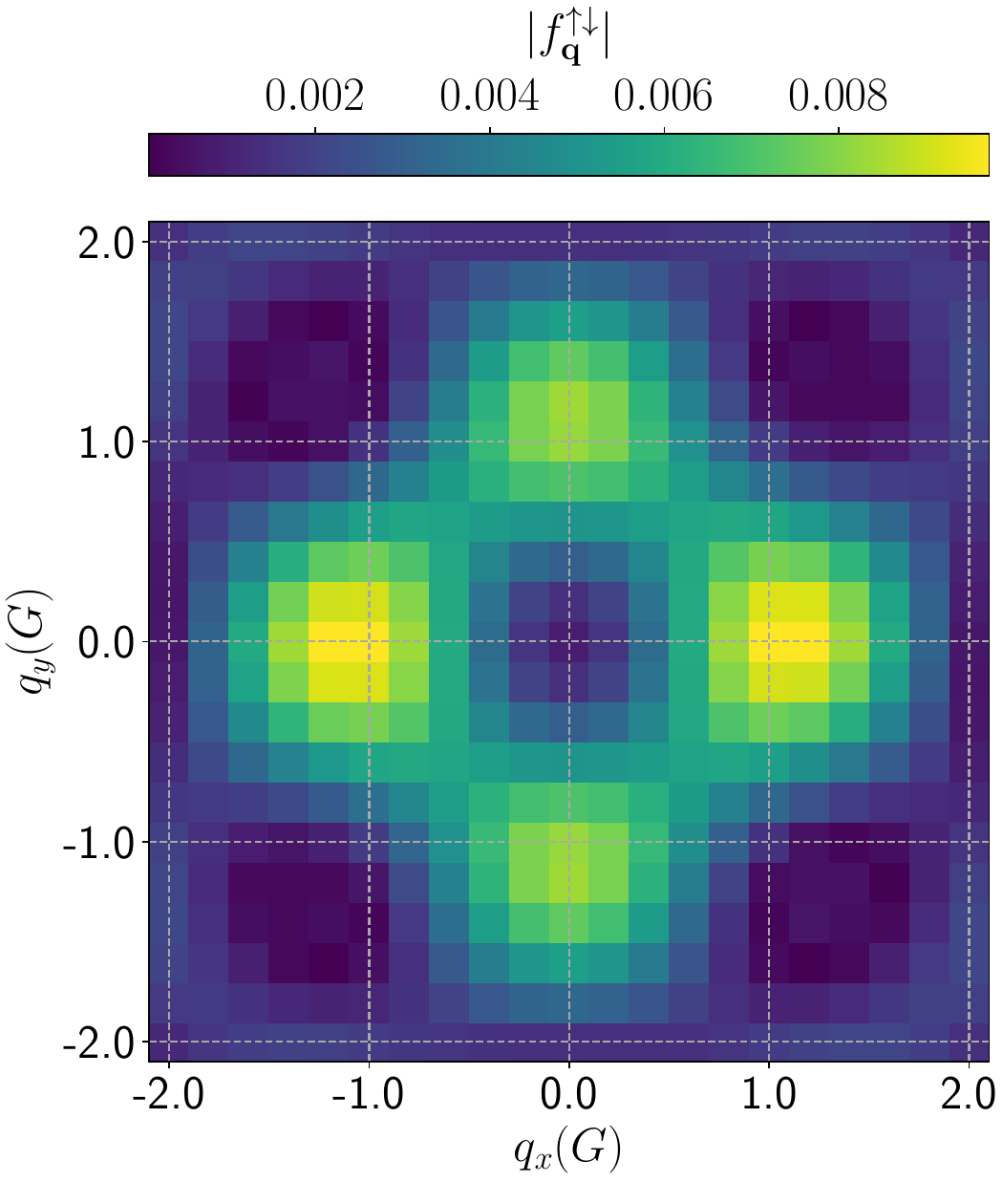}
    \caption{$v_0=0.15$}
  \end{subfigure}
  \hfill
  \begin{subfigure}[b]{0.32\columnwidth}
    \centering
    \includegraphics[width=\columnwidth]{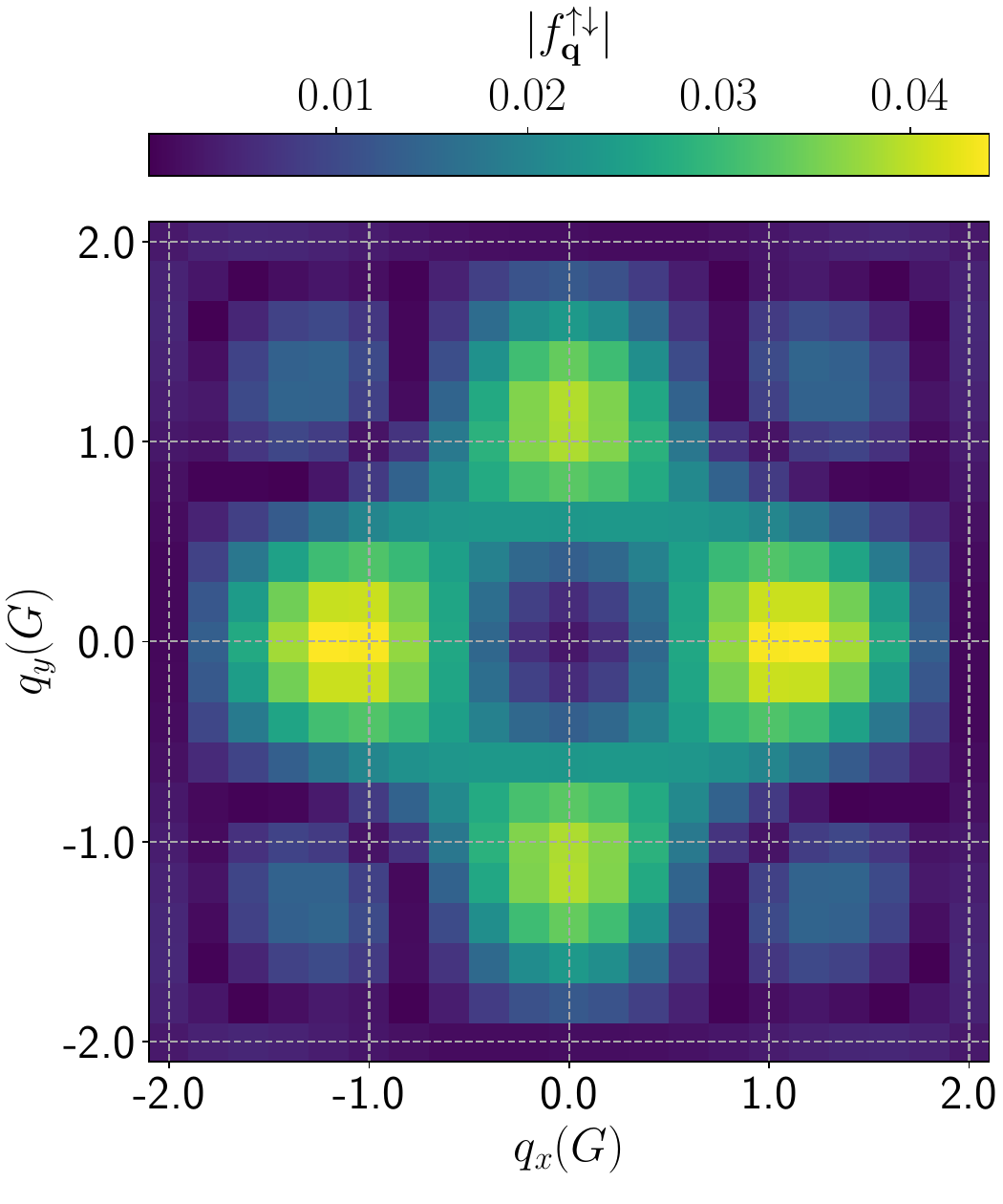}
    \caption{$v_0=0.2$}
  \end{subfigure}
  \hfill
  \begin{subfigure}[b]{0.32\columnwidth}
    \centering
    \includegraphics[width=\columnwidth]{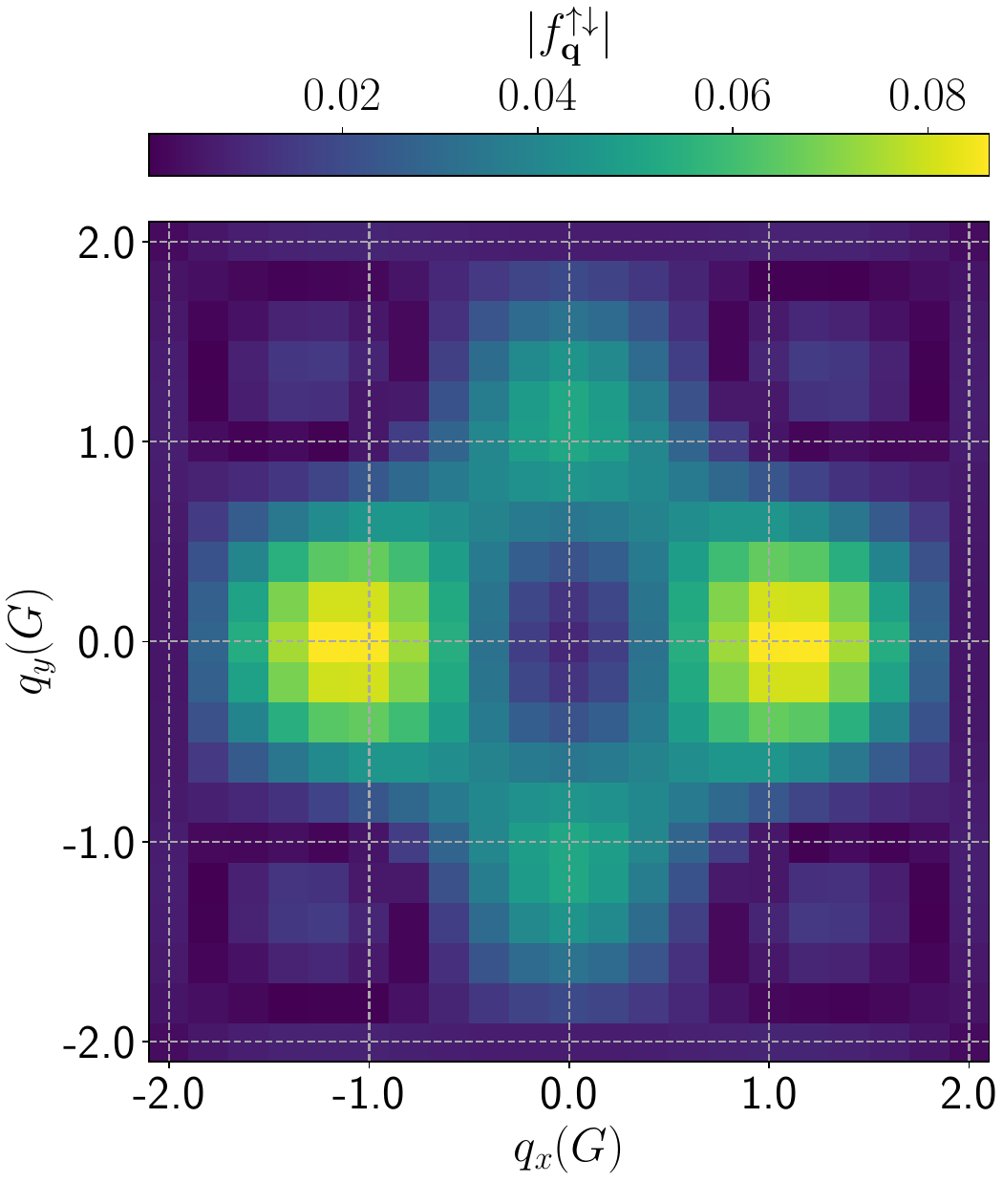}
    \caption{$v_0=0.25$}
  \end{subfigure}
  \hfill
  \begin{subfigure}[b]{0.32\columnwidth}
    \centering
    \includegraphics[width=\columnwidth]{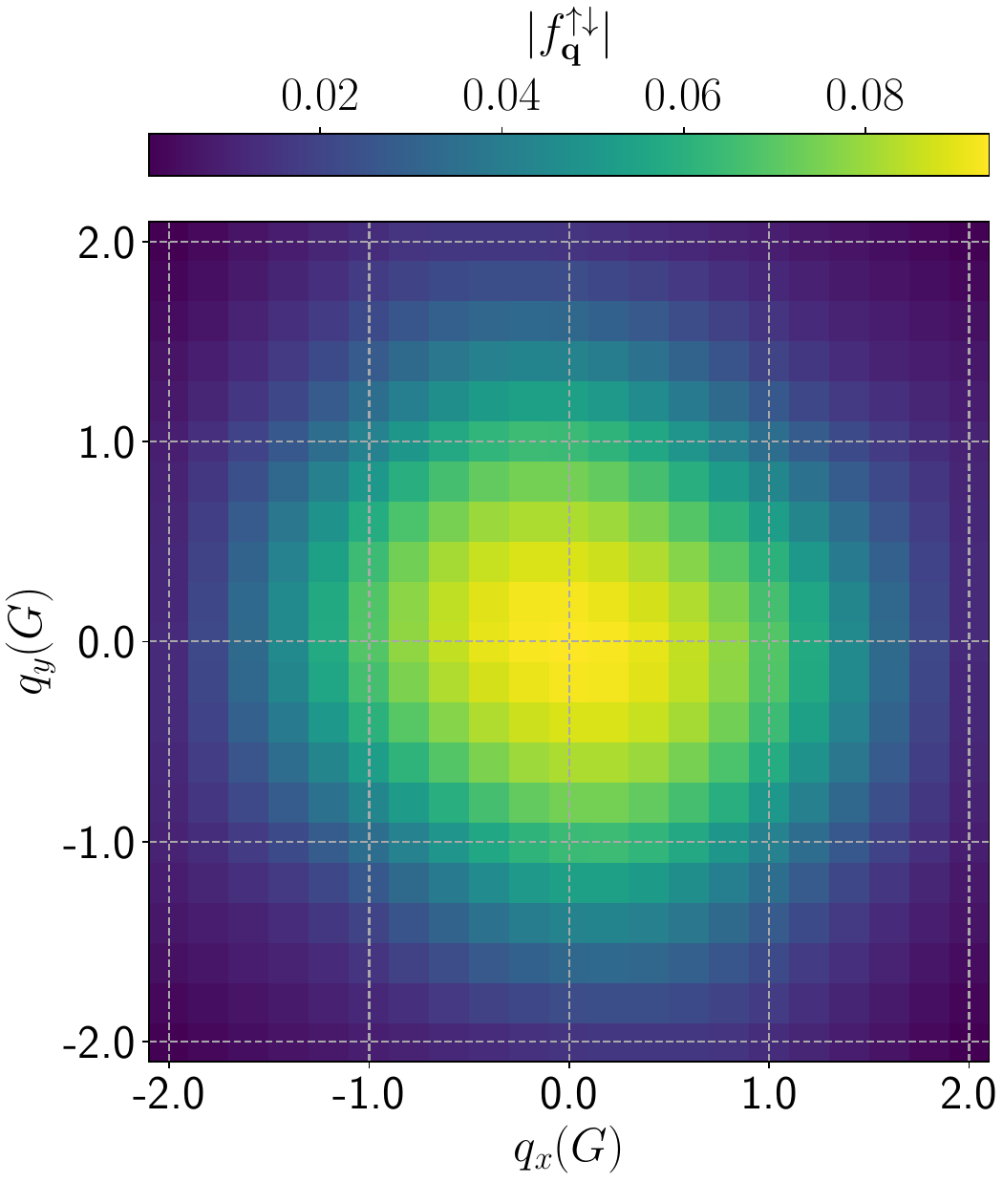}
    \caption{$v_0=0.3$}
  \end{subfigure}
  \hfill
  \begin{subfigure}[b]{0.32\columnwidth}
    \centering
    \includegraphics[width=\columnwidth]{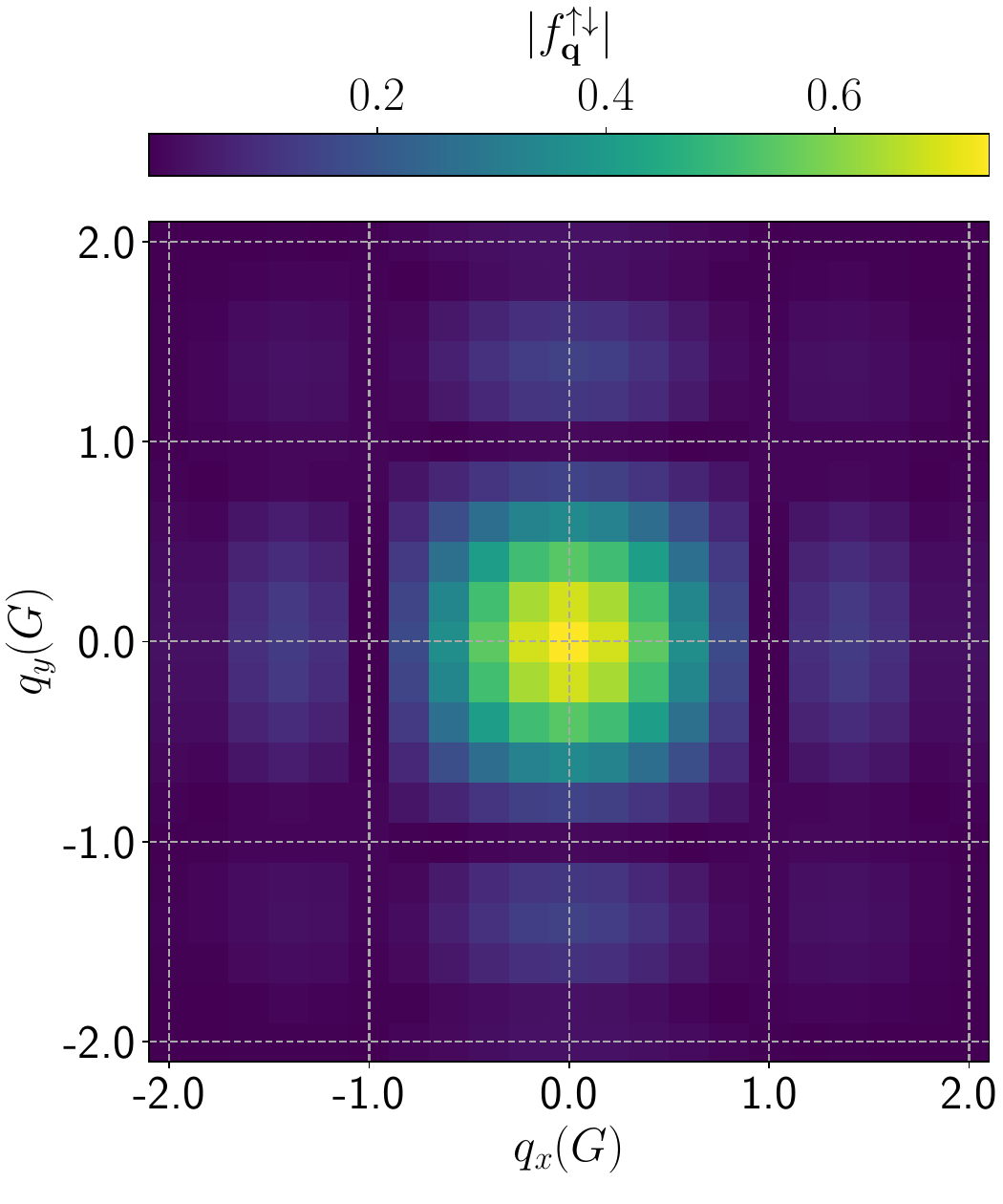}
    \caption{$v_0=0.35$}
  \end{subfigure}
  \hfill
  \begin{subfigure}[b]{0.32\columnwidth}
    \centering
    \includegraphics[width=\columnwidth]{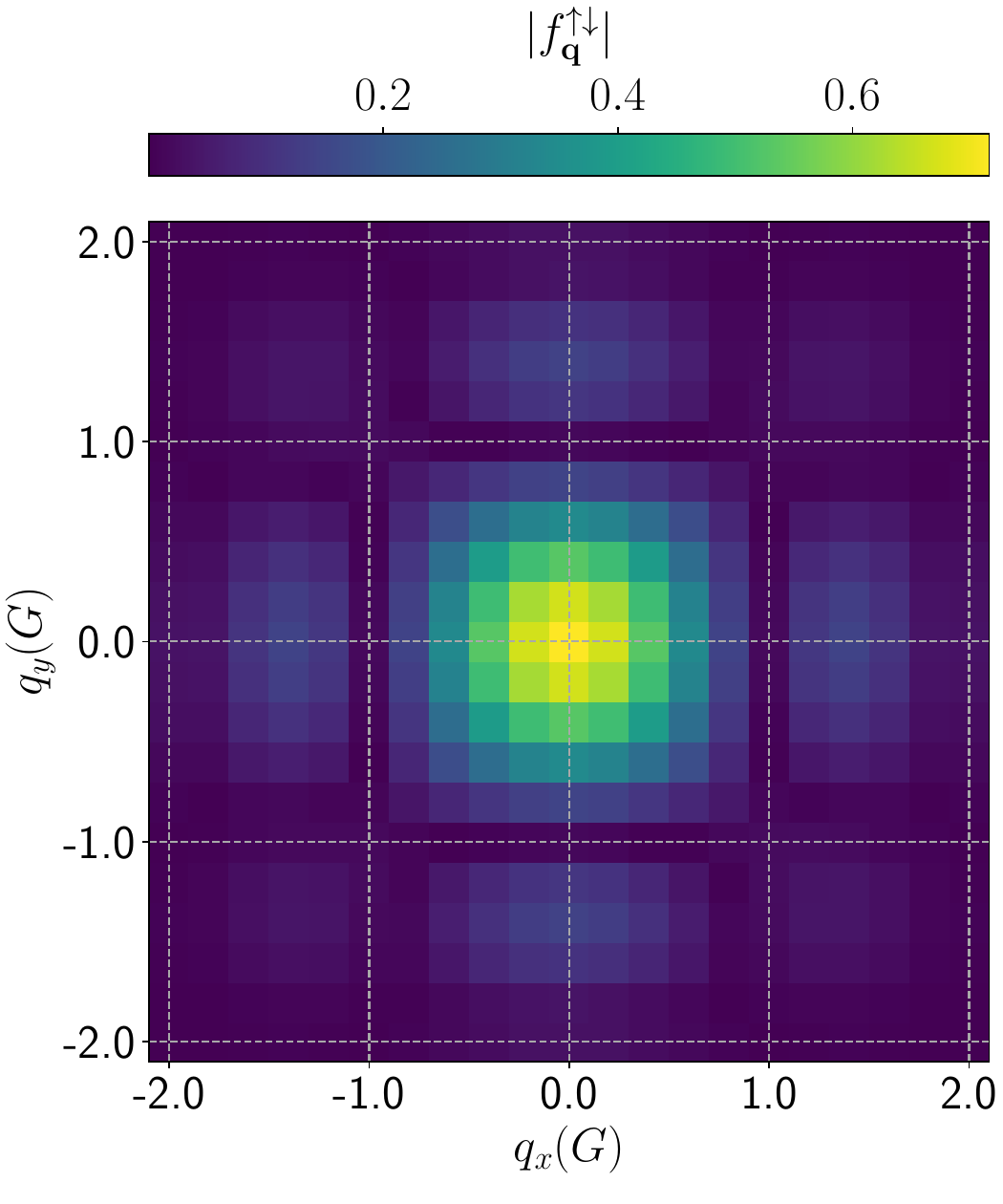}
    \caption{$v_0=0.4$}
  \end{subfigure}
  \hfill
  \begin{subfigure}[b]{0.32\columnwidth}
    \centering
    \includegraphics[width=\columnwidth]{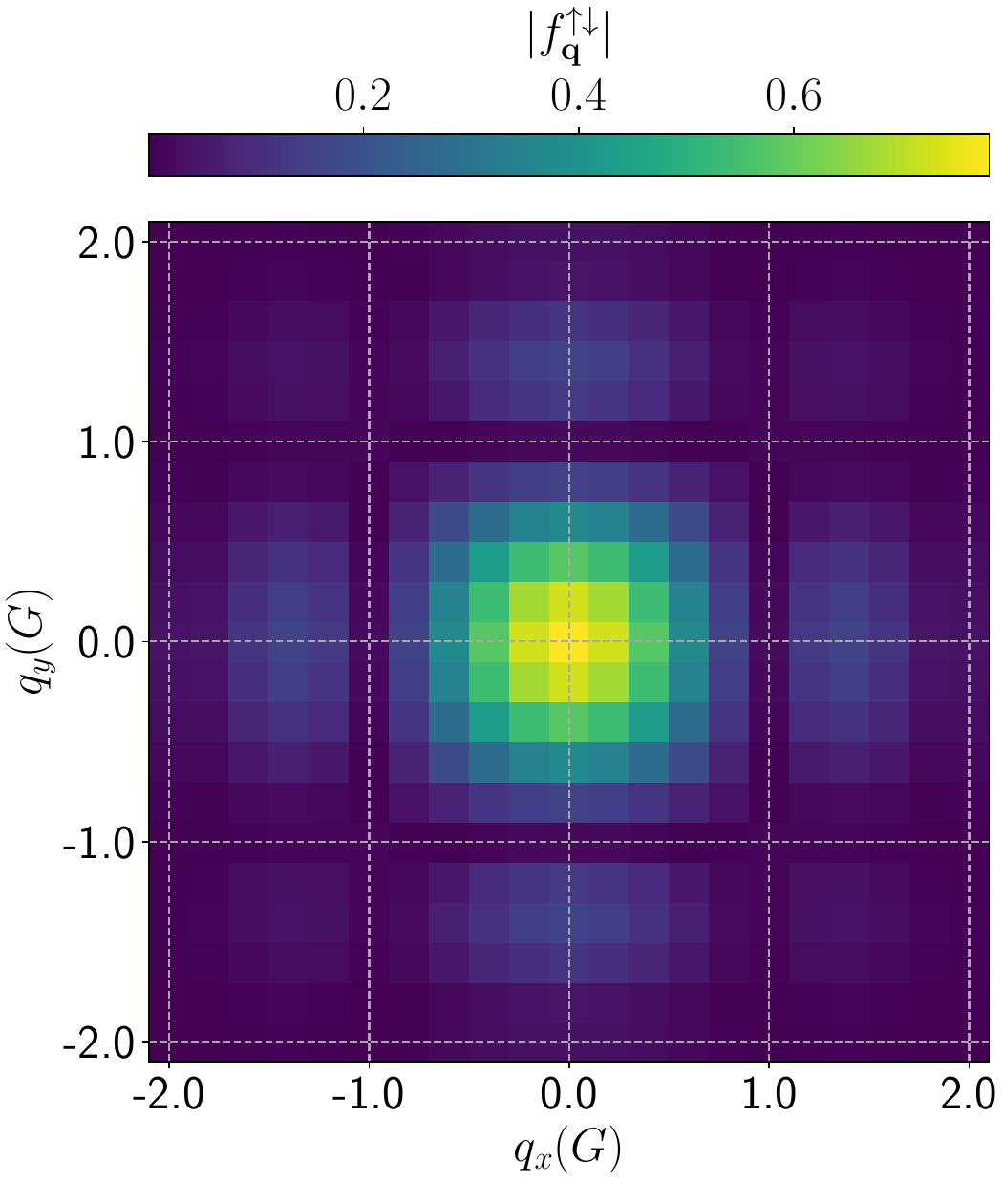}
    \caption{$v_0=0.45$}
  \end{subfigure}
  \hfill
  \begin{subfigure}[b]{0.32\columnwidth}
    \centering
    \includegraphics[width=\columnwidth]{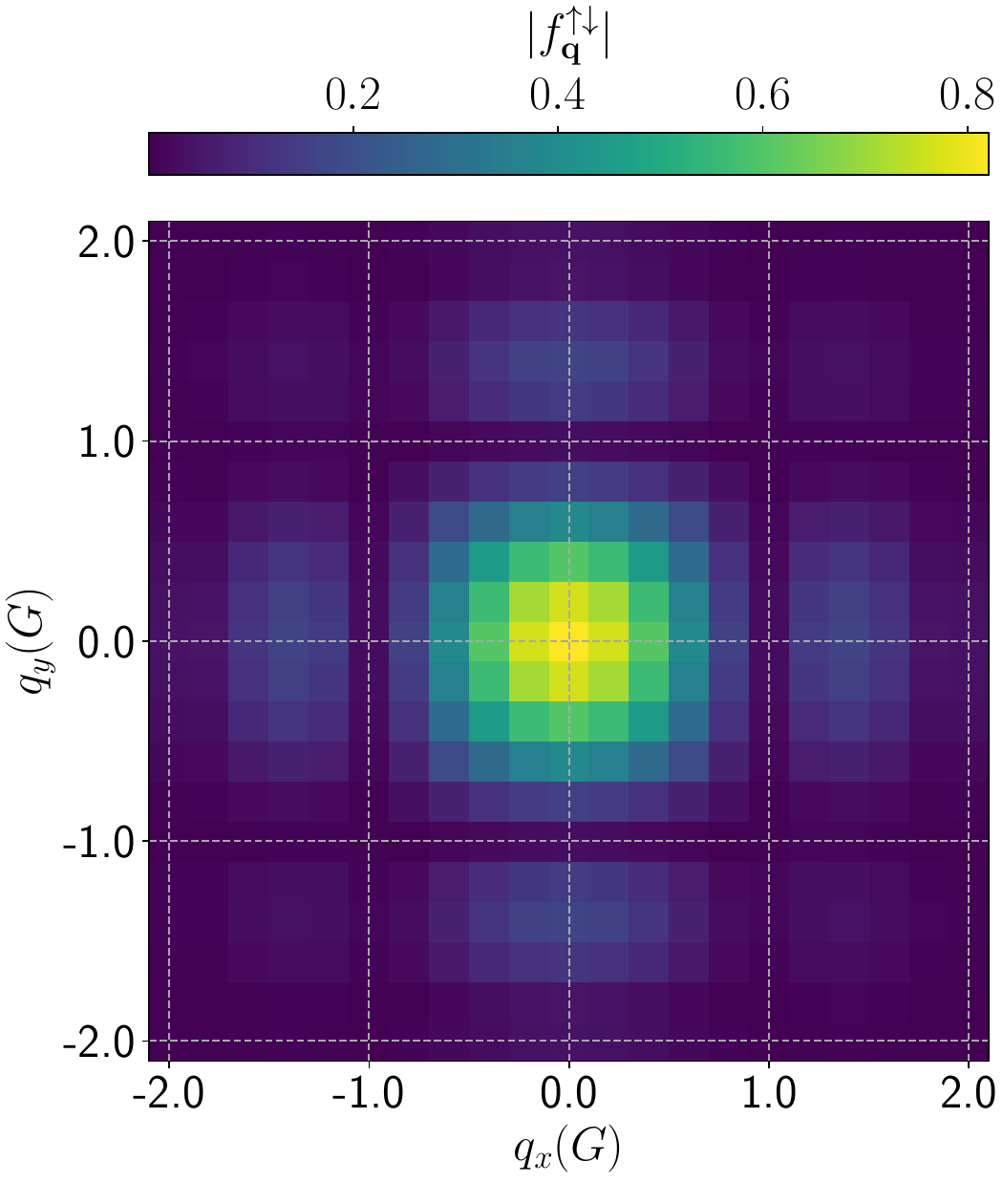}
    \caption{$v_0=0.5$}
  \end{subfigure}
  \caption{Condensate fractions of the $(N^\uparrow, N^\downarrow) = (13, 5)$ 2D spin-imbalanced Fermi gas across the BCS-BEC crossover.}
  \label{fig:13_5-pmd}
\end{figure}

\begin{figure}[ht!]
  \centering
  \begin{subfigure}[b]{0.32\columnwidth}
    \centering
    \includegraphics[width=\columnwidth]{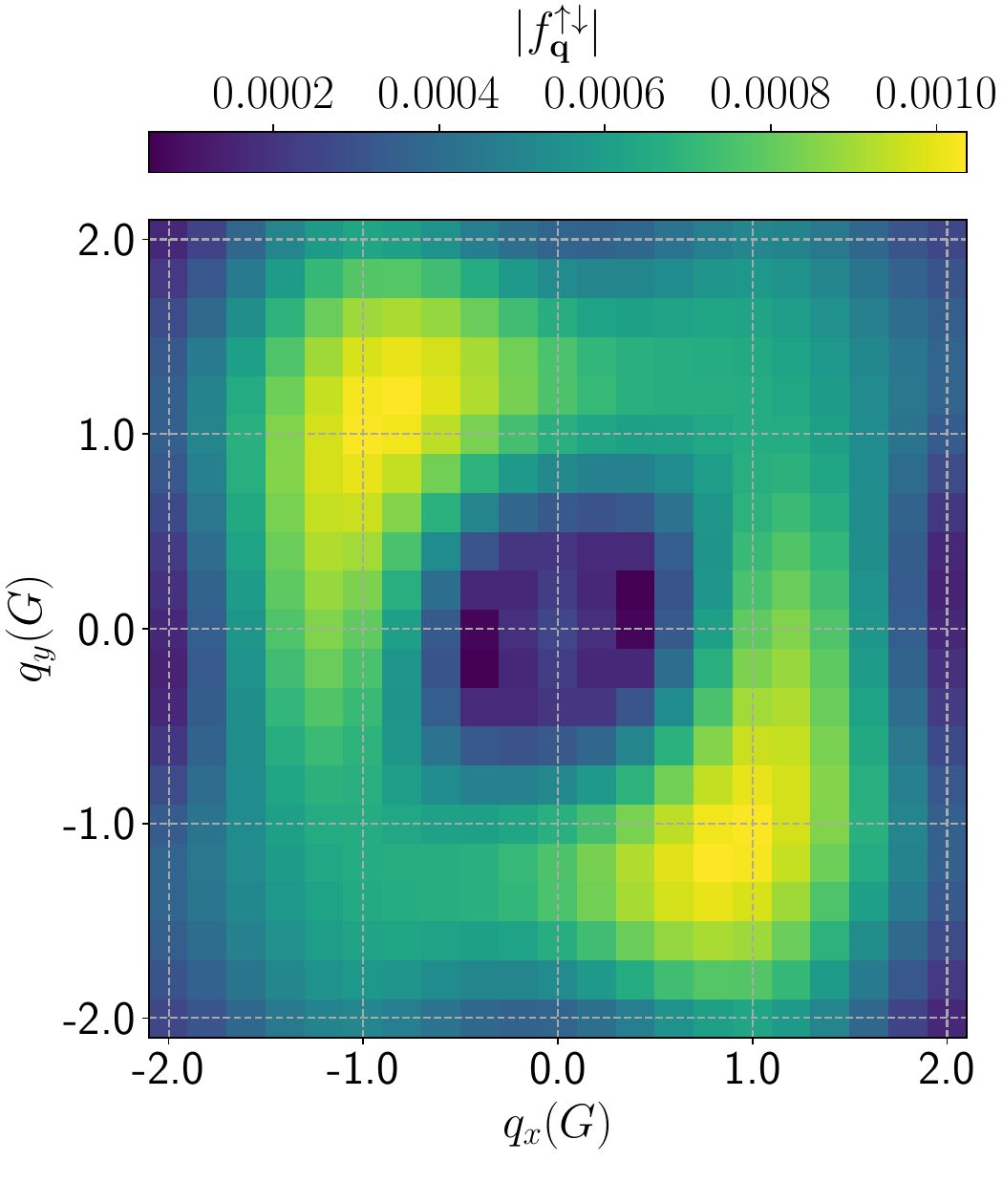}
    \caption{$v_0=0.1$}
  \end{subfigure}
  \hfill
  \begin{subfigure}[b]{0.32\columnwidth}
    \centering
    \includegraphics[width=\columnwidth]{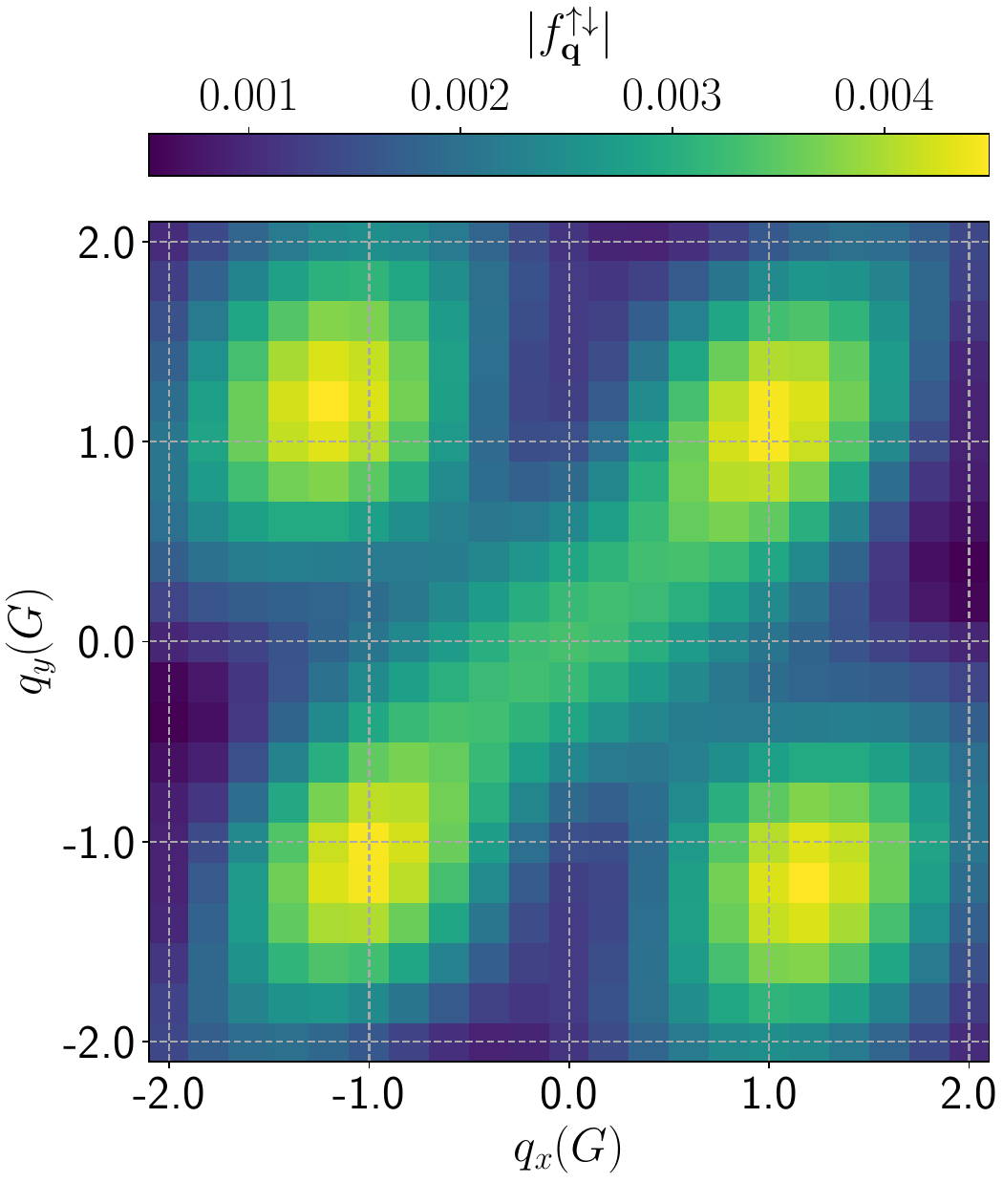}
    \caption{$v_0=0.15$}
  \end{subfigure}
  \hfill
  \begin{subfigure}[b]{0.32\columnwidth}
    \centering
    \includegraphics[width=\columnwidth]{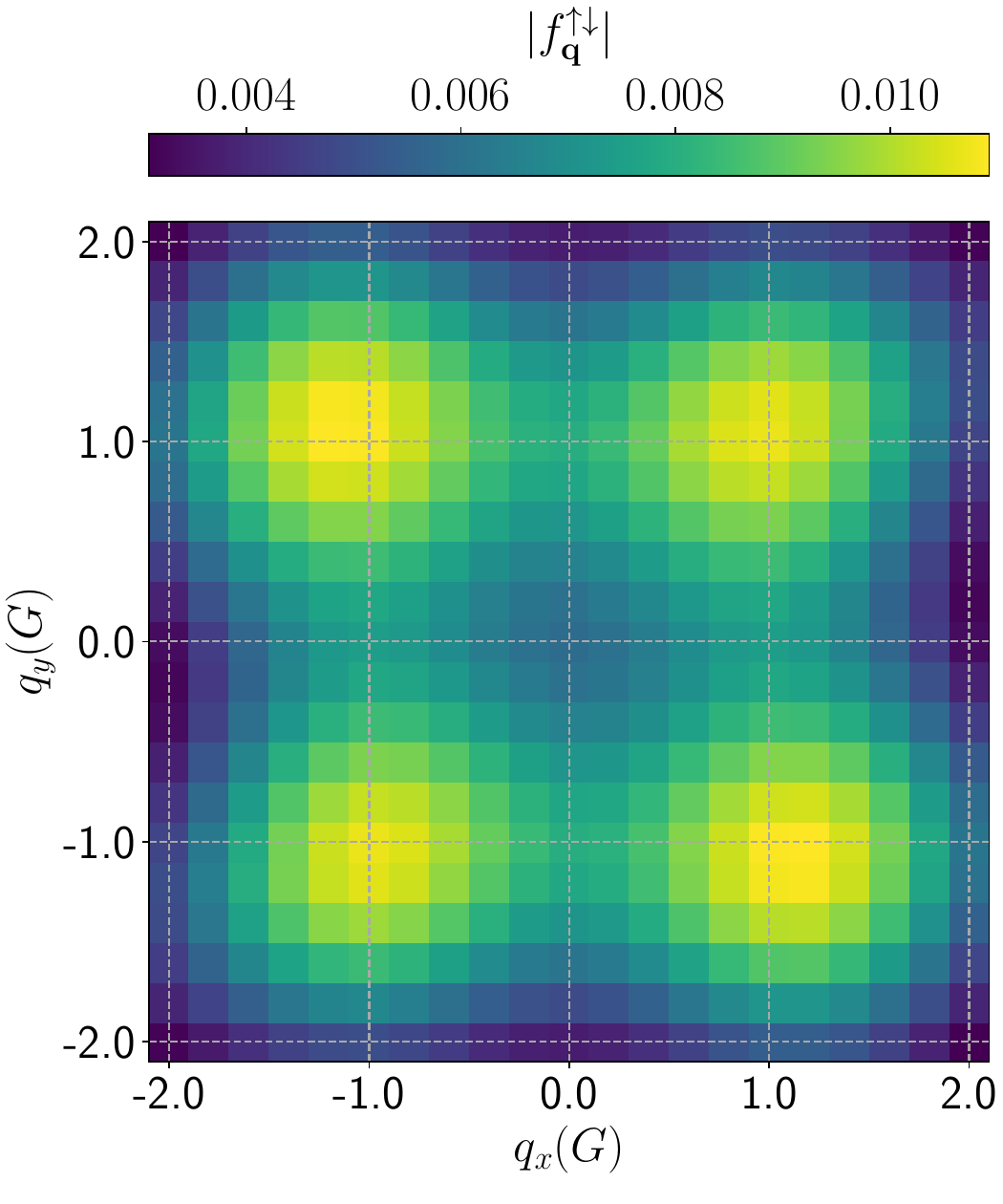}
    \caption{$v_0=0.2$}
  \end{subfigure}
  \hfill
  \begin{subfigure}[b]{0.32\columnwidth}
    \centering
    \includegraphics[width=\columnwidth]{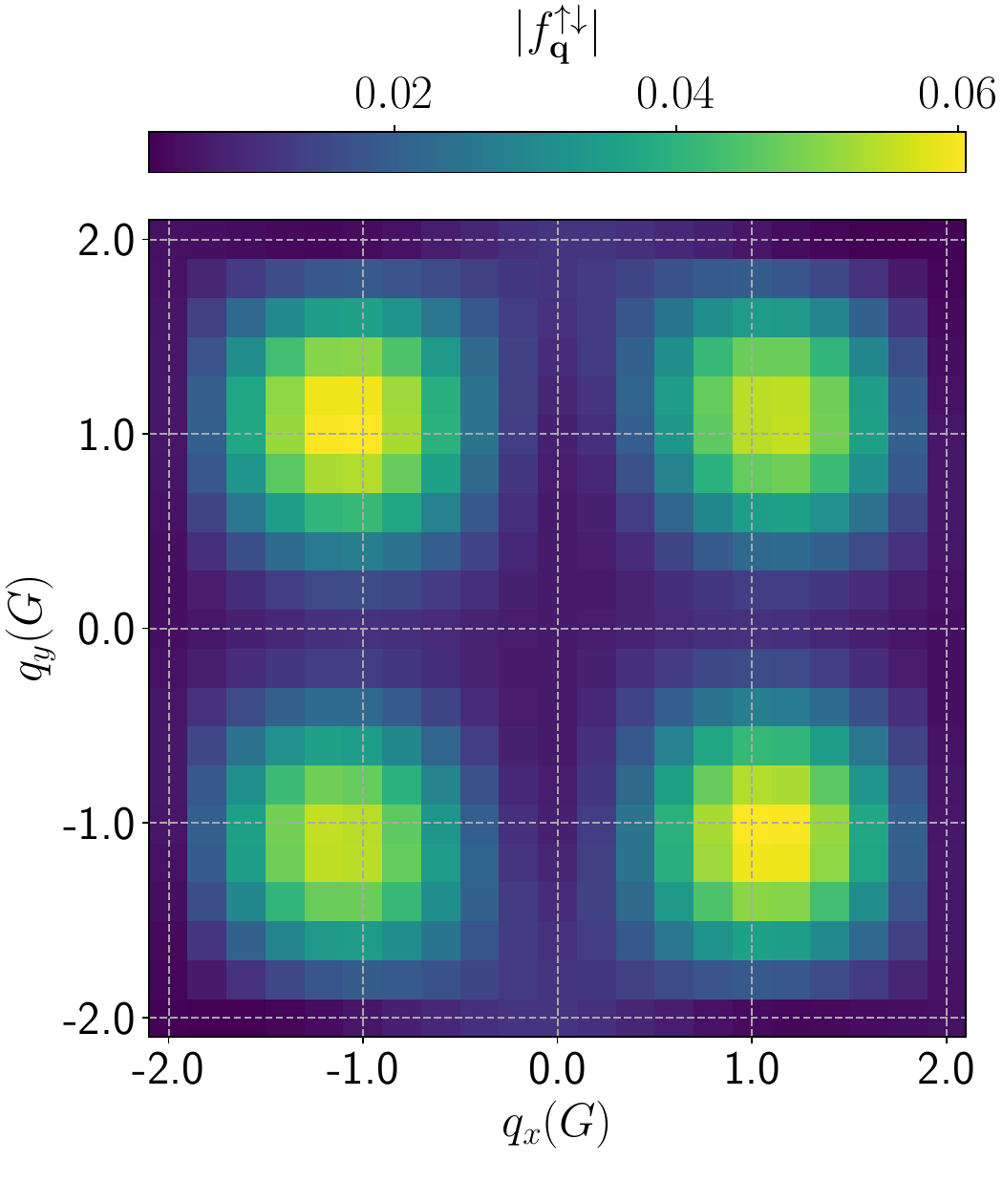}
    \caption{$v_0=0.25$}
  \end{subfigure}
  \hfill
  \begin{subfigure}[b]{0.32\columnwidth}
    \centering
    \includegraphics[width=\columnwidth]{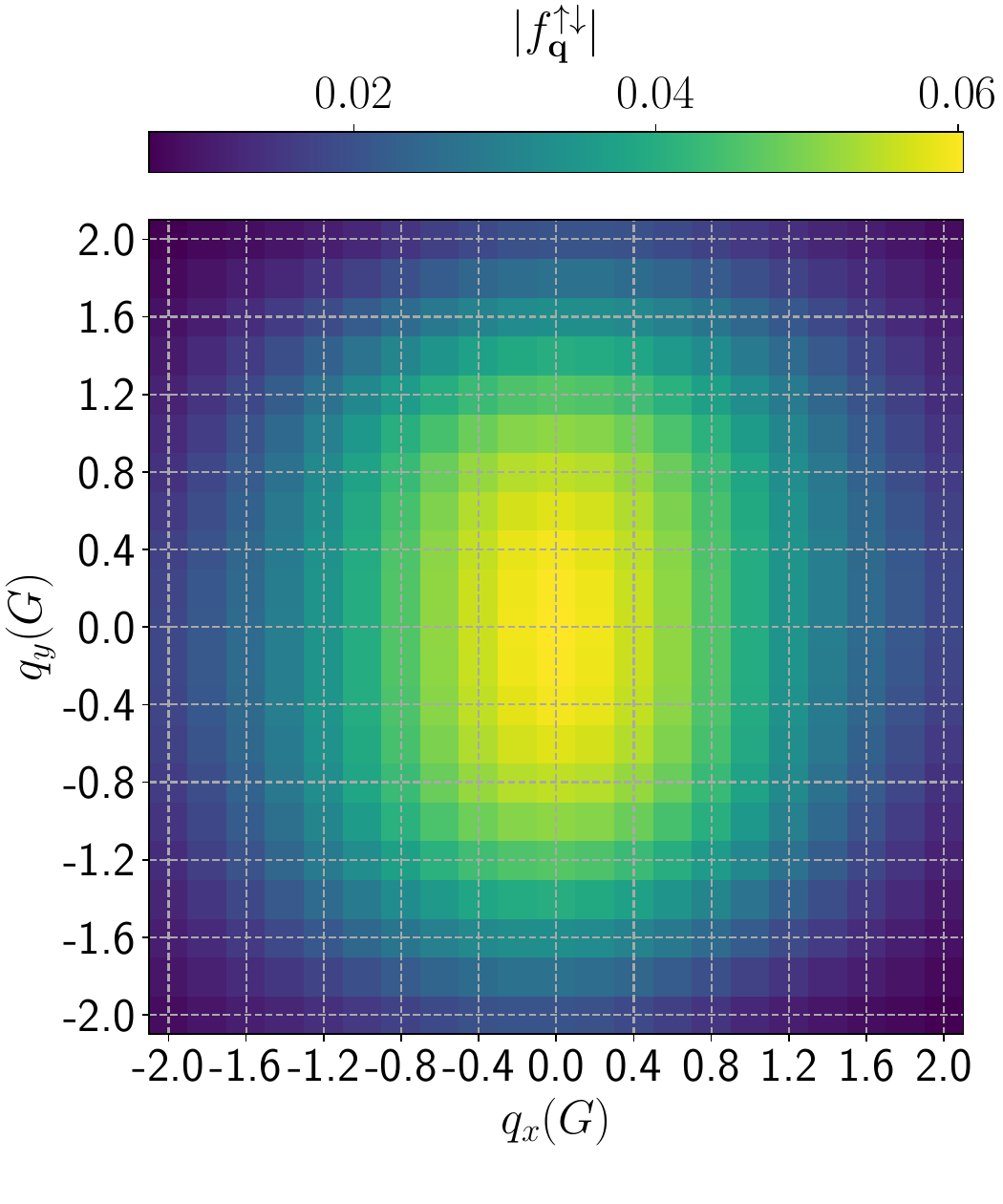}
    \caption{$v_0=0.3$}
  \end{subfigure}
  \hfill
  \begin{subfigure}[b]{0.32\columnwidth}
    \centering
    \includegraphics[width=\columnwidth]{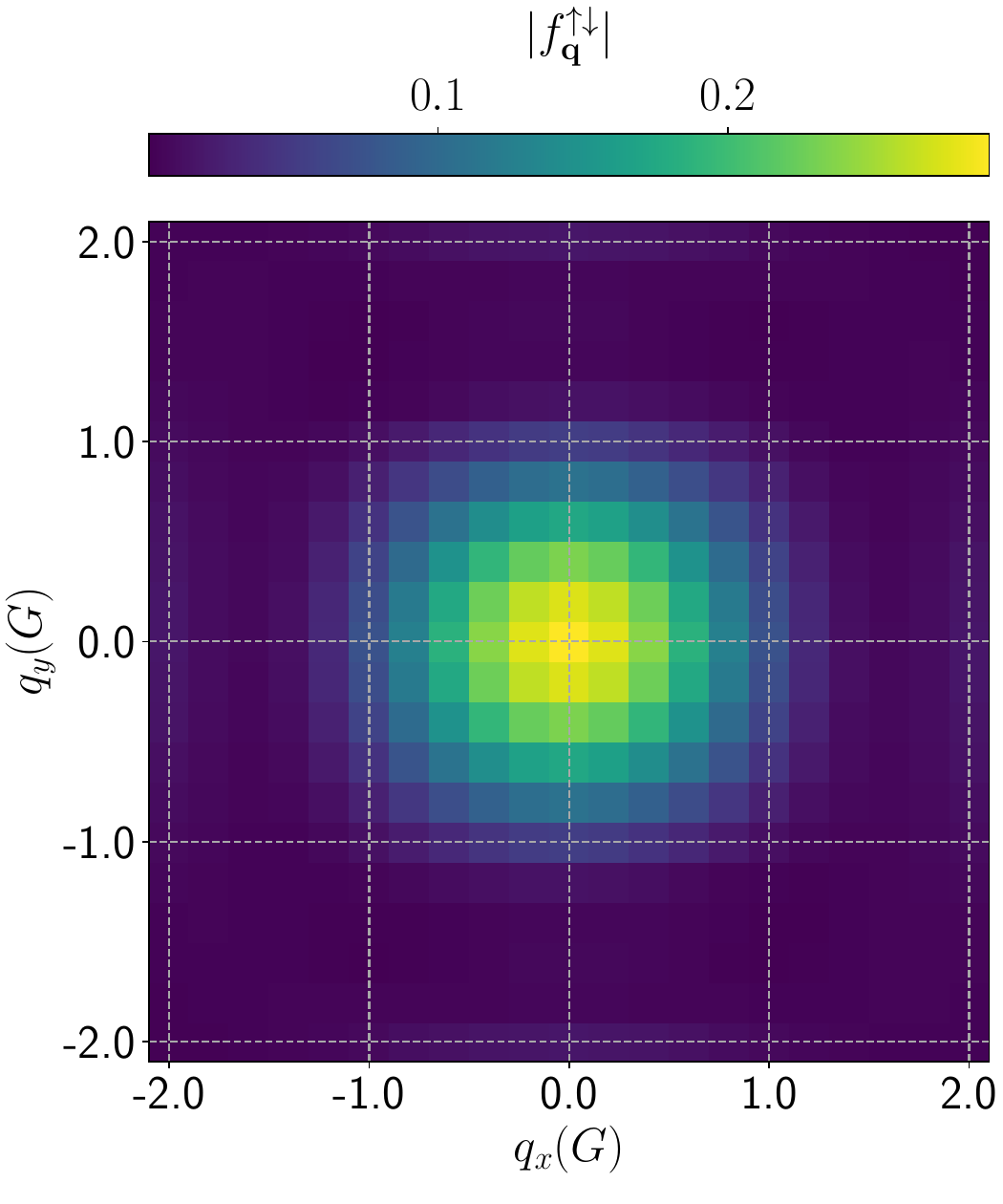}
    \caption{$v_0=0.35$}
  \end{subfigure}
  \hfill
  \begin{subfigure}[b]{0.32\columnwidth}
    \centering
    \includegraphics[width=\columnwidth]{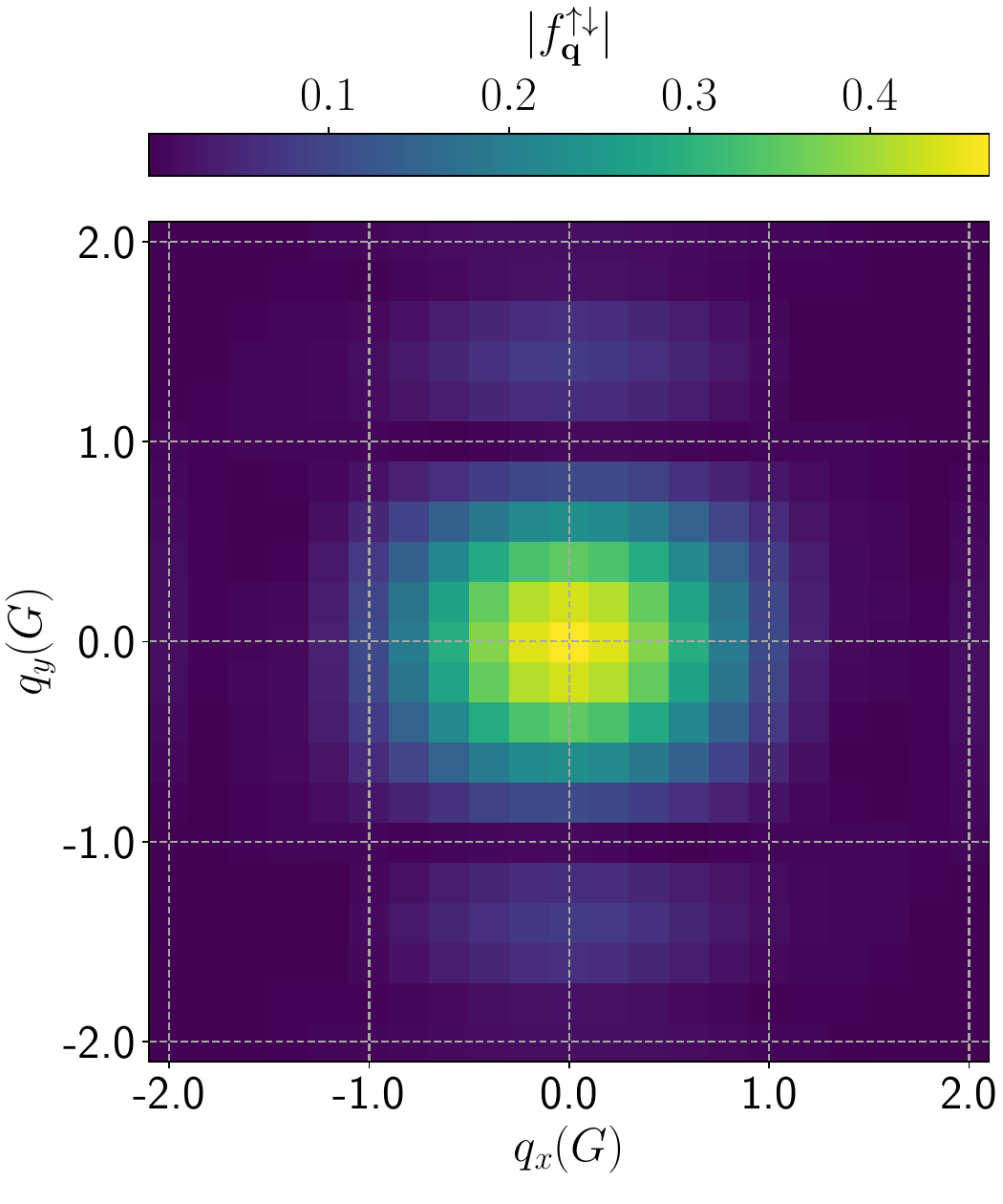}
    \caption{$v_0=0.4$}
  \end{subfigure}
  \hfill
  \begin{subfigure}[b]{0.32\columnwidth}
    \centering
    \includegraphics[width=\columnwidth]{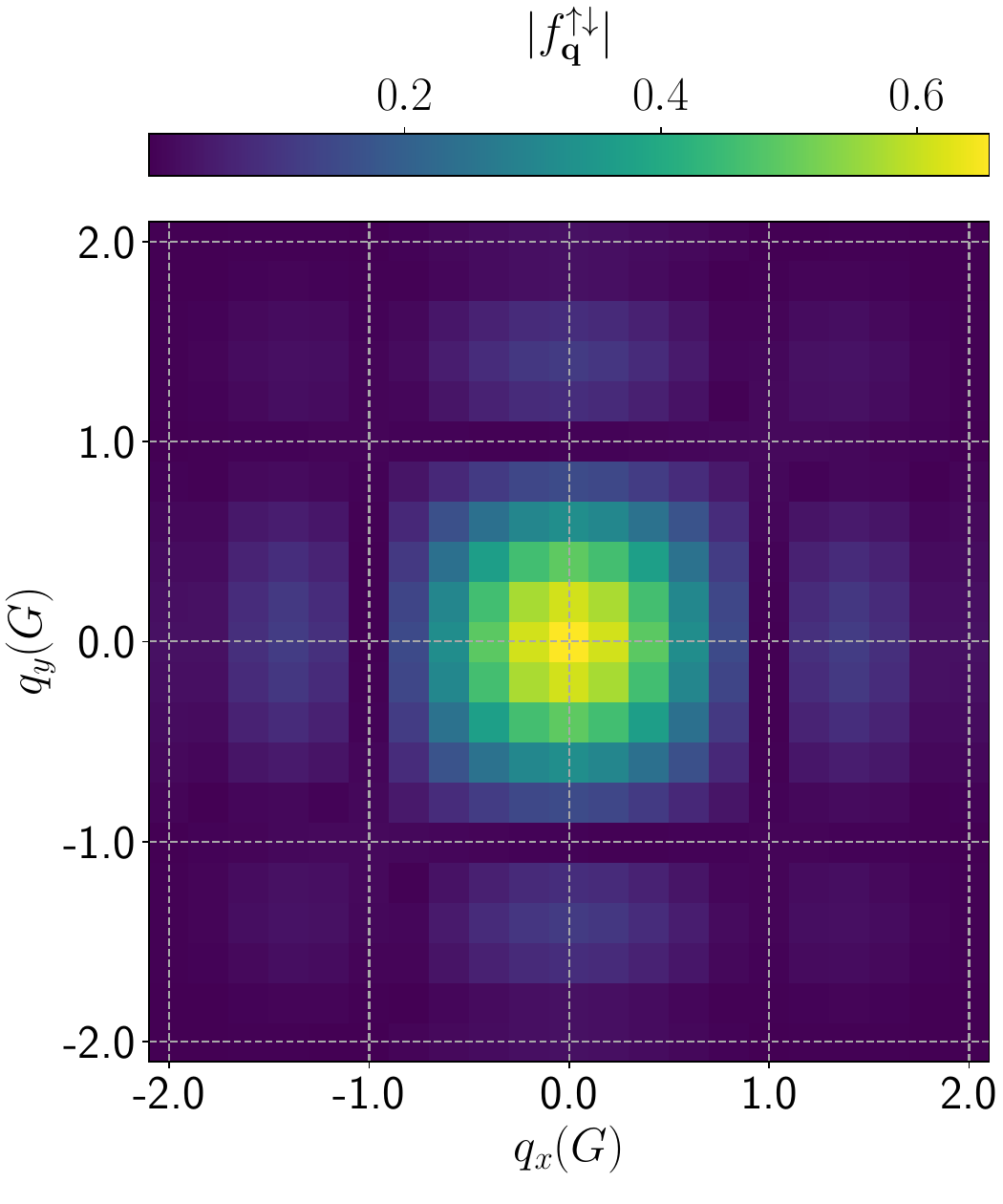}
    \caption{$v_0=0.45$}
  \end{subfigure}
  \hfill
  \begin{subfigure}[b]{0.32\columnwidth}
    \centering
    \includegraphics[width=\columnwidth]{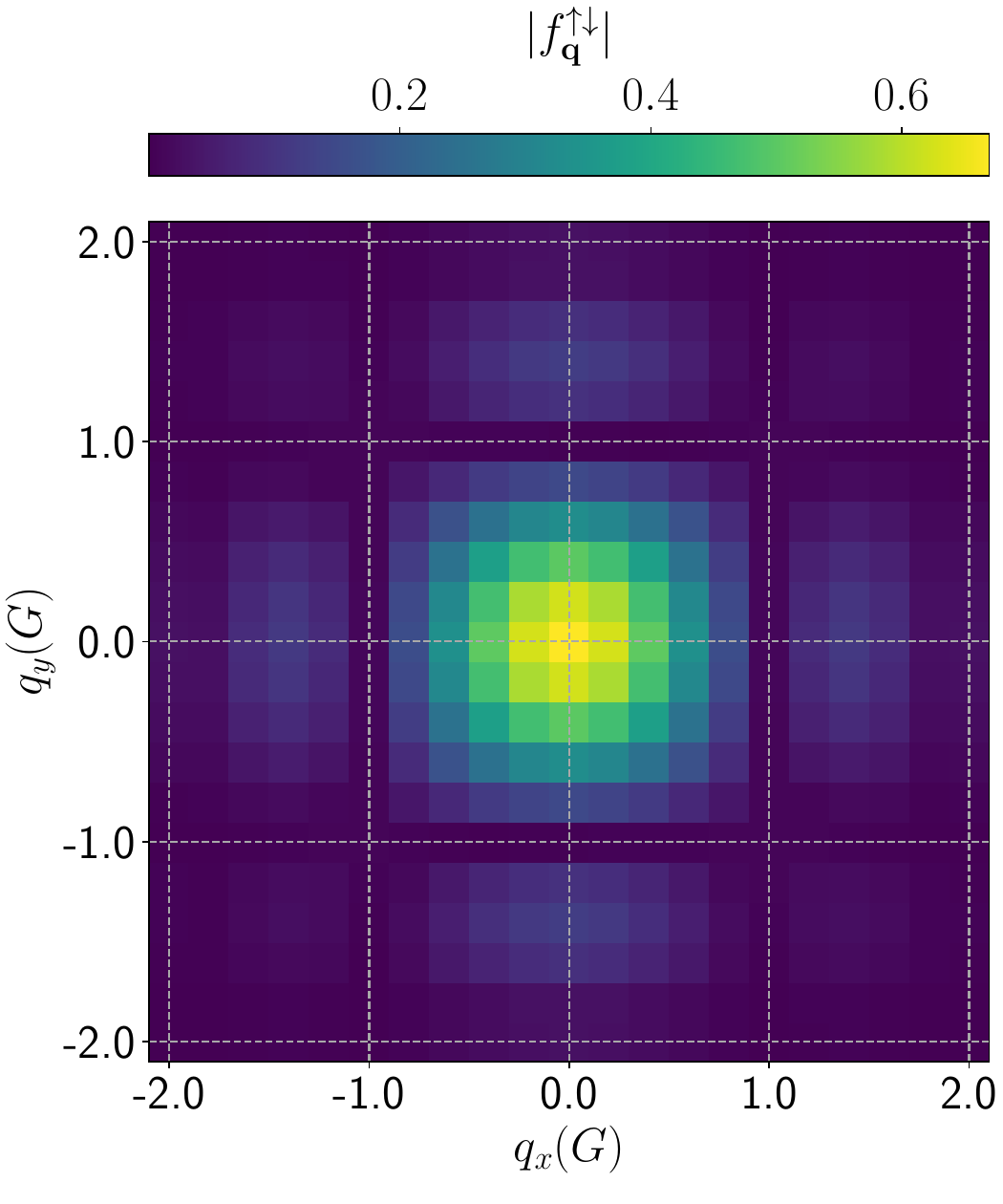}
    \caption{$v_0=0.5$}
  \end{subfigure}
  \caption{Condensate fractions of the $(N^\uparrow, N^\downarrow) = (25, 9)$ 2D spin-imbalanced Fermi gas across the BCS-BEC crossover.}
  \label{fig:25_9-pmd}
\end{figure}

In this subsection, we present the condensate fraction of the 2D spin-imbalanced Fermi gas with $(N^\uparrow, N^\downarrow) = (13, 5)$ and $(N^\uparrow, N^\downarrow) = (25, 9)$, respectively, across the BCS-BEC crossover.
Results of the $(N^\uparrow, N^\downarrow) = (13, 5)$ system are shown in \cref{fig:13_5-pmd} and the results of the $(N^\uparrow, N^\downarrow) = (25, 9)$ system are shown in \cref{fig:25_9-pmd}.
The strength of interactions, $v_0$, are mentioned in the caption of each subfigure.

\subsection{One-Particle Density}
\begin{figure}[ht!]
  \begin{subfigure}[b]{0.45\columnwidth}
    \centering
    \includegraphics[width=\columnwidth]{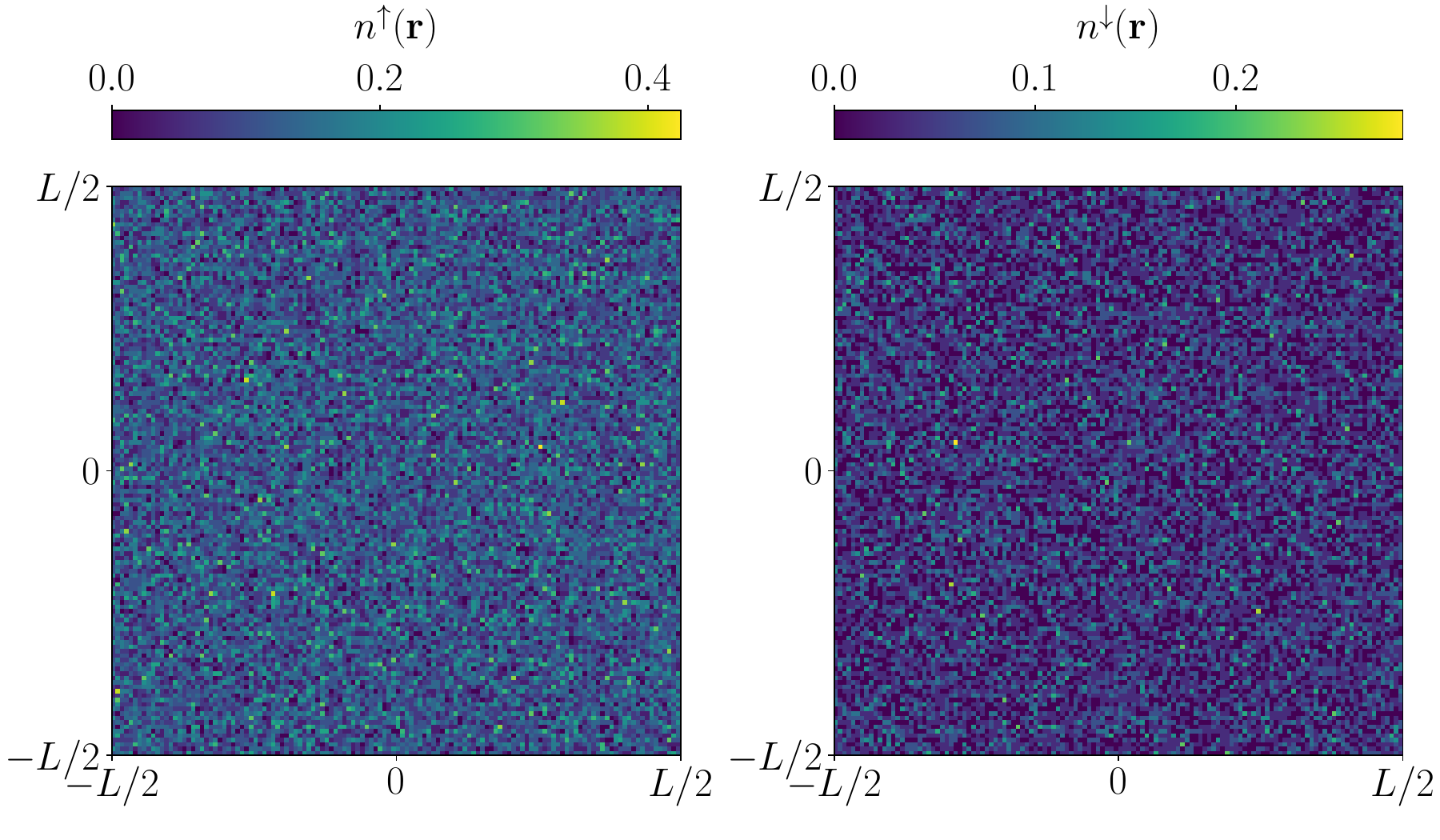}
    \caption{$v_0=0.15$}
  \end{subfigure}
  \hfill
  \begin{subfigure}[b]{0.45\columnwidth}
    \centering
    \includegraphics[width=\columnwidth]{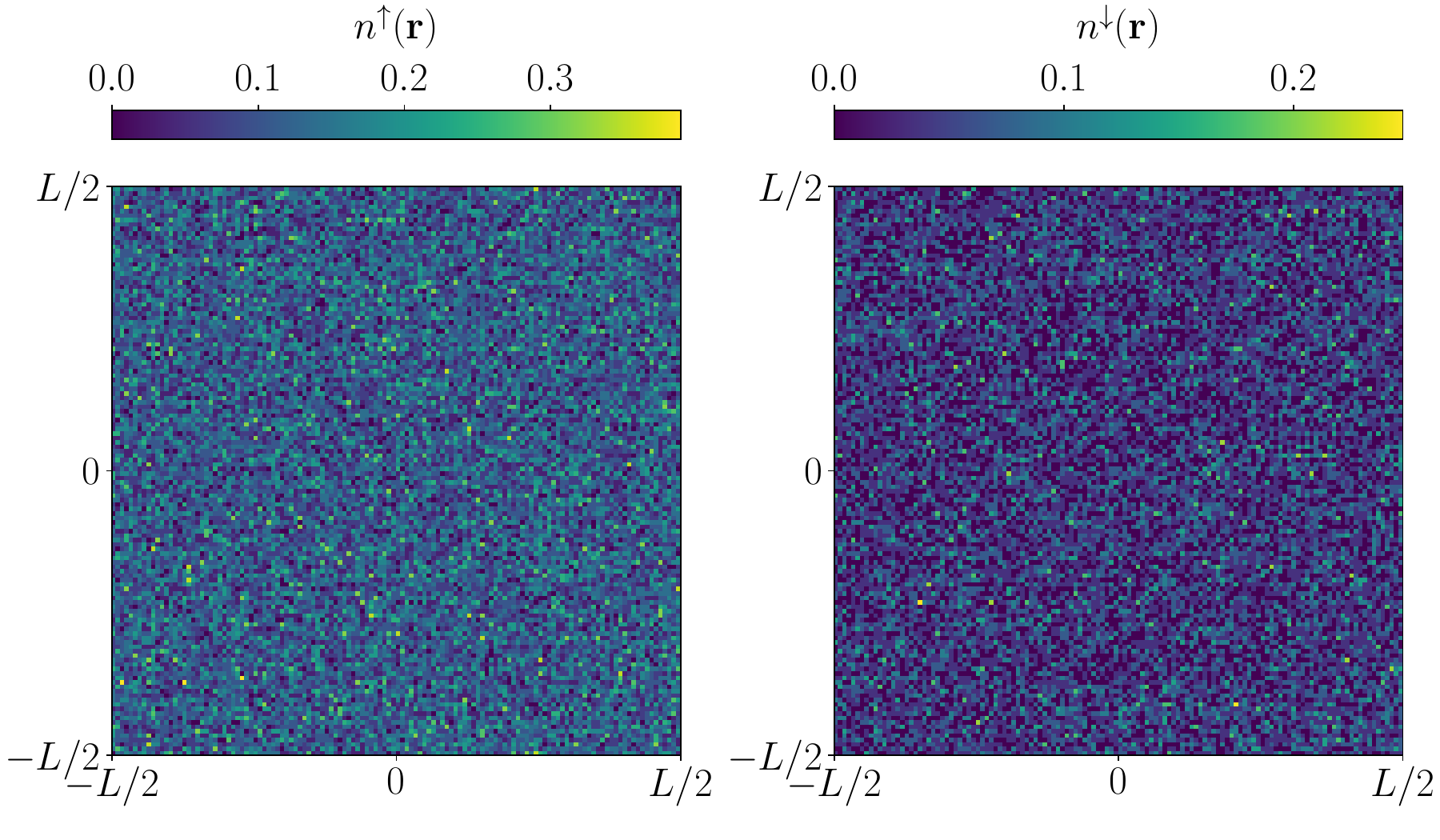}
    \caption{$v_0=0.2$}
  \end{subfigure}
  \hfill
  \begin{subfigure}[b]{0.45\columnwidth}
    \centering
    \includegraphics[width=\columnwidth]{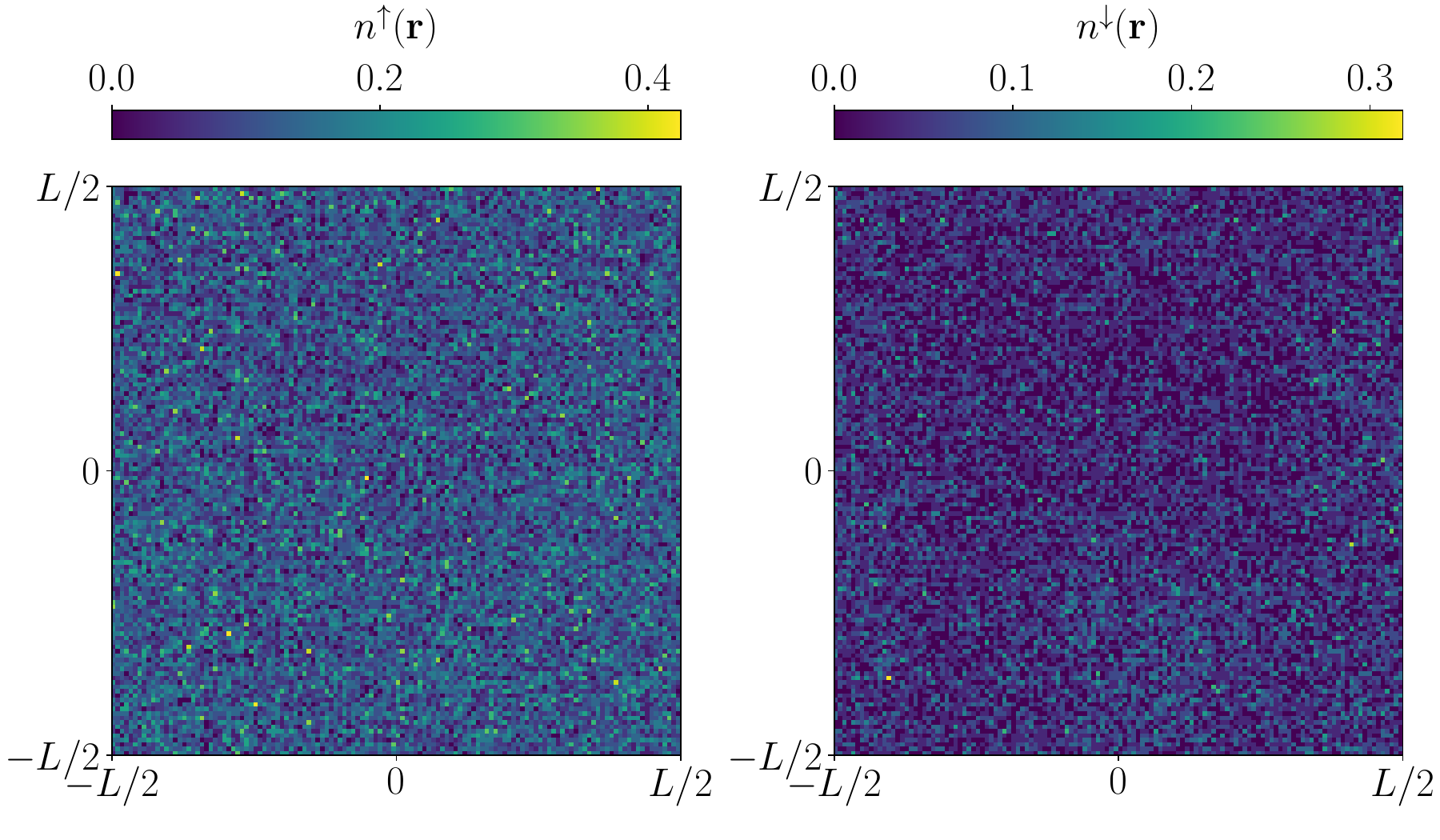}
    \caption{$v_0=0.25$}
  \end{subfigure}
  \hfill
  \begin{subfigure}[b]{0.45\columnwidth}
    \centering
    \includegraphics[width=\columnwidth]{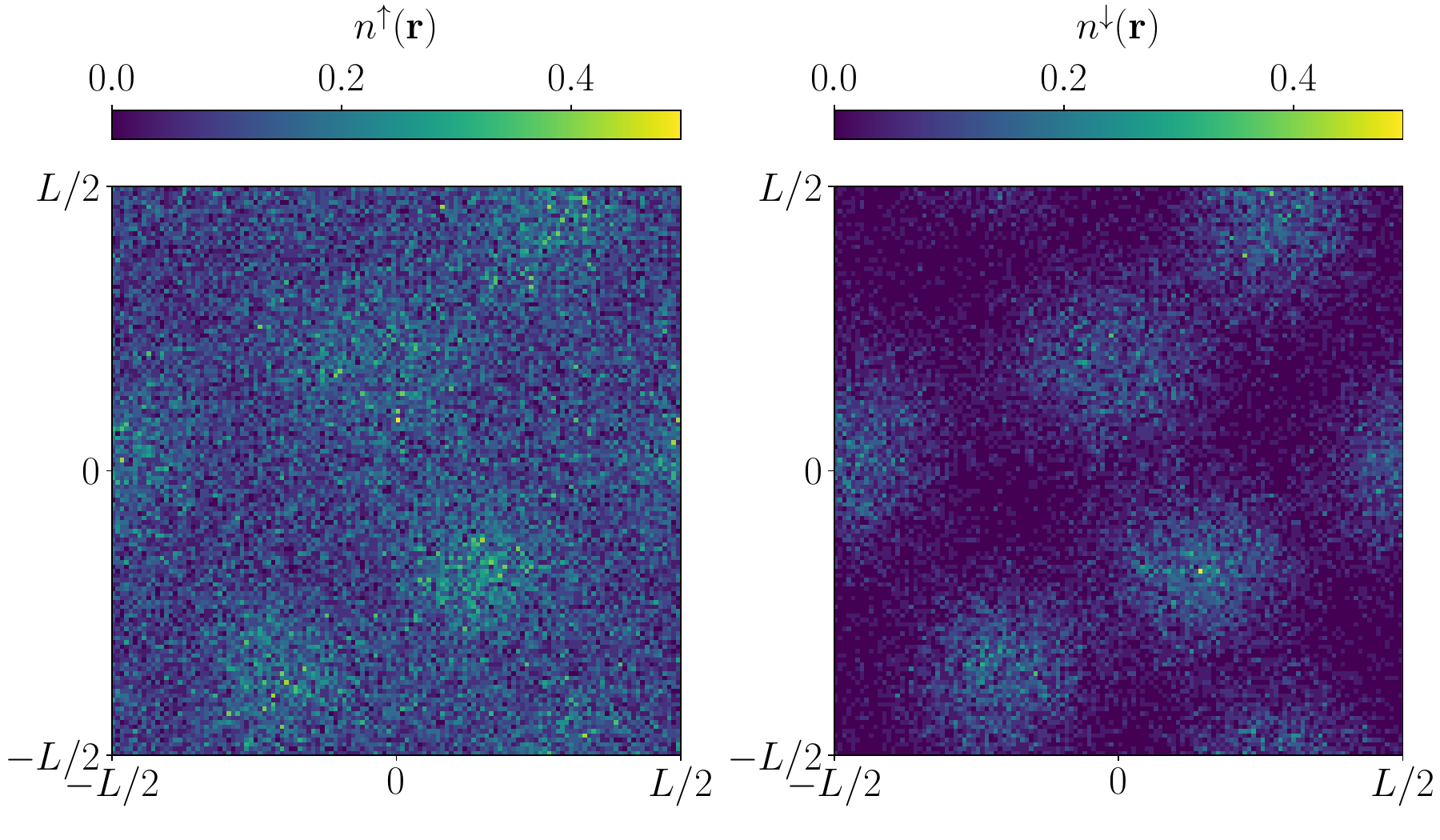}
    \caption{$v_0=0.3$}
  \end{subfigure}
  \hfill
  \begin{subfigure}[b]{0.45\columnwidth}
    \centering
    \includegraphics[width=\columnwidth]{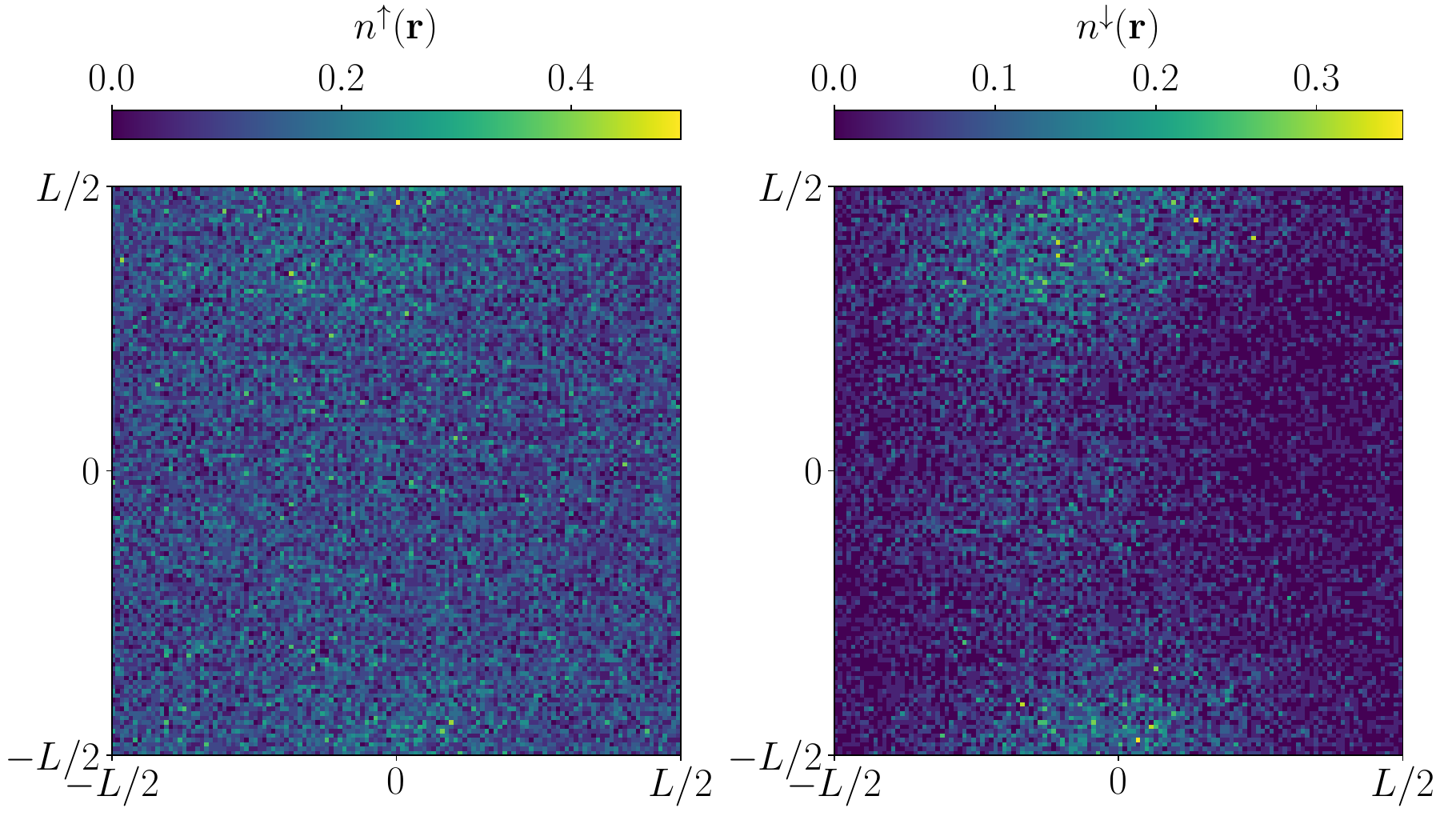}
    \caption{$v_0=0.35$}
  \end{subfigure}
  \hfill
  \begin{subfigure}[b]{0.45\columnwidth}
    \centering
    \includegraphics[width=\columnwidth]{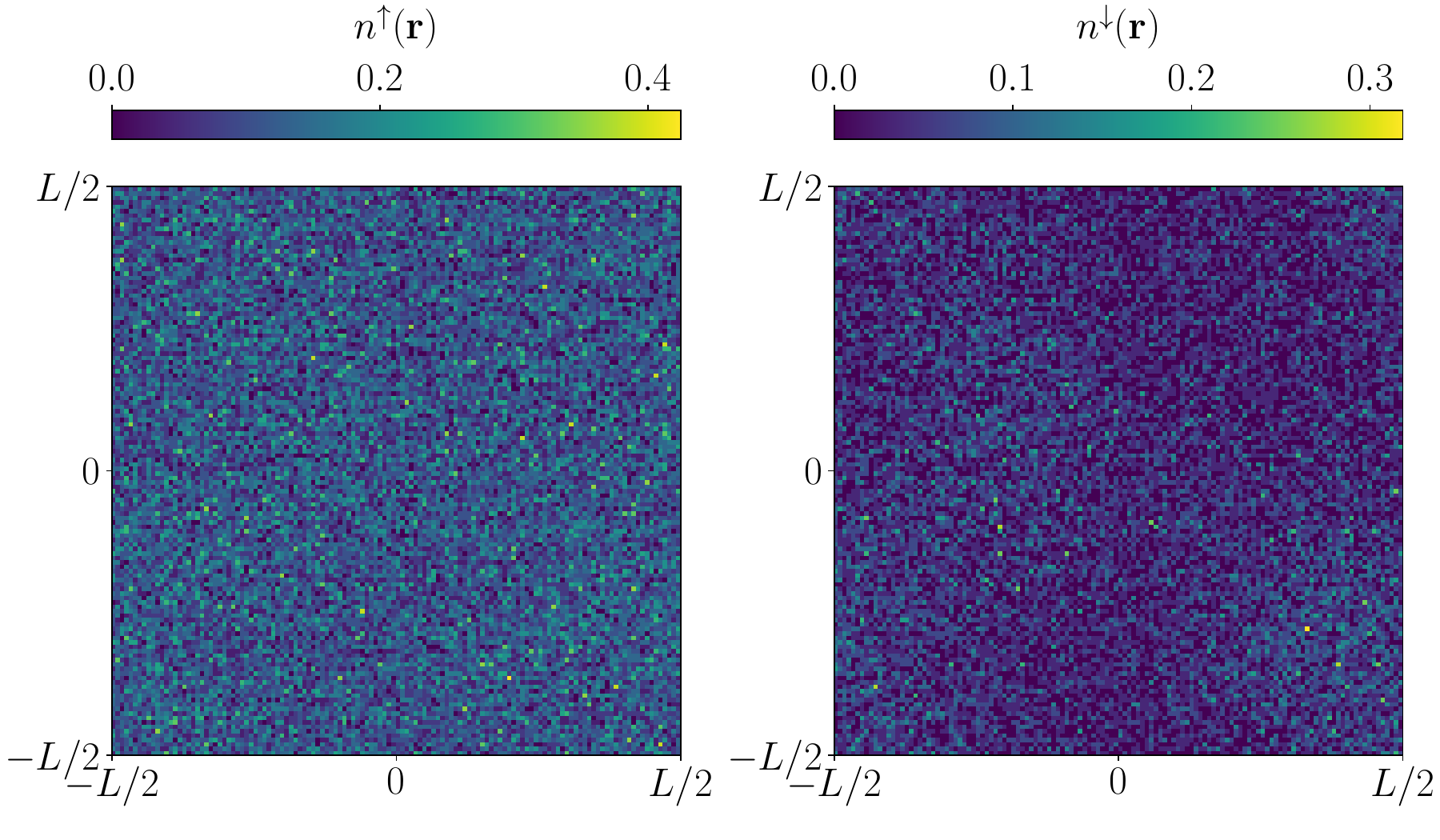}
    \caption{$v_0=0.4$}
  \end{subfigure}
  \hfill
  \begin{subfigure}[b]{0.45\columnwidth}
    \centering
    \includegraphics[width=\columnwidth]{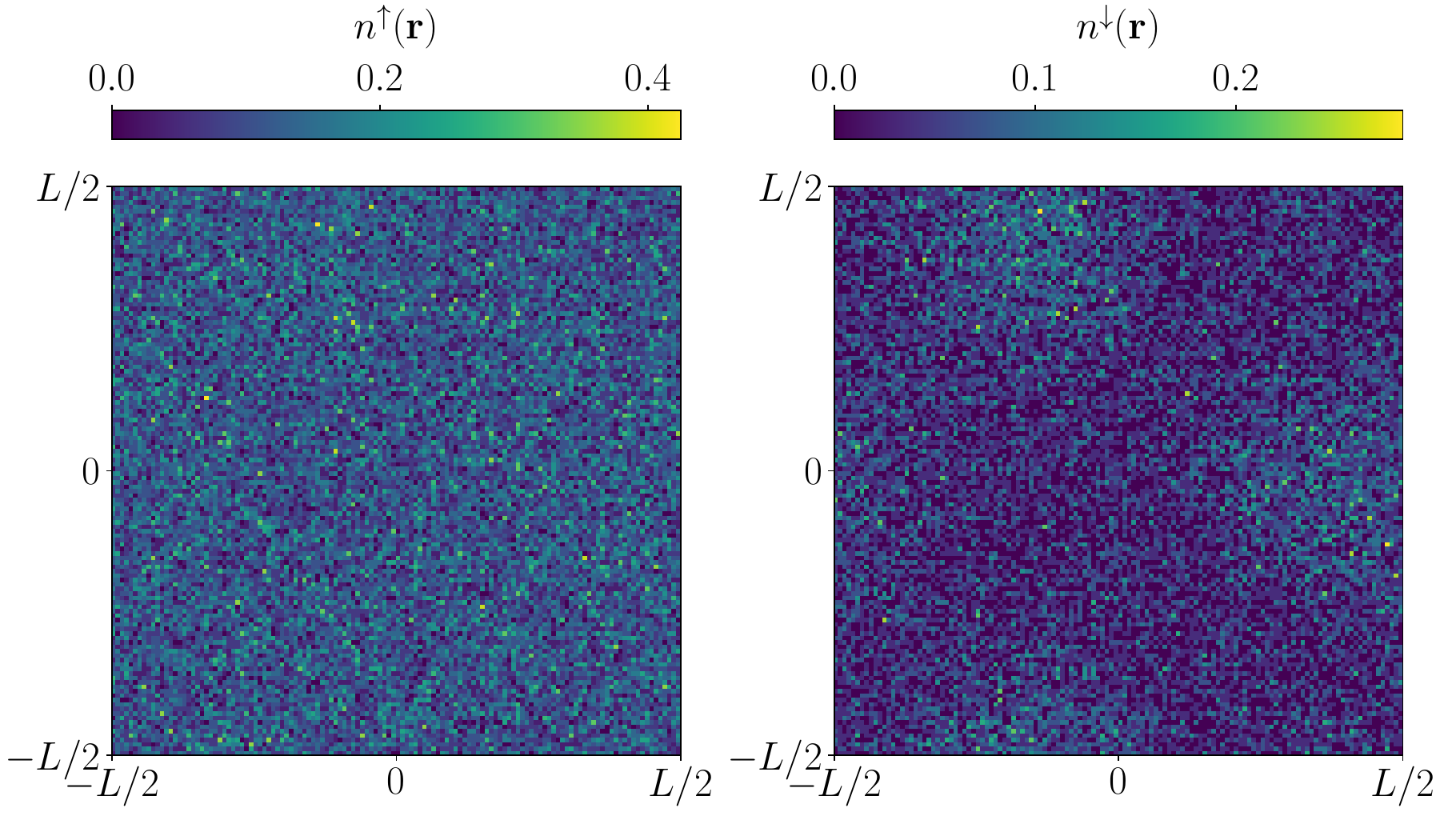}
    \caption{$v_0=0.45$}
  \end{subfigure}
  \hfill
  \begin{subfigure}[b]{0.45\columnwidth}
    \centering
    \includegraphics[width=\columnwidth]{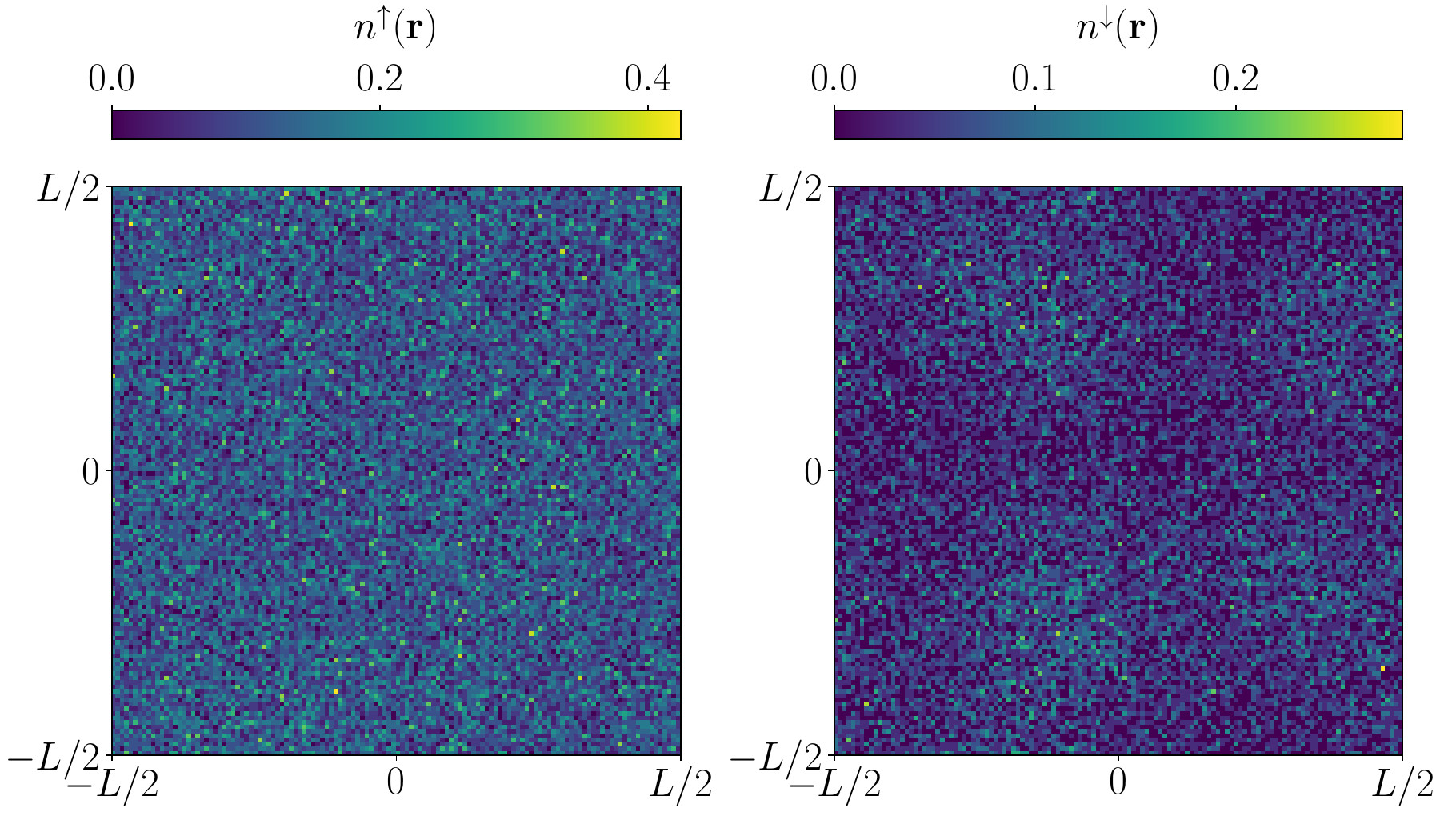}
    \caption{$v_0=0.5$}
  \end{subfigure}
  \caption{
    One-particle density of the 2D spin-imbalanced Fermi gas with $(N^\uparrow, N^\downarrow) = (13, 5)$ across the BCS-BEC crossover, plotted separately for different spin channels.
    For each subfigure, the plot on the left-hand side is the density of spin-up particles and the plot on the right-hand side is the density of the spin-down particles.
  }
  \label{fig:13_5-one_density}
\end{figure}

\begin{figure}[ht!]
  \begin{subfigure}[b]{0.45\columnwidth}
    \centering
    \includegraphics[width=\columnwidth]{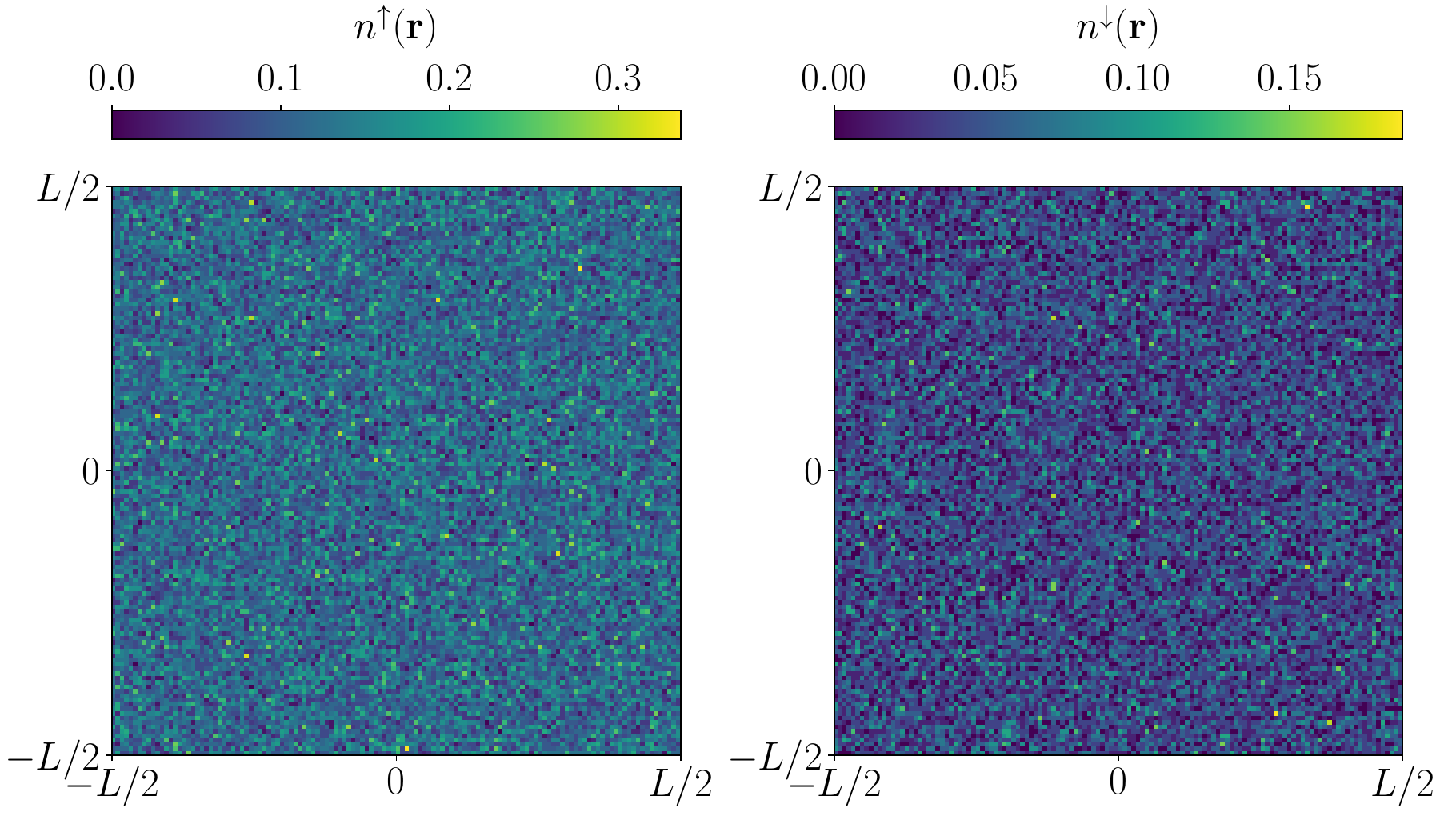}
    \caption{$v_0=0.15$}
  \end{subfigure}
  \hfill
  \begin{subfigure}[b]{0.45\columnwidth}
    \centering
    \includegraphics[width=\columnwidth]{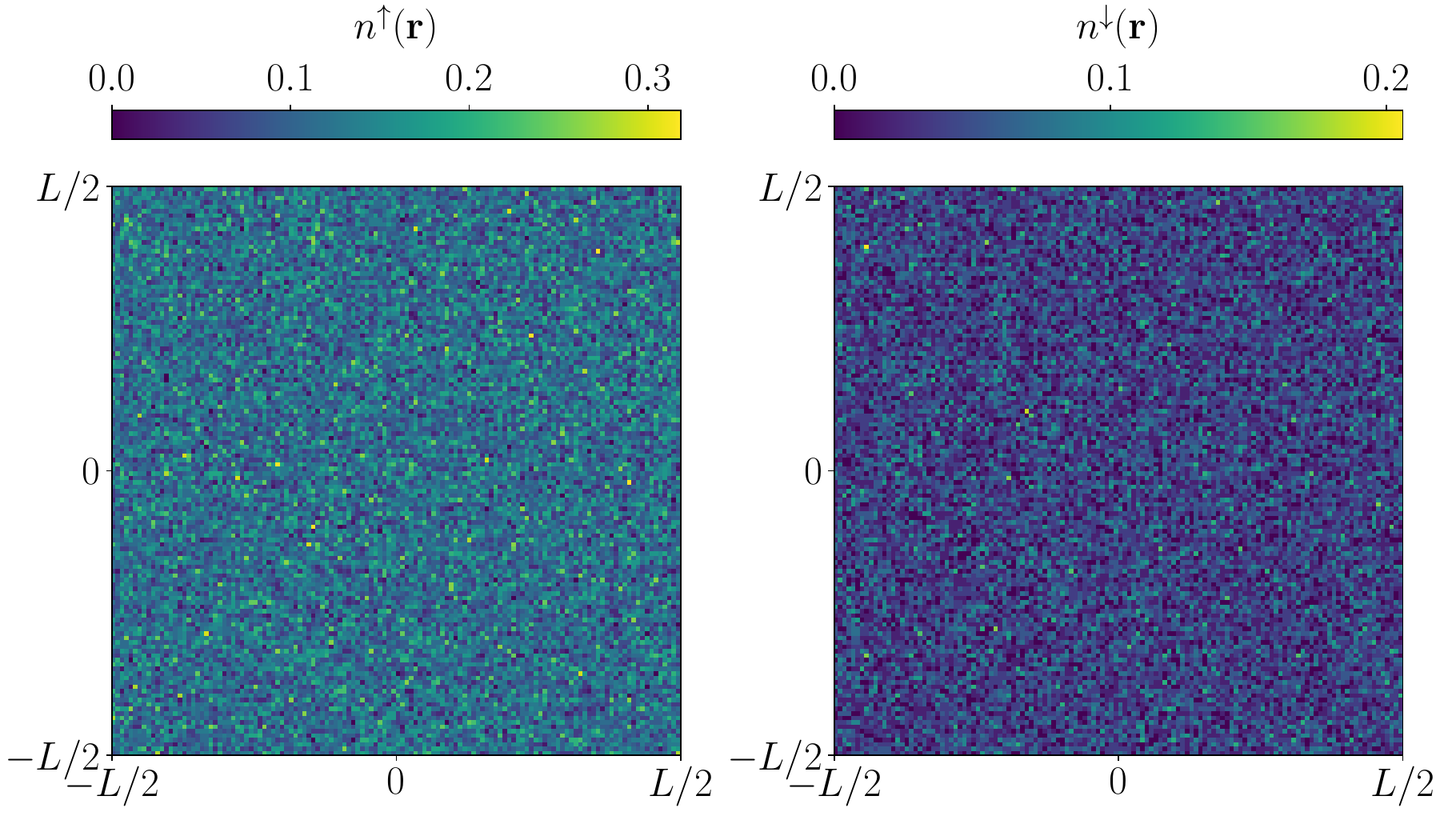}
    \caption{$v_0=0.2$}
  \end{subfigure}
  \hfill
  \begin{subfigure}[b]{0.45\columnwidth}
    \centering
    \includegraphics[width=\columnwidth]{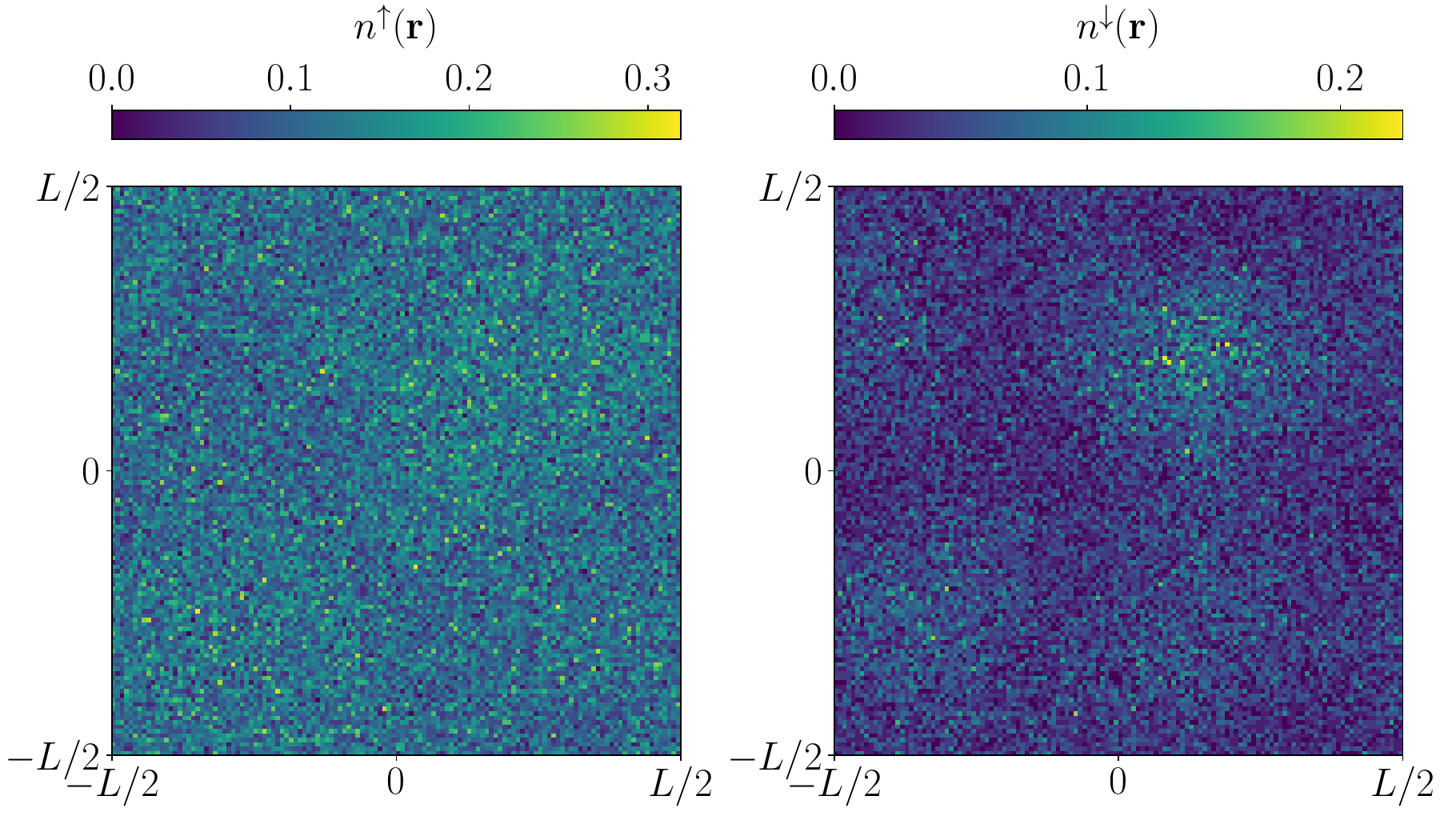}
    \caption{$v_0=0.25$}
  \end{subfigure}
  \hfill
  \begin{subfigure}[b]{0.45\columnwidth}
    \centering
    \includegraphics[width=\columnwidth]{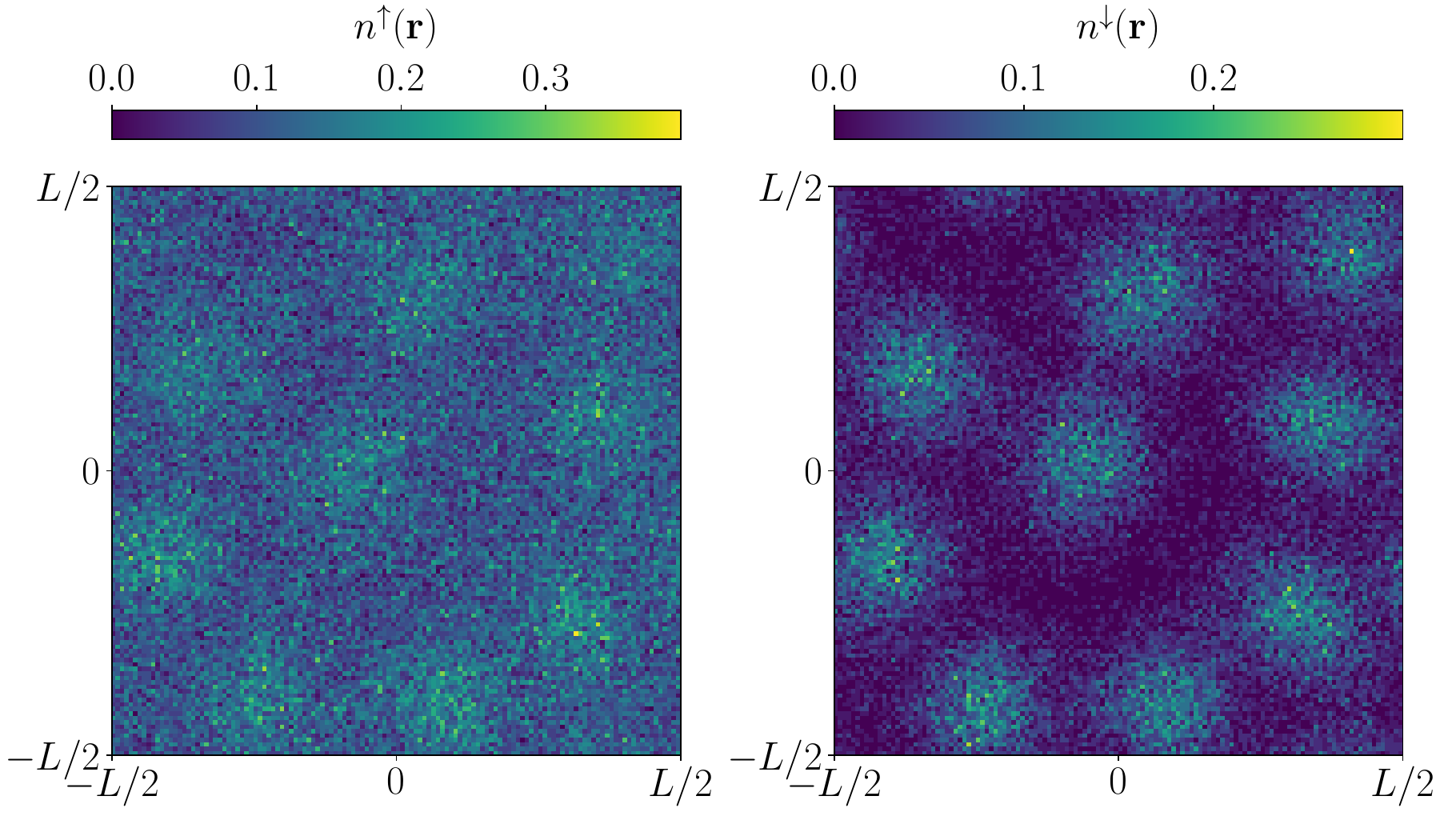}
    \caption{$v_0=0.3$}
  \end{subfigure}
  \hfill
  \begin{subfigure}[b]{0.45\columnwidth}
    \centering
    \includegraphics[width=\columnwidth]{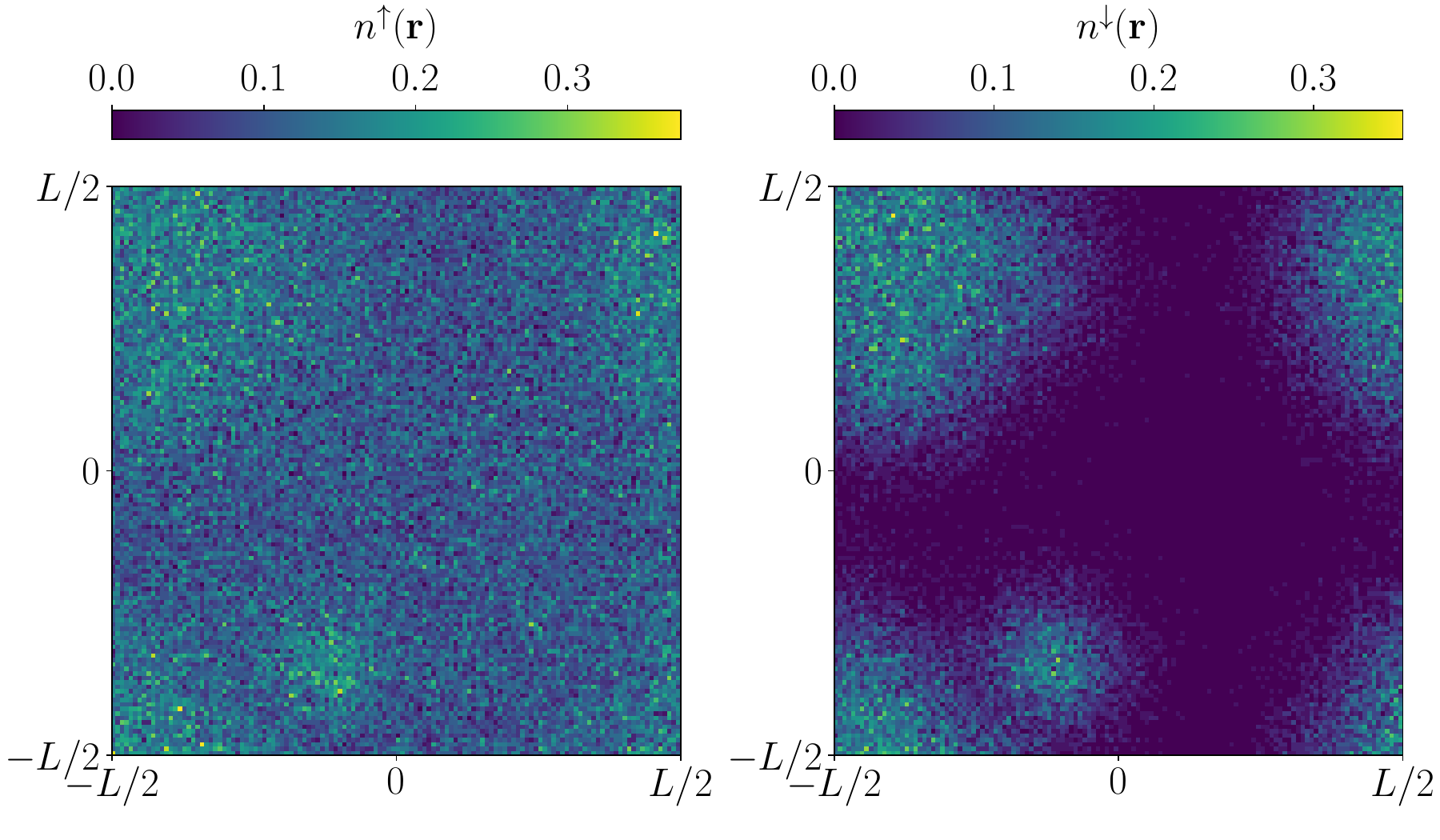}
    \caption{$v_0=0.35$}
  \end{subfigure}
  \hfill
  \begin{subfigure}[b]{0.45\columnwidth}
    \centering
    \includegraphics[width=\columnwidth]{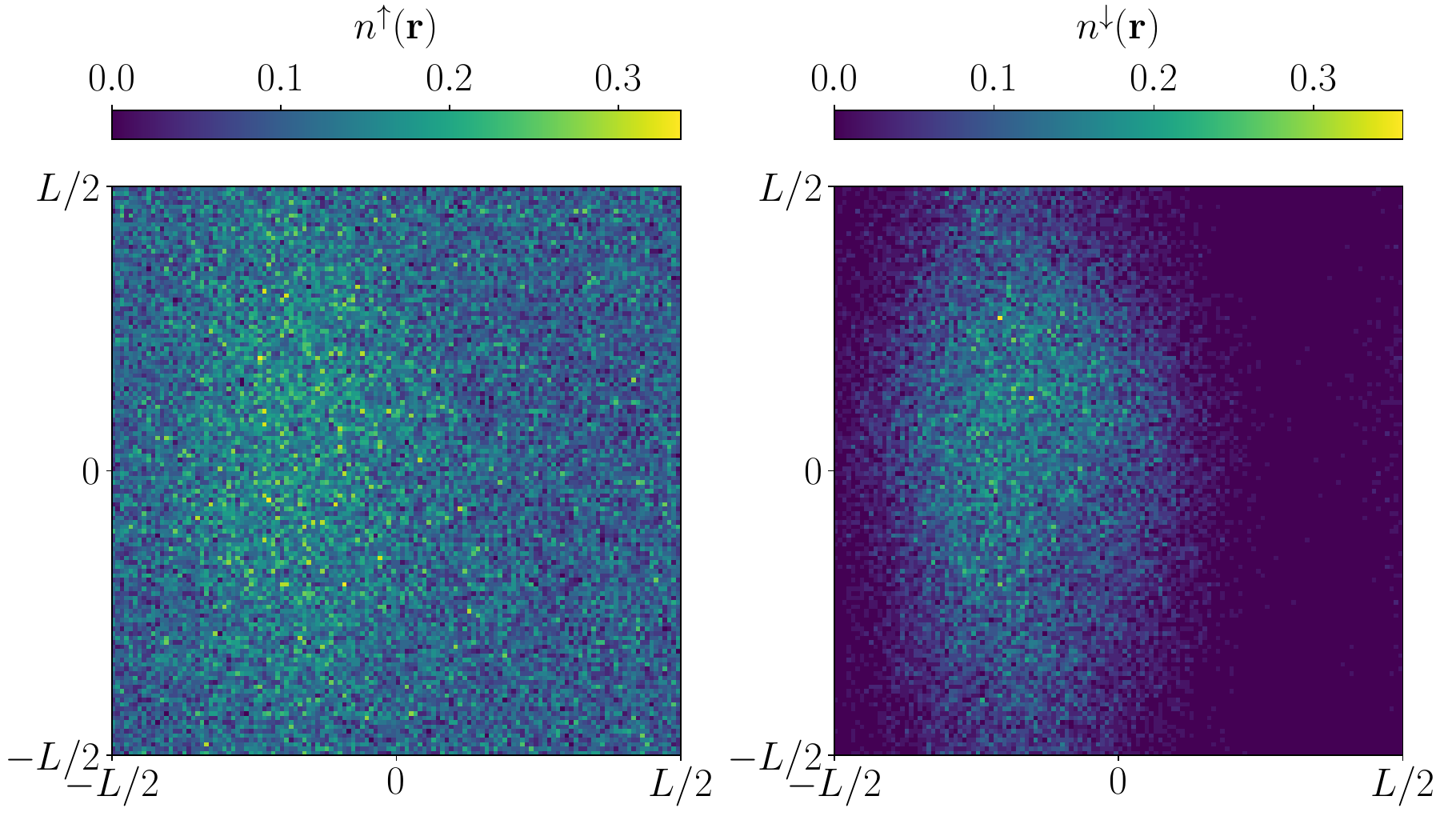}
    \caption{$v_0=0.4$}
  \end{subfigure}
  \hfill
  \begin{subfigure}[b]{0.45\columnwidth}
    \centering
    \includegraphics[width=\columnwidth]{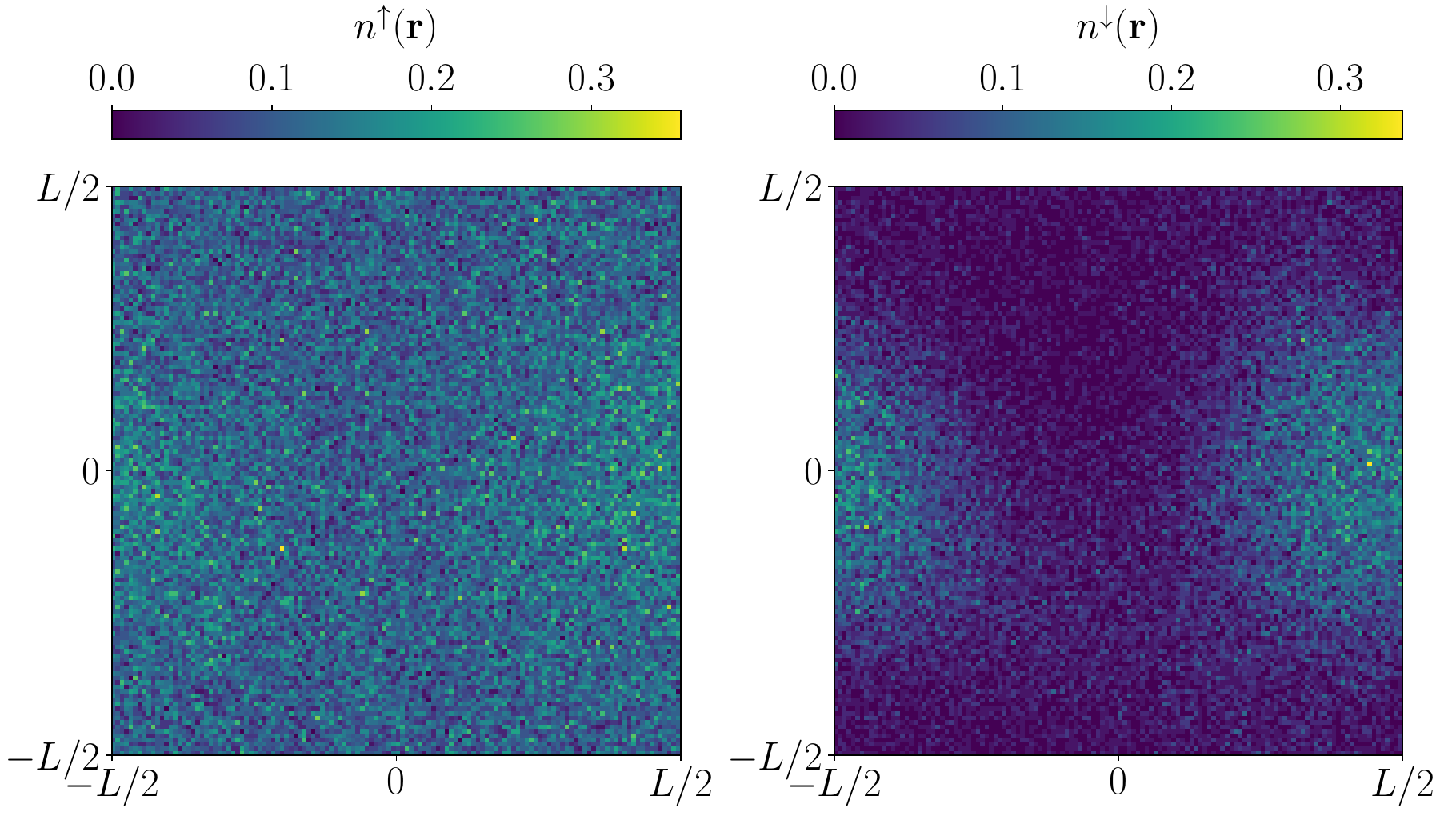}
    \caption{$v_0=0.45$}
  \end{subfigure}
  \hfill
  \begin{subfigure}[b]{0.45\columnwidth}
    \centering
    \includegraphics[width=\columnwidth]{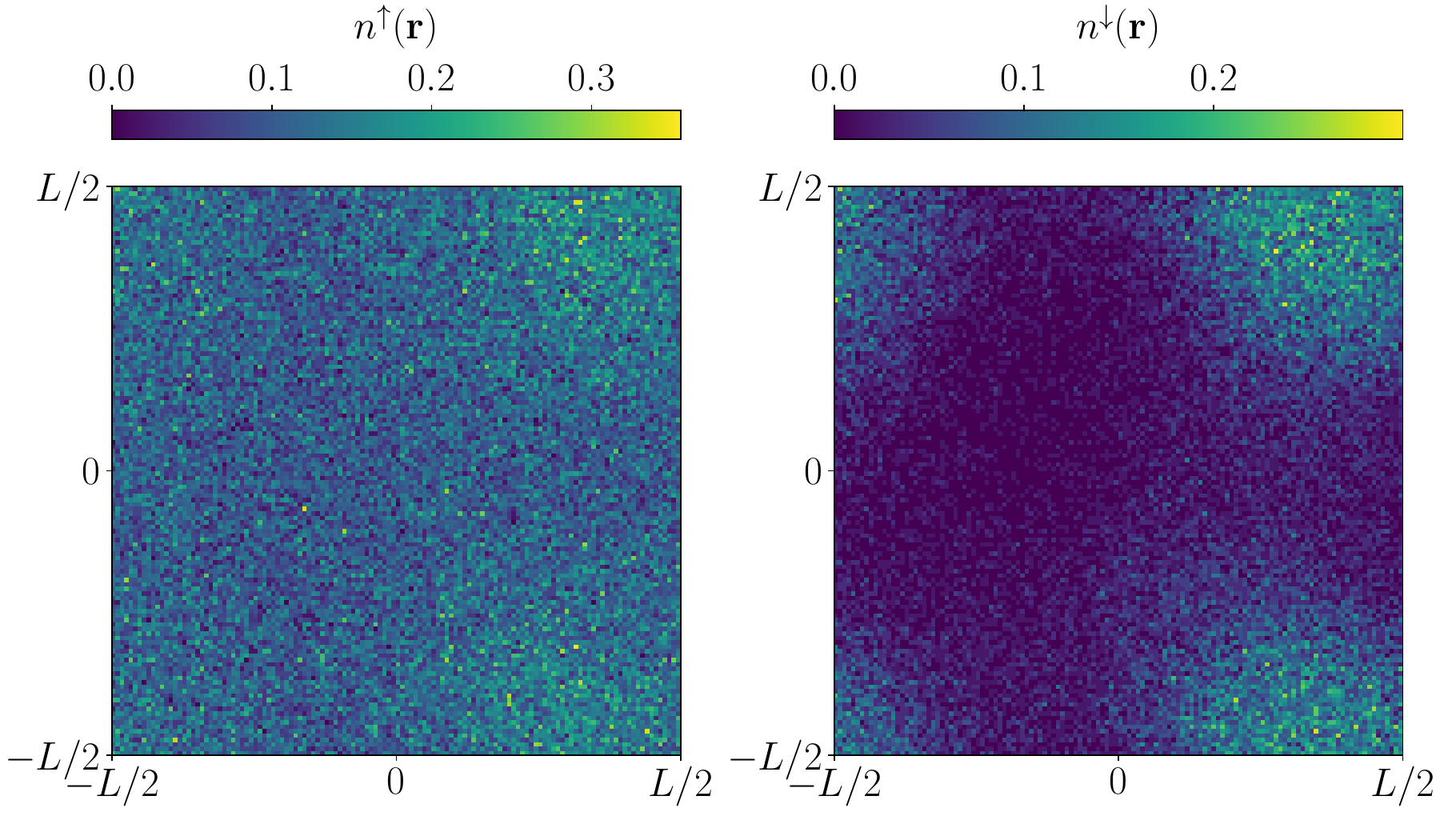}
    \caption{$v_0=0.5$}
  \end{subfigure}
  \caption{
    One-particle density of the 2D spin-imbalanced Fermi gas with $(N^\uparrow, N^\downarrow) = (25, 9)$ across the BCS-BEC crossover, plotted separately for different spin channels.
    For each subfigure, the plot on the left-hand side is the density of spin-up particles and the plot on the right-hand side is the density of the spin-down particles.
  }
  \label{fig:25_9-one_density}
\end{figure}

\begin{figure*}[ht!]
  \centering
  \includegraphics[width=\textwidth]{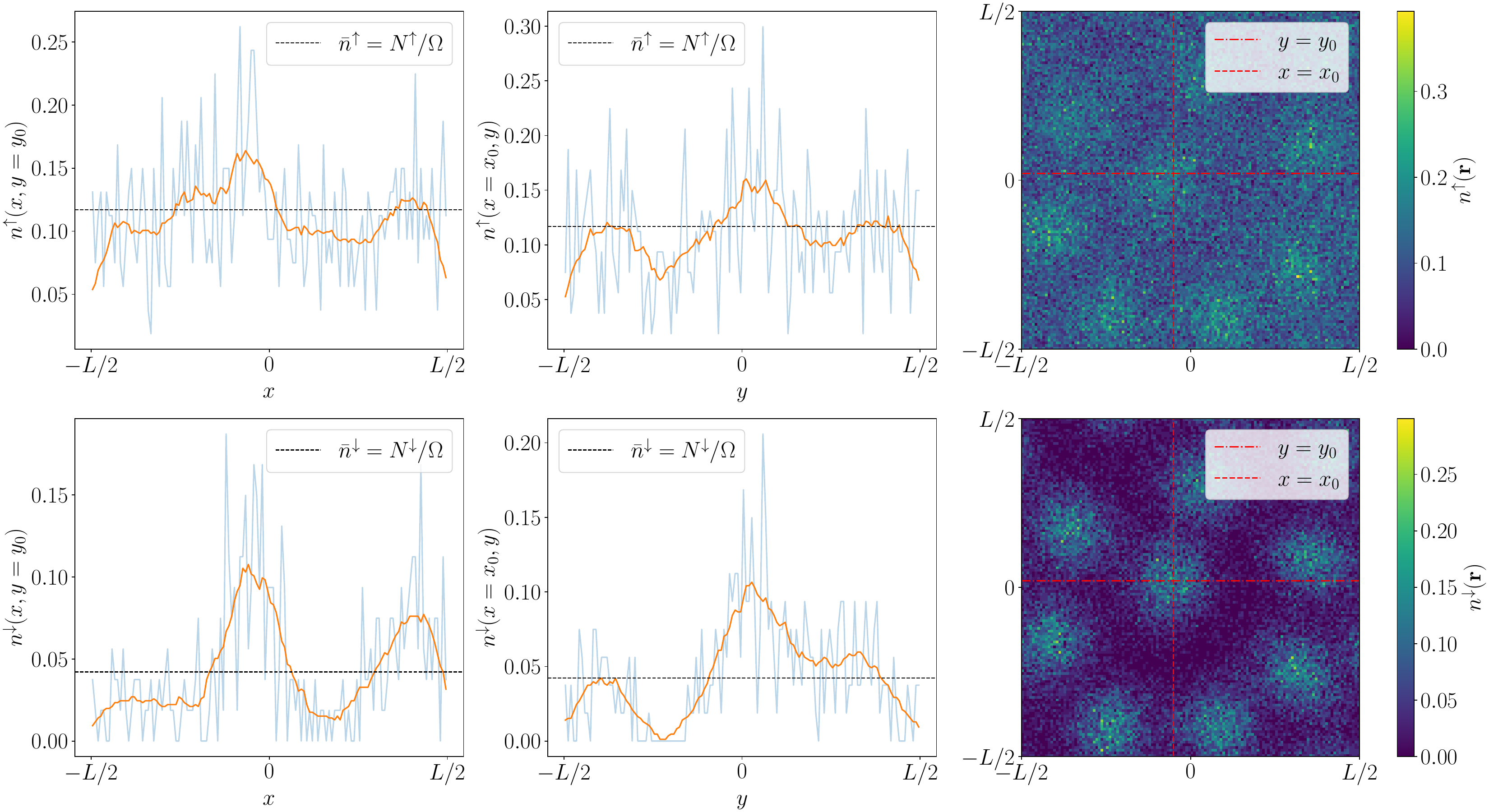}
  \caption{
    One-particle densities and their 1D projections for the $(25, 9)$ spin-imbalanced Fermi gas at $v_0=0.3$, plotted separately for different spin channels.
    The first row shows the majority spin (spin-up) density and the second row represent the minority spin (spin-down) density.
    The first column shows a horizontal slice of the 2D density with a fixed $y=y_0 \approx -0.06 L$, indicated by the \emph{red dash-dotted horizontal} line on the figure in the third column.
    The second column shows a vertical slice of the 2D density with a fixed $x=x_0 \approx 0.17 L$, indicated by the \emph{red dashed vertical} line on the figure in the third column.
    In the 1D projections, the black dashed horizontal line represents the average number density, $\bar{n}^\sigma = \frac{N^\sigma}{\Omega}$, of fermions of spin $\sigma$ in the simulation cell with volume $\Omega$.
    The blue curves are the actual values of the density across the cell, with the orange curves being running average of the blue curves.
  }
  \label{fig:v0.3-25_9-one_density_projection}
\end{figure*}

In this subsection, we present the one-particle real-space density of the 2D spin-imbalanced Fermi gas with $(N^\uparrow, N^\downarrow) = (13, 5)$ and $(N^\uparrow, N^\downarrow) = (25, 9)$, respectively, across the BCS-BEC crossover.
Results of the $(N^\uparrow, N^\downarrow) = (13, 5)$ system are shown in \cref{fig:13_5-one_density} and the results of the $(N^\uparrow, N^\downarrow) = (25, 9)$ system are shown in \cref{fig:25_9-one_density}.
The strength of interactions, $v_0$, are mentioned in the caption of each subfigure.
As our results of the real-space densities of the $v_0=0.1$ systems are uniform and show no qualitative difference compared to the densities of the $v=0.15$ systems, we chose to neglect the former and only show results for $v_0 \ge 0.15$ here.
In addition, we present the 1D projections of the real-space density of the $(25,9)$ SIFG at the crystalline phase ($v_0=0.3$) in Fig.~\ref{fig:v0.3-25_9-one_density_projection}.

